\documentclass[a4paper,10pt]{article}
\usepackage[pdftex, final]{graphicx}

\usepackage[T1]{fontenc}
\usepackage[utf8]{inputenc}
\usepackage[default]{lato}
\usepackage[a4paper, total={174mm,257mm}, left=18mm, top=20mm]{geometry}
\usepackage{natbib}
\usepackage{enumitem}
\usepackage{collect}
\usepackage{xspace}
\usepackage{colortbl}
\usepackage{tikz}
\usepackage{rotating}
\usepackage{amsmath}
\usepackage{longtable}
\usepackage{float}
\usepackage[breaklinks, colorlinks, linkcolor = blue, urlcolor = blue, citecolor = blue, anchorcolor = blue]{hyperref}
\usepackage{amssymb}
\usepackage[hang,flushmargin]{footmisc}

\usepackage{setspace}
\usepackage{caption}
\input{aas_defs}

\newcommand\reddps{Rubin-Euclid Derived Data Products\,\,}

\renewcommand{\arcsec}{$^{\prime\prime}$\xspace}

\newcommand{\contact}[1]{}
\newcommand{\contributors}[1]{ {\bf Contributors:} \textit{#1} }

\newcommand{\secref}[1]{\hyperref[#1]{Section~\ref{#1}}}

\newcommand{\tabref}[1]{\hyperref[#1]{Table~\ref{#1}}}
\newcommand{\figref}[1]{\hyperref[#1]{Figure~\ref{#1}}}

\newcommand{\fndccref}[1]{\hyperref[fnd:#1]{FND-\ref{fnd:#1}-CC}}
\newcommand{\recccref}[1]{\hyperref[rec:#1]{REC-\ref{rec:#1}-CC}}
\newcommand{\ddpccref}[1]{\hyperref[ddp:#1]{DDP-\ref{ddp:#1}-CC}}

\newcommand{\fndssref}[1]{\hyperref[fnd:#1]{FND-\ref{fnd:#1}-SS}}
\newcommand{\recssref}[1]{\hyperref[rec:#1]{REC-\ref{rec:#1}-SS}}
\newcommand{\ddpssref}[1]{\hyperref[ddp:#1]{DDP-\ref{ddp:#1}-SS}}     
                                                     
\newcommand{\fndlvref}[1]{\hyperref[fnd:#1]{FND-\ref{fnd:#1}-LV}}
\newcommand{\reclvref}[1]{\hyperref[rec:#1]{REC-\ref{rec:#1}-LV}}
\newcommand{\ddplvref}[1]{\hyperref[ddp:#1]{DDP-\ref{ddp:#1}-LV}}
                                                     
\newcommand{\fndgpref}[1]{\hyperref[fnd:#1]{FND-\ref{fnd:#1}-GP}}
\newcommand{\recgpref}[1]{\hyperref[rec:#1]{REC-\ref{rec:#1}-GP}}
\newcommand{\ddpgpref}[1]{\hyperref[ddp:#1]{DDP-\ref{ddp:#1}-GP}}
                                                     
\newcommand{\fndluref}[1]{\hyperref[fnd:#1]{FND-\ref{fnd:#1}-LU}}
\newcommand{\recluref}[1]{\hyperref[rec:#1]{REC-\ref{rec:#1}-LU}}
\newcommand{\ddpluref}[1]{\hyperref[ddp:#1]{DDP-\ref{ddp:#1}-LU}}
                                                    
\newcommand{\fndtrref}[1]{\hyperref[fnd:#1]{FND-\ref{fnd:#1}-TR}}
\newcommand{\rectrref}[1]{\hyperref[rec:#1]{REC-\ref{rec:#1}-TR}}
\newcommand{\ddptrref}[1]{\hyperref[ddp:#1]{DDP-\ref{ddp:#1}-TR}}
                                                     
\newcommand{\fndgeref}[1]{\hyperref[fnd:#1]{FND-\ref{fnd:#1}-GE}}
\newcommand{\recgeref}[1]{\hyperref[rec:#1]{REC-\ref{rec:#1}-GE}}
\newcommand{\ddpgeref}[1]{\hyperref[ddp:#1]{DDP-\ref{ddp:#1}-GE}}
                                                     
\newcommand{\fndanref}[1]{\hyperref[fnd:#1]{FND-\ref{fnd:#1}-AN}}
\newcommand{\recanref}[1]{\hyperref[rec:#1]{REC-\ref{rec:#1}-AN}}
\newcommand{\ddpanref}[1]{\hyperref[ddp:#1]{DDP-\ref{ddp:#1}-AN}}
                                                     
\newcommand{\fndscref}[1]{\hyperref[fnd:#1]{FND-\ref{fnd:#1}-SC}}
\newcommand{\recscref}[1]{\hyperref[rec:#1]{REC-\ref{rec:#1}-SC}}
\newcommand{\ddpscref}[1]{\hyperref[ddp:#1]{DDP-\ref{ddp:#1}-SC}}
                                                     
\newcommand{\fndslref}[1]{\hyperref[fnd:#1]{FND-\ref{fnd:#1}-SL}}
\newcommand{\recslref}[1]{\hyperref[rec:#1]{REC-\ref{rec:#1}-SL}}
\newcommand{\ddpslref}[1]{\hyperref[ddp:#1]{DDP-\ref{ddp:#1}-SL}}
                                                     
\newcommand{\fndpuref}[1]{\hyperref[fnd:#1]{FND-\ref{fnd:#1}-PU}}
\newcommand{\recpuref}[1]{\hyperref[rec:#1]{REC-\ref{rec:#1}-PU}}
\newcommand{\ddppuref}[1]{\hyperref[ddp:#1]{DDP-\ref{ddp:#1}-PU}}

\newcommand{\fndsoref}[1]{\hyperref[fnd:#1]{FND-\ref{fnd:#1}-SO}}
\newcommand{\recsoref}[1]{\hyperref[rec:#1]{REC-\ref{rec:#1}-SO}}
\newcommand{\ddpsoref}[1]{\hyperref[ddp:#1]{DDP-\ref{ddp:#1}-SO}}

\newcounter{fndcount} 

\newenvironment{fndenv}[1]
    {
    \refstepcounter{fndcount}  
    #1
    }

\definecollection{nfindings}

\newcommand{\fnd}[4]{
	\begin{fndenv}
	\vspace{5pt}
	\noindent\textbf{FND-\thefndcount-{#4}:~#2}\newline
	\label{fnd:#1}\noindent {#3}
	\vspace{5pt}
	\end{fndenv}
}
\newcounter{reccount}   

\newenvironment{recenv}[1]
    {
    \refstepcounter{reccount}  
    #1
    }

\definecollection{nrecommendations}

\newcommand{\rec}[4]{
	\begin{recenv}
	\vspace{5pt}
	\noindent\textbf{REC-\thereccount-{#4}:~#2}\newline
	\label{rec:#1}\noindent {#3}
	\vspace{5pt}
	\end{recenv}
}

\newcounter{ddpcount} 

\newenvironment{ddpenv}[1]
    { 
    \refstepcounter{ddpcount}  
    #1
    }

\definecollection{nddps}

\newcommand{\ddp}[5]{
	\begin{ddpenv}
	\vspace{5pt}
	\noindent\textbf{DDP-\theddpcount-{#5}:~{#2}}\newline
	\label{ddp:#1}\noindent{#3}\newline
    \textbf{Timescale:}~{#4}
	\vspace{5pt}
	\end{ddpenv}
}

\newcounter{authorcount} 

\newenvironment{authorenv}[1]
    { 
    \refstepcounter{authorcount}  
    #1
    }

\definecollection{nauthors}
\newcommand{\ddpauthor}[3]{
    \begin{authorenv}
    \vspace{-22pt}
	\label{author:#3}\noindent{\sc {\href{https://orcid.org/#1}{\small{#2} \hskip2pt \includegraphics[width=8pt]{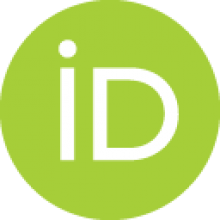} }}}\vspace{-27pt}\newline
	\end{authorenv}
}
\newcommand{\ddpaffiliation}[1]{
	{\small{\it {#1}}}\newline
}

\definecolor{cb-cats}{RGB}{153,221,255}
\definecolor{cb-phz}{RGB}{238,136,102}
\definecolor{cb-stamps}{RGB}{88,207,173}

\graphicspath{{./}{figures/}}

\date{}   
\title{}  
\author{} 

\begin{document}

\begin{titlepage}
   \begin{center}
      \tikz[remember picture,overlay] \node[opacity=1,inner sep=0pt] at (current page.center){\includegraphics[width=\paperwidth,height=\paperheight]{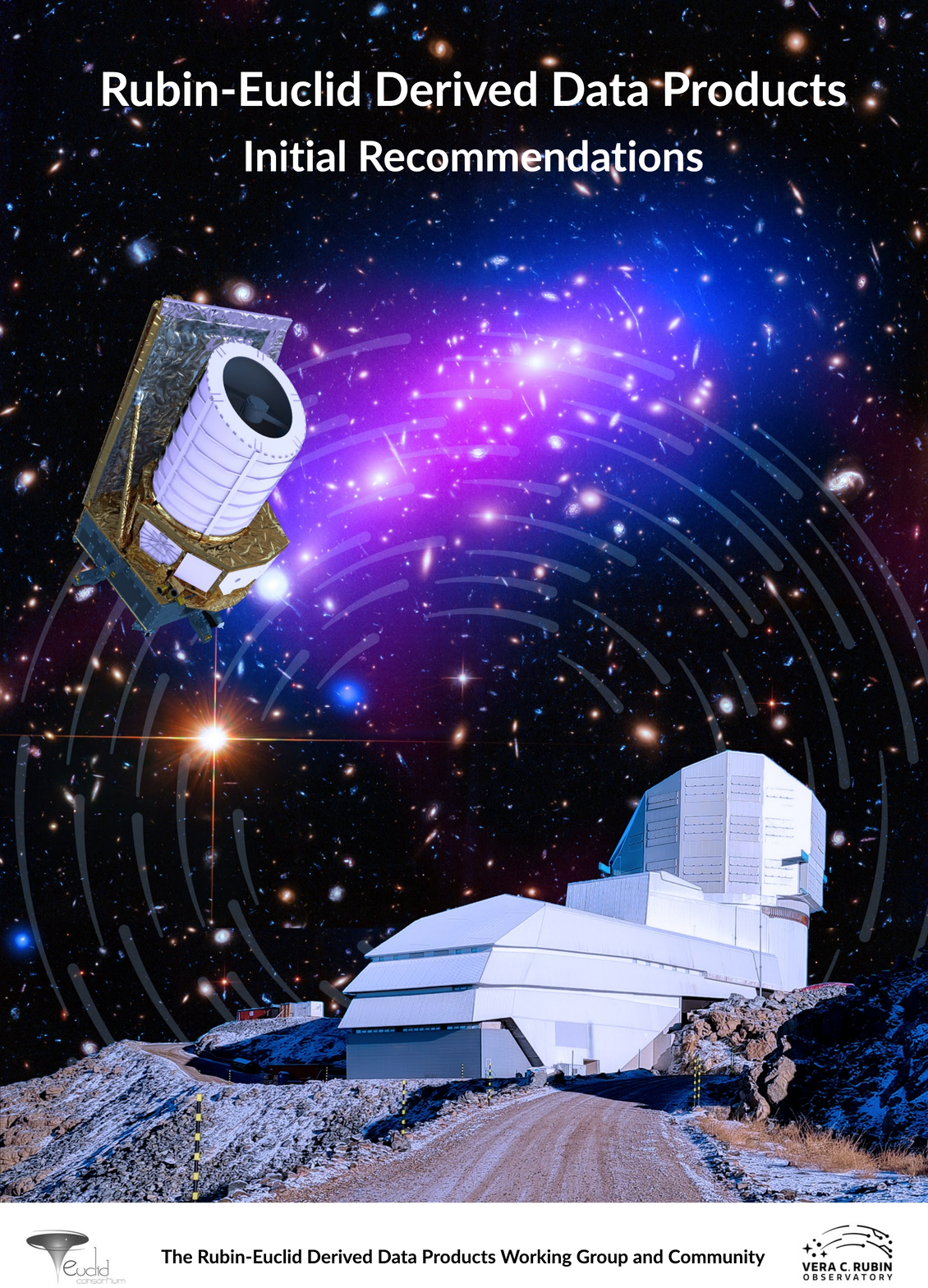}};
 \vspace*{3.5cm}
 \label{TitlePage}
 {\let\newpage\relax}
 \vspace*{\fill}
   \end{center}
\end{titlepage}

\section*{The Rubin-Euclid Derived Data Products (DDPs) Working Group}
\label{sec:workinggroup}

\textbf{Date}: 7 October 2022 
\newline
\textbf{Version}: V1.1\\

\noindent This report is the result of a joint discussion between the Rubin and the Euclid scientific communities (the DDP Community) coordinated by the DDP Working Group under the guidance of the DDP Working Group oversight committee.


\textbf{Working Group Members:} [dual project membership is indicated with an asterisk$^*$]
\vspace{-0.3cm}
\begin{itemize}
    \item {\bf Chairs:} Leanne Guy for Rubin \& Jean-Charles Cuillandre for Euclid
  
    \item {\bf Rubin:} Yusra Alsayyad, Etienne Bachelet, Manda Banerji$^*$, Franz Bauer, Jim Bosch, Tom Collett$^*$, Siegfried Eggl, Catherine Heymans$^*$, François Lanusse$^*$, Peter Melchior, Dara Norman, Michael Troxel
  
    \item {\bf Euclid:} Eric Aubourg$^*$, Herv\'e Aussel, Christopher Conselice$^*$, Adriano Fontana$^*$, Henk Hoekstra$^*$, Isobel Hook$^*$, Konrad Kuijken$^*$, Joe Mohr, Michele Moresco, Reiko Nakajima, St\'ephane Paltani, Daniel Stern$^*$
    
    \item {\bf Guest specialists from the community:} Benoit Carry, Annette Ferguson$^*$
\end{itemize}

\textbf{Corresponding authors:} Leanne Guy (\href{mailto:leanne.guy@lsst.org}{leanne.guy@lsst.org}) \& 
Jean-Charles Cuillandre (\href{mailto:jc.cuillandre@cea.fr}{jc.cuillandre@cea.fr})
\newline

\textbf{DDP Working Group Oversight Committee Members:}
\vspace{-0.3cm}
\begin{itemize}
    \item {\bf Rubin:} Robert Blum, Phil Marshall, \v{Z}eljko Ivezi\'{c}
    \vspace{-0.2cm}
    \item {\bf Euclid:} Yannick Mellier, Jason Rhodes, Ren\'e Laureijs
\end{itemize}

\vspace{9cm}
\textbf{Cover image credits:} ESA/ATG medialab (\href{https://sci.esa.int/web/euclid/-/artist-s-impression-of-euclid-2}{spacecraft}); NASA, ESA, CXC, C. Ma, H. Ebeling and E. Barrett (University of Hawaii/IfA), et al. and STScI (\href{https://hubblesite.org/contents/news-releases/2009/news-2009-17.html}{background}); Rubin Obs./NSF/AURA (observatory)
\vspace{0.5cm}
\newline
\textbf{Cover page credits:} E. Acosta, Rubin Obs/NSF/AURA

\thispagestyle{empty}
\newpage

{
  \hypersetup{linkcolor=black}
  \tableofcontents
  \thispagestyle{empty} 
}
\newpage

\addcontentsline{toc}{section}{Executive Summary}
\section* {Executive Summary} \label{sec:summary}

The Vera C. Rubin Observatory Legacy Survey of Space and Time (imaging) and the Euclid survey (imaging and spectroscopy) will each deliver groundbreaking astronomical datasets over this decade in the optical and near-infrared. Both surveys will map thousands of square degrees of sky from the ground and space respectively, with an overlap area of approximately 9000 square degrees at high galactic latitudes. The combination of Euclid's high spatial resolution imaging in the optical and near-infrared photometry with Rubin's densely sampled deep multi-band optical imaging will greatly enhance the science yield of both surveys.

The work of the \textit{Rubin-Euclid Derived Data Products Working Group} presented in this report was focused on designing and recommending an initial set of \textit{Derived Data products} (DDPs) that could realize the science goals enabled by joint processing. 
All interested Rubin and Euclid data rights holders were invited to contribute via an online discussion forum and a series of virtual meetings.
Strong interest in enhancing science with joint DDPs emerged from across a wide range of astrophysical domains: Solar System, the Galaxy, the Local Volume, from the nearby to the primaeval Universe, and cosmology. The working group operated by consensus; the chairs worked with the relevant experts to review and consolidate all proposed DDPs to recommend the initial short list presented herein.

Our initial set of recommended DDPs can be broadly grouped into two categories:
\vspace{-0.4cm}
\begin{enumerate}[noitemsep]
\item Cross-cutting DDPs (5), which will enable a wide range of complementary science goals,
\item Science-specific DDPs (58), which will enhance the science yield for a specific science case.
\end{enumerate} 

\vspace{-0.2cm}
Associated with each of the science-specific recommended DDPs is a description of the science case it enables, an estimate of the algorithmic and computing needs, and an indication of the timescale on which the DDP would be scientifically useful. The two most impactful cross-cutting DDPs were found to be multi-band Rubin+Euclid photometric catalogs, and the exchange of image stamps of small overlapping areas of the sky. The former will enable an improved estimate of photometric redshifts, enhancing all areas of galactic and extragalactic science, and the latter would enable key scientific investigations for transient, strong lensing, and drop-out science. 
At the sensitivity of Rubin and Euclid, blending is a noticeable systematic for all measurements. Joint pixel-level modeling  will significantly improve the detection and deblending of sources in both Rubin and Euclid images, and result in superior photometric catalogs. Moreover, Rubin's model for correcting differential chromatic refraction can be improved by a factor of $\approx$\,2 by incorporating pixel-level morphological information from Euclid’s high resolution VIS band, which would drive improvements in essentially all downstream DDPs. We recommend that DDPs that will enable transient science should materialize on a short timescale, e.g. 24\,hr, and should be based on a fast joint processing of the data, whereas more complex DDPs would fit better in the context of the annual Rubin data release scenario.

A key factor in maximizing the impact of Rubin-Euclid DDPs is the coordination of each survey's observing strategy;  maximizing the spatial and temporal overlap of the two surveys will enhance almost all science domains. Openly sharing all imaging pixel data over a common small area of the sky across both projects will enable the early development of algorithms and  software that will be beneficial to all DDPs. We recommend for this purpose the proposed Rubin Deep Drilling Field over the 23 square degree Euclid Deep Field South.

As both projects approach first light, we recommend the prompt creation of working groups and task forces charged with producing the DDPs recommended in this report. This should include a task force to explore the detailed software and computing requirements, a joint simulations group, and a joint implementation group to realize the recommendations of this report. Furthermore, we strongly advocate for a tiered approach over the lifetime of both surveys to producing the DDPs.

Finally, we note that the model for defining DDPs presented in this report can be thought of as a pathfinder to extend the Rubin-Euclid model of jointly derived data products into the Rubin-Euclid-Roman domain.
\newpage

\section{Introduction to the Rubin-Euclid Derived Data Products}\label{sec:intro}

\noindent\contributors{\hyperref[author:lguy]{Leanne Guy (WG)}, \hyperref[author:jcuillandre]{Jean-Charles Cuillandre (WG)}}

\subsection*{Nature and goals of the DDP effort}

Following a series of reflections over the years on the scientific synergy between Rubin-LSST and Euclid \citep{Rhodes_2017, capak/etal:2019, Capak_2019b}, the Vera C. Rubin Observatory Director and the Euclid Consortium Board launched an effort mid-2020 acknowledging that both the Rubin and the Euclid science communities would benefit from the Rubin and Euclid datasets being jointly processed to produce shared ``Derived Data Products'' (DDPs). They put in place a DDP Working Group that should recommend an initial set of DDPs,  which would be shared promptly and simultaneously with both the Euclid Consortium and all Rubin Observatory data rights holders for scientific use in a way that protects the unique science of each collaboration. The DDP Working Group's focus is on designing DDPs; it is a standing committee that can recommend revisions to DDPs or further DDPs as both the Euclid and Rubin survey progress. It is not however the group that will decide who makes the DDPs, where they are made, how they are made, nor what funding mechanism shall pay for that effort. This group also does not focus on issues of data rights nor potential scientific collaborations between Rubin/Euclid. The DDP Working Group consists of an equal number of representatives from the Vera Rubin Observatory data rights holders and the Euclid Consortium.

The initial charge to the DDP Working Group consisted of setting up a consultation open to all interested Rubin and Euclid data rights holders, and focused on gathering community input for potential DDPs. 
Based on the input, the DDP Working Group proceeded with designing an initial set of desired DDPs that could be shared with both the Euclid Consortium and the LSST Science Community in a way that protects the unique science of each collaboration, and is consistent with both communities’ data policies. This report presents those DDPs and outlines the scientific justification for each, quantifies its impact, and issues an initial set of recommendations to each project's respective management. 
If approved, the respective projects must then come to an agreement about where, how, by whom, on what time scale, and with what funding the DDPs will be created.

While the primary goal of the present effort is to define an initial set of derived data products that both sides would like to see shared during the proprietary period, an equally important goal is to create a lasting framework for the justification and definition of derived data products and the inter-project coordination required to produce them, without regard to whether the data was in the proprietary period.

\subsection*{DDP definition}

It became apparent early in the process that a precise definition of a \textit{Derived Data Product} was needed in order to define the scope and formulate ideas compatible with the charge. On the Rubin side, the data policy states that in the case of large collaborations sanctioned by Rubin, joint DDPs may be created and may include proprietary data if it would significantly enhance the science enabled for the Rubin community. This means sharing certain data between collaborations so as to balance the science opportunity of the Rubin community with Rubin community science priorities. For the purposes of a Rubin-Euclid joint DDP program, as envisioned in this document, the joint data products would be produced under the Rubin Data Rights policy DPOL-602: In-kind Data Sets and Derived Data Products \citep{RDO-013}.

Inspired by the Rubin Data Policy DPOL-601, which defines an LSST derived data product (DDP) as being a data product derived from LSST proprietary data, but that cannot be used to recreate any proprietary LSST data product(s), we defined a Rubin-Euclid DDP as follows: 

``{\it Data are considered a Rubin-Euclid Derived Data Product (Rubin-Euclid DDP) if the data product is derived from Rubin and Euclid proprietary data for the joint benefit of those two communities while preserving aspects of each community’s “proprietary” science.}''

For example, DDPs that match only those sources visible above 5\,$\sigma$ in the Euclid VIS and NISP will protect the unique science that derives Rubin's from unprecedented depth. Similarly, Euclid unique science that derives from  the high spatial resolution (VIS and NISP imaging) is protected by exporting photometric measurements of isolated or deblended Rubin sources only. 

Furthermore, it is important to stress that, while access to pixels and some ancillary data will be needed to produce the DDPs, it is the DDPs themselves that will be shared to both collaborations. The Euclid consortium will not obtain full access to all of the Rubin pixels, nor vice versa.

While the aim is to share a common set of fully open DDPs between the two collaborations, special cases that enhance a given science topic for either project might require specific measurements to be made available separately for each collaboration, for example Rubin image cut-outs (vignettes) for the drop-out science for Euclid and vice versa for some of the Rubin transient science cases. Mutual agreements on specific scientific niches that do not necessitate a dataset-wide effort could be proposed to enhance targeted scientific return.  

\subsection*{Nature of the consultation}

A traditional in-person workshop over several days in late 2020 was initially envisaged as the path forward for both projects to lay down together the foundations for an initial set of recommended DDPs.
The global Covid-19 pandemic however meant that such an in-person meeting was not possible. 
Instead, we adopted a novel approach using an online discussion forum open to both Rubin and Euclid members. 
The \href{https://community.rubin-euclid-ddp.org/}{Rubin-Euclid Derived Data Products Forum} enabled extended virtual continuous and asynchronous discussion across all time zones and will continue to support future exchanges for years to come as the DDP Working Group continues to tackle future charges. 
At the core of the DDP Working Group charter (\secref{sec:ddpcharter}) was ensuring that both communities were consulted together, and at large, to gather science-driven input on the desired DDPs.
This modified format not only enabled us to achieve this, but turned out to be more inclusive, enabling far more scientists to engage than an in-person workshop would have allowed for. 
Additionally, we were able to  avoid highly disruptive and costly international travel, and drastically reducing the carbon footprint and waste associated with in-person meetings. 
The forum is based on the \href{https://www.discourse.org/}{"Discourse - Civilized Discussion"} platform, the same platform adopted for the \href{http://community.lsst.org}{Rubin Community forum} 

Due to the challenges posed by the pandemic, the online discussion was extended over a 5-month period to ensure everyone had a chance to participate.  
Two initial virtual kick-off meetings were held on 24 and 26 February 2021, with approximately 200 attendees in total. The aim of these meetings was to set the context and timeline for the discussion, which continued to evolve until June 2021. Over the course of the discussion, 350 members have joined the forum and many have contributed, leading to nearly 50 independent science discussions distributed across the main science themes of the forum: {\it Solar System, Milky Way, Transients, Nearby Universe, AGN, Galaxy evolution, Clusters of galaxies, Galaxy clustering, Strong lensing, Weak lensing, Primaeval Universe}.

An additional prerequisite to producing DDPs is for both Rubin and Euclid to observe common parts of the sky. We explore how to maximize the spatial and temporal overlap between the respective wide and deep surveys in \secref{sec:observing} (surveys parameters).

The various scientific topics explored during the discussion are grouped into the following categories, each represented by a two-letter acronym throughout this whole document. It is used together with a unique identification number assigned to each Finding (FND), Recommendation (REC), or Derived Data Product (DDP) to keep track of the scientific origin, e.g. FND-10-SS is the tenth finding of the report, originating from the Solar System discussion. 

\begin{center}
\begin{tabular}{ c l } 
    CC & Cross--Cutting (Findings, Recommendations and DDPs applicable across multiple scientific categories)\\
    SO & Surveys Optimisation (maximization of both spatial and temporal overlap)\\
    SS & Solar System\\
    LV & Local Volume (Milky Way halo \& Local Group)\\
    GP & Galactic Plane\\
    LU & Local Universe\\
    TR & Transients\\
    GE & Galaxy Evolution\\
    AN & Active Galactic Nuclei\\
    SC & Static Cosmology (weak lensing, clustering, clusters)\\
    SL & Strong Lensing\\
    PU & Primaeval Universe \\
\end{tabular}
\end{center}

\subsection*{The 5 high-level questions to help define a DDP}

To focus the community effort to contribute in a consistent manner to the DDP discussion, the DDP Working Group posed 5 high-level questions to guide the discussion. The idea was for any interested Rubin or Euclid scientist to develop answers to these 5 questions and post them on the forum in the relevant category or categories to stimulate discussion. The forum posts could either be a new topic if clearly disconnected from others, or an addition to an ongoing joint discussion between scientists within a developing topic. 
A second virtual meeting took place in June 2021 where Working Group members summarized the outcome per science category based on this community input. All presentations are posted on the forum. 
This material became the basis for the science Sections authored in this report by Working Group members; they are the source of the initial set of recommendations in this DDP Working Group report. 
All 10 science Sections in this report follow a similar structure.\\

\noindent The 5 high-level DDP questions are:
\begin{itemize}
\item {\bf Q1 -- Science Case} : Considering DDP use cases from joint pixel-level processing, or input prior information exchange (list driven), or catalog-level processing, what science would be enabled or enhanced with a Rubin-Euclid joint processing?
\item {\bf Q2 -- Nature of the Derived Data Products} : Which survey specific data products are needed to realize your science cases outlined in Question 1?
\item {\bf Q3 -- Algorithms} : Do the algorithms already exist to carry out the necessary processing to deliver the data products described in Question 2?
\item {\bf Q4 -- Computing resources} : What level of critical resources would be needed to develop and operate the approaches and data volume described in Questions 2 and 3?
\item {\bf Q5 -- Timescale} : On what timescales would the DDPs described in Question 2 be useful?
\end{itemize}

This report proposes recommendations related to both the development of DDPs and joint processing in general, and proposals for specific DDPs that were reviewed and accepted by the working group. The following 5 Sections focus on the cross-cutting elements from the 10 science Sections (findings, recommendations, derived data products, summary table), and the surveys optimization Section.

\section{Cross-Cutting findings}\label{findings}

In this Section we detail the cross-cutting findings that emerged from the joint Rubin-Euclid discussions on maximizing the scientific output of both projects. 
These findings lead to cross-cutting recommendations (\secref{recommendations}) and then to recommended DDPs (\secref{ddps}).

\fnd{ddp}{Broad scientific impact of shared products derived from a joint dataset}{The most significant finding is that the combination of the Rubin data with the Euclid data will result in significant scientific gains for both missions independently. This can be achieved through various means of varying complexity, implying a broad timescale for implementation all the while protecting the science that is unique to each project. This top finding implies we should adopt a tiered approach to DDPs. This holds for all astrophysical fields explored throughout the 5-month-long open forum discussion across the Rubin and the Euclid scientific communities.}{CC}

\fnd{keys}{Key factors in optimizing the science harvest}
{Optimization of the science harvest for both surveys requires: {\bf a)} Coordination of each survey's observing strategy, e.g. maximizing the overlap area and matching the depths for wide-field surveys, and synchronizing the observation of deep fields;  {\bf b)} Joint processing at the pixel level, e.g. to produce image cutouts for transients or high-$z$ drop-outs, and joint photometric catalogs; {\bf c)} Exploration of solutions for data co-location so that codes for joint pixel-level processing can easily and tightly couple to both survey's data sources, and to enable analysis of the resulting DDPs by both communities independently.}{CC}

\fnd{catphot}{Highest impact cross-cutting DDPs}{The generation of precise multi-band photometric catalogs through the combined analysis of Rubin and Euclid datasets in particular will result in improved estimation of photometric redshifts, which will significantly enhance all extragalactic science areas for both missions. Image cutouts exchanges (pixels) on tiny areas of the sky will enable key scientific investigations for  transient, strong lensing, and drop-out science.}{CC}

\fnd{blending}{Impact of blending}{At the sensitivity of Rubin and Euclid, blending is a noticeable systematic for all measurements, but it affects Rubin more strongly. Joint modeling at the pixel level will significantly improve the detection and deblending of sources in both Rubin and Euclid images,  and result in superior photometric catalogs.}{CC}

\fnd{transients}{Transient science}{Unique transient science can be enabled by coupling the Rubin-LSST and the Euclid surveys, pending in particular coordinated observations of deep fields. Of key importance, the timely reporting of astrometry and activity of transients will enable rapid follow up.}{CC}

\fnd{timescale}{DDP production timeframes}{The timeframes in which DDPs would be relevant and hence should be produced varies and can be broadly grouped in to 3 periods: {\bf a)} Short term (24\,hrs), this timeframe is particularly important for transient alerts; {\bf b)} Medium term (following 1-3 years of Rubin/LSST data). The majority of the DDPs recommended in this report can be produced on this time scale; {\bf c)} Following 2 to 3 years of data taking. In all cases, where possible, DDPs should be investigated as soon as Rubin and Euclid observe a common part of the sky to optimize scientific exploitation of the data.}{CC}

\fnd{software}{Need for dedicated software for joint processing}{Complex dedicated software will be needed to jointly process data from Rubin and Euclid. The nature of the differences in the telescopes and instruments pose challenges for the joint pixel processing. To aid the adoption of DDPs for high-precision science cases, the algorithms and software package used in their creation will go through an extensive reviewing and approval process before implementation. Also, openly sharing imaging data across both projects over a small sky area would benefit to software development while also allowing both communities to investigate delivered DDPs in depths (from source pixel to the DDP) and propose paths for further improvements in future releases.}{CC}

\fnd{dcr-improvements}
{Differential Chromatic Refraction (DCR) model improvements}
{Strong spectral features such as broad emission lines, Lyman/Balmer breaks or steep continuum gradients lead to unique relative astrometric shifts as a function of airmass, primarily affecting Rubin's u and g bands, due to differential chromatic refraction (DCR). Rubin plans to model the underlying spectrum at the pixel level and correct for these shifts in individual exposures, leading to improved spatial \textit{and} spectral resolution (effectively permitting sub-band images to be constructed). This will in turn improve photometric accuracy and enhance photometric redshifts and SED modeling by appreciable factors. Using the pixel-level morphological information from Euclid's high resolution VIS band as priors in conjunction with the individual ground-based resolution ugrizy images from Rubin at different airmasses could help to improve Rubin's DCR correction model by up to a factor of $\approx$\,2 for all overlapping areas compared to using the Rubin data alone. This would drive improvements in essentially all downstream DDPs (in particular the cross-cutting \ddpccref{listdriven-catalogs}, \ddpccref{mw-fp-catalogs}, \ddpccref{mw-db-catalogs}). The DCR model can only be improved over the region where the VIS data are available and is fundamentally rooted in joint-pixel level processing.
}{CC}


\section{Cross-Cutting Recommendations} \label{recommendations}

In this Section we detail the cross-cutting recommendations deriving from the cross-cutting findings and some of the recommendations from the individual science Sections.

\rec{resources}
{Instigate a computing task force to explore details of producing DDPs in a tiered approach}
{Production of DDPs will require access to both datasets, both of which are at the petabyte scale, significant computing resources, and dedicated algorithm development. Considering the complexity of the task, adopting a tiered approach to the development of DDPs will enable science from day one with pragmatic approaches and ensure the very best DDPs are eventually produced through a more complex path. The recommendation is to set up a
computing task force to consider computing \& infrastructure requirements for the creation of all recommended DPPs. This taskforce should build upon the efforts of \citet{chary/etal:2020}.}{CC}

\rec{implementation}{Define timescales for investigating and implementing DDPs}
{In order to meet the scientific expectations, DDPs ought to be implemented as soon as Rubin and Euclid observe a common part of the sky, as early as 2023 (\recsoref{surveys-5}). This means starting the investigation on methods and infrastructure (\recccref{resources}) now. 
DDPs that will enable transient science should materialize on a short timescale, e.g. 24\,hr, and should be based on a fast joint processing of the data, whereas more complex DDPs would fit better in the context of the annual Rubin data release scenario. (\recsoref{surveys-4}).}
{CC}

\rec{sims}
{Instigate a simulations group}
{Fund and support the development of a joint simulations group to better quantify the scientific gain of many of the recommended DDPs (\recccref{DCR-joint-model}, \reclvref{local-volume-1}, \recluref{lusim}, \recscref{static-cosmo-sims}, \recslref{sl-sims}, \recpuref{evaluate}). Additional Euclid participation in an ongoing Rubin/Roman joint simulation effort would satisfy this recommendation; appropriate augmentations to that effort on all sides should be explored.
This effort should build on previous efforts in this area, such as those of \cite{chary/etal:2020}.}{CC}

\vspace{1cm} 
\rec{implementationwg}{Instigate a shared implementation group}
{Complementing the current DDP working group in charge of gathering community input and formulating recommended DDPs, a new group under similar management ought to be created to advise on the implementation phase. This group would oversight targeted activities by the groups recommended above (\recccref{resources}, \recccref{sims}).}{CC}

\rec{rec:data-access}{Enable access to survey datasets}{Both the Rubin and Euclid data products will be in the petabyte size range. Effective joint analysis will require the datasets to be co-hosted and for resources to exist to support joint processing at the location of the data.}{CC}  

\rec{software}{Fund an effort to explore the development of software for joint pixel processing}{Given the very different nature of the two telescopes and instruments, there are many challenges to combining data at the pixel level. 
This recommendation is to initiate an investigative effort to assess the requirements on software for joint pixel level processing to produce the DDPs reported herein, and on the timescales indicated. This effort should build on previous efforts in this area, such as those of \cite{chary/etal:2020}.
}{CC} 

\rec{blending}
{Establish the methodology for joint detection and deblending}
{Initiate studies on the best way to jointly detect, model, and deblend sources. These studies should make use of both  simulations (\recccref{sims}), and 
images taken in a region that reaches or exceeds the expected final wide survey depths and is designated for full open sharing by both imaging surveys, such as the EDF-South ( \recccref{shared-data}, \recgeref{De-blender} \recanref{AGNdeblender}).
If joint models are deemed to infringe on the distinctive advantages of one survey over the other, they can either be withheld from the DDP data releases or degraded, e.g. in sensitivity or spatial resolution, to match the individual surveys.}{CC}

\rec{widesurveys}
{Maximize the overlap area of wide surveys and near-simultaneous observations on deep fields}
{The production of DDPs requires that the two wide surveys observe the same area of the sky; it is therefore beneficial to try to maximize the overlap area (\recsoref{surveys-1}, \recsoref{surveys-2}, \recgpref{galactic-plane-1}). Tweaking the LSST scheduler to maximise near-simultaneous observations of the two southern Euclid deep fields will enhance many science domains (\recsoref{surveys-2}, \recssref{solar-system-2}, \recgpref{galactic-plane-1}, \rectrref{rec-transients-1}, \recanref{AGN-SimultaneousObs}).}
{CC}

\rec{shared-data}{Openly share photometric observations of a deep field}{For software development and calibration purposes, an overlapping sky area observed to depths exceeding the nominal limits of the Euclid and Rubin wide surveys is critical. The eventual production of all DDPs will benefit from a photometric dataset (imaging) of a small area of the sky shared openly across both Euclid and Rubin. 
We recommend 1) Rubin to select the Euclid Deep Field South as the fifth Deep Drilling Field, and 2) openly sharing the joint photometric observations considering the mutual engagement
(\recsoref{surveys-3}, \ddpscref{static-cosmo-ambitious-photom-deep}).}{CC}

\rec{DCR-joint-model}{Establish the methodology for using Euclid VIS band data to improve Rubin's DCR correction model}
{Using Euclid's high resolution VIS band as priors in conjunction with Rubin's ugrizy images taken at different airmasses will improve Rubin's DCR correction model for all overlapping areas (\fndccref{dcr-improvements}). Work will be required to understand how to incorporate Euclid morphological and spectral input into the Rubin DCR correction model. Initially, when just a few Rubin images exist, the DCR model will be poor and incorporating the Euclid morphological information along with relatively agnostic spectral priors should lead to immediate gains in Rubin u and g-band imaging quality. As more Rubin data accrues, the individual Rubin images will add statistical weight to constrain more accurately the underlying spectral shape within each pixel and refine the Rubin images further. We recommend this to be investigated initially over the EDF-South DDF (\recccref{shared-data}) as a demonstration test case and, if a significant improvement is demonstrated, ultimately applied to the entire overlapping wide area in the long term.}{CC}

\section{Cross-Cutting Derived Data Products} \label{ddps}

In this Section we make recommendations for cross-cutting \reddps requested across several science discussions. These are generic DDPs that will enhance science across both collaborations in a balanced way, and as such they cover several recommended DDPs across several science Sections (c.f. dynamic links hereby). All individual recommended DDPs are first listed in a synthetic table (\secref{ddpstable}), each of them being fully justified in its science Section.


\ddp{listdriven-catalogs}{Multi-band Rubin+Euclid list-driven photometry catalogs}{Photometric redshifts are at the heart of the high-profile cosmology science cases of both surveys, with stringent accuracy requirements that cannot be met using a combination of two independent photometry catalogs. At a minimum, a list-driven photometry source exchange is required in association with photometric extraction software and apertures configuration (an enabling element not part of the released DDP itself). This outsourcing of forced photometry (galaxies = \ddpscref{static-cosmo-euclid-source}$+$\ddpscref{static-cosmo-rubin-source}, but this also applies to point sources = \ddplvref{lv-catalog-list}$+$\ddpluref{lu4}$+$\ddppuref{IRdetcat}) will result in separate catalogs, each consortium releasing the photometry outsourced to the other consortium. Euclid would provide a list of sources detected in its detection bands (VIS and NISP) and obtain Rubin's u,g,r,i,z photometry, and Rubin would obtain Euclid's Y,J,H near-infrared photometry from sources selected in its bands, all the while respecting the data policy driven source selection rule, i.e. only release the point sources and distant galaxies detected in all r,i,z and Y,J,H bands across both catalogs above 5-sigma, see \secref{sec:intro}.}
{As soon as the two surveys overlap, and within the \recccref{implementation} recommended timescales.  The benefits for cosmology from Rubin+Euclid photometric redshifts are expected to be seen on a later-deeper Rubin timescale (see discussion in \ddpscref{static-cosmo-rubin-source}).}{CC}

\ddp{mw-fp-catalogs}{Multi-band Rubin+Euclid forced photometry catalog based on joint-pixel processing}{Starting with object detections in one survey, measure PSF, aperture and total fluxes and/or upper limits across all bands using matched images in the other survey, while respecting the data policy driven \ddpccref{listdriven-catalogs} source selection function. This approach is the equivalent of \ddpccref{listdriven-catalogs} with co-located pixels and a consolidated software which will result in two separate u,g,r,i,z,y,Y,J,H photometric catalogs (point sources = \ddplvref{lv-catalog-force}, galaxies = \ddpscref{static-cosmo-full-depth-rubin}, star/galaxy classification = \ddplvref{lv-star-galaxy}). Essential for the production of high-quality photometric redshifts, object classifications and robust characterisation of the physical properties of galaxies, AGN and transients (\ddptrref{transient-lightcurves}, \ddptrref{transient-host-phot}, \ddpgeref{gal1}, \ddpanref{agn-joint-phot-cat}). A slew of physical parameters would further boost the scientific return (\ddpgeref{gal4}, \ddpgeref{gal5}, \ddpgeref{gal6}, \ddpanref{agn-joint-prop-cat}, \ddpslref{sl-2}, \ddpslref{sl-3}).}{Post Euclid and Rubin DR1. Incrementally increasing in area/depth/complexity through the lifetime of both surveys from the early days of mutual overlap. 
}{CC}

\ddp{mw-db-catalogs}{Multi-band Rubin+Euclid deblended photometry catalog from joint-pixel processing}{Starting with object detections across both surveys based on the \ddpccref{listdriven-catalogs} selection function, measure deblended component with VIS and total fluxes and/or upper limits across all u,g,r,i,z,y,Y,J,H bands using matched images in both survey datasets, while respecting the data policy driven \ddpccref{listdriven-catalogs} source selection function (point sources = \ddplvref{lv-catalog-force-deblend}, galaxies = \ddpscref{static-cosmo-ambitious-photom-wide}, star/galaxy classification = \ddplvref{lv-star-galaxy}). A further useful variation on this DDP could be to force nuclear and extended component fits or limits for all (or subsets) of objects. All science aspects listed in \ddpccref{mw-fp-catalogs} would be further enhanced ({e.g.} \ddpanref{agn-joint-forced-deblend-phot-cat}). This represents the most complex approach. }{Post Euclid and Rubin DR1. Incrementally increasing in area/depth/complexity through the lifetime of both surveys from the early days of mutual overlap.}{CC}


\ddp{photoz}{Galaxy ``pixel'' photometric redshifts with machine learning}{Full probability distributions for the photometric redshift estimates are required for all science cases which need to propagate errors into physical parameters using a range of algorithms incorporating both empirical/training-set based methods and template-fitting run on the joint multi-wavelength catalogs (\ddptrref{transient-host-z}, \ddpgeref{gal3}, \ddpanref{agn-joint-photoz-cat}). Photometric redshift estimates will be produced independently by each consortium based on the \ddpccref{listdriven-catalogs}, \ddpccref{mw-fp-catalogs}, \ddpccref{mw-db-catalogs} catalogs, however a joint-pixel analysis with machine learning will further benefit photometric redshift estimates at both surveys depth limits in particular when deblending becomes an issue for Rubin. Similar selection function as the above photometric catalogs DDPs.}
{Post Euclid and Rubin DR1. Incrementally increasing in area/depth/complexity through the lifetime of both surveys from the early days of mutual overlap.}{CC}

\ddp{cutouts}{Image cutout delivery service}
{An Image Cutout Service will provide a one-time only traceable delivery mechanism to exchange pixels on small areas of the sky that will enable the approved scientific investigations (\fndccref{catphot}) described in this report. There will be no general repository of cutouts and the use of this data for other scientific applications will be prohibited. The number of pixels to be shared will be driven by the angular size of the source of interest. Cutouts should be exchanged only for areas with matched depths in both surveys (with the exception by nature of the \ddppuref{Stamps}'s drop-out science) for the given scientific application. The number of cutouts will range from 1) unlimited for rare objects, e.g. transients, to 2) determined by conflicts with another DDP science case (e.g. strong lensing versus high-z drop-outs in \ddppuref{Stamps}, to 3) limited by the need to construct representative samples (AGNs, low-z galaxies, etc.) following criteria to be delivered by both projects. The generic rules of this cross-cutting DDP apply to all cutout precursors: \ddpssref{solar-system-stamps}, \ddpgpref{mw-euclid-cutouts}, \ddpluref{lu1}, \ddptrref{transient-cutouts}, \ddpgeref{gal2}, \ddpslref{sl-5}, \ddppuref{Stamps}.}
{As soon as the two surveys overlap.}{CC}


\newpage

\section{Summary table and schedule of all recommended Derived Data Products} \label{ddpstable}

We present all DDPs recommended by the Working Group in this report in a summary 2-page table with priority flags leading to a tiered prioritization. The five columns of each table per science area consist of:
\begin{center}
\begin{tabular}{| p{\dimexpr0.1\textwidth-2\tabcolsep-\arrayrulewidth\relax}|
                p{\dimexpr0.78\textwidth-2\tabcolsep-\arrayrulewidth\relax} |
              }
\hline
\bf Column & \bf Description\\
\hline 
1 & DDP code name, the dynamic link leads to the relevant location in this report\\
\hline 
2 & DDP benefit, B = the DDP benefits equally to both Rubin and Euclid communities, R or E = the DDP serves primarily the Rubin or the Euclid community\\
\hline 
3 & DDP relative scientific priority in the sense of scientific importance among the proposed DDPs within the specific science Section (P1$>$P2)\\
\hline 
+ & DDP urgency level in the sense of being needed quickly (U1$>$U2). For example for time-sensitive science it may well be worth first doing a quick job to share data on a small area of sky and then spend longer doing a full analysis of a larger area for other science, even if considered of higher scientific priority in the long-term\\
\hline 
+ & DDP desired production timescale: RT = ``Real-Time'' ($\sim$\,day) for transients, YR = ``Yearly Release'' matching the Rubin-LSST releases starting Year 1 (meaning from the start of the LSST), DR = ``Main Data Releases'' for products that can wait for longer timescales, such as Euclid DR3\\
\hline
4 & DDP Tier (see below)\\
\hline
5 & DDP high-level descriptive name\\
\hline 
\end{tabular}
\end{center}

The combination of the three elements from column 3 illustrates the interplay between scientific priority, urgency of data access, and timescale for accessing the given DDP. This combination reflects the tiered and iterative approach throughout the lifetime of both surveys and translates the scientific need into the  major DDP timescales (Figure \ref{fig:data-release-schedule} presents the matched schedules of the respective data releases). The main combinations of the DDPs lead to four tiers listed (with associated yellow color code for use in the DDP tables):

\begin{center}
\begin{tabular}{| p{\dimexpr0.15\textwidth-2\tabcolsep-\arrayrulewidth\relax}|
                p{\dimexpr0.07\textwidth-2\tabcolsep-\arrayrulewidth\relax} |
                p{\dimexpr0.65\textwidth-2\tabcolsep-\arrayrulewidth\relax} |
              }
\hline
\bf Column 3 &\bf{Tier} & \bf Description\\
\hline 
P1 + U1 + RT & \centering\cellcolor{yellow!30} T0 & Ready when both telescopes observe the same sky in 2023\\
\hline 
P1 + U1 + YR & \centering\cellcolor{yellow!50} T1 & In conjunction with the Rubin-LSST Year 1 release in 2025\\
\hline 
P1 + U2 + DR & \centering\cellcolor{yellow!70} T2 & In conjunction with the Euclid DR2 and LSST Year 3 in 2027\\
\hline 
P2 + U2 + DR & \centering\cellcolor{yellow!90} T3 & In conjunction with the Euclid DR3 and LSST Year 4 in 2029\\
\hline 
Non-baseline & \centering\cellcolor{brown} T4 & Pending definition of Euclid's non-allocated time (illustrative DDPs)\\
\hline 
\end{tabular}
\end{center}

\vspace{0.4cm}
As a simple visual referential, the DDPs are color coded in the summary table, allowing at first glance to see which class of cross-cuttings DDPs they relate to:
\begin{center}
\begingroup
\setlength{\tabcolsep}{8pt} 
\renewcommand{\arraystretch}{0.6} 
\begin{tabular}{l l}
\cellcolor{cb-cats}\phantom{x} & Multi-band Rubin+Euclid photometry catalogs\\
& \\
\cellcolor{cb-phz}\phantom{x} & Galaxy photometric redshifts\\
& \\
\cellcolor{cb-stamps}\phantom{x} & Image cutouts/stamps\\
& \\
\cellcolor{black!15}\phantom{x} & Standalone DDP not realized through a Cross-Cutting DDP\\
\end{tabular}
\endgroup
\end{center}

\setlength\tabcolsep{3pt}

\newpage
\begin{center}
\hphantom{na}

\vspace{-1.3cm}
{\bf  \Large Recommended Rubin-Euclid Derived Data Products summary table [1/2]}
\vspace{-0.3cm}
\par Acronyms/Codes per column: 1) DDP code name; 2) Community served: B(oth), E(uclid), R(ubin); 3) Priority\linebreak (P1 to P2) + Urgency (U1 to U3) + Timescale (Real Time, Yearly, Data Releases); 4) Production tier (T0 to T3) 
\end{center}
\vspace{-0.2cm}

\noindent {\bf  \underline{Cross-Cutting (CC) }}\\
\begin{tabular}{ l l l l l l }
\cellcolor{cb-cats}\ddpccref{listdriven-catalogs} & B & P1+U1+YR & \cellcolor{yellow!50} T1 & Multi-band Rubin+Euclid photometry list-driven catalogs\\
\cellcolor{cb-cats}\ddpccref{mw-fp-catalogs} & B & P1+U2+DR & \cellcolor{yellow!70} T2 & Multi-band Rubin+Euclid forced photometry catalog from joint-pixel processing\\
\cellcolor{cb-cats}\ddpccref{mw-db-catalogs} & B & P2+U2+DR & \cellcolor{yellow!70} T3 & Multi-band Rubin+Euclid deblended photometry catalog from joint-pixel processing\\
\cellcolor{cb-phz}\ddpccref{photoz} & B & P2+U2+DR & \cellcolor{yellow!50} T3 & Galaxy ``pixel'' photometric redshifts\\
\cellcolor{cb-stamps}\ddpccref{cutouts} & B & P1+U1+RT & \cellcolor{yellow!30} T0 & Image cutouts/stamps delivery service\\
\end{tabular}

\vspace{0.4cm}
\noindent {\bf  \underline{Solar System (SS) }}\\
\begin{tabular}{ c l l l l l l}
\cellcolor{black!15}\ddpssref{solar-system-astrometry} & B & P1+U1+RT & \centering\cellcolor{yellow!30} T0 & Timely Solar System Object astrometry \\
\cellcolor{cb-stamps}\ddpssref{solar-system-stamps} & B & P1+U1+RT & \centering\cellcolor{yellow!30} T0 & Stamps for Solar System Objects \\
\cellcolor{black!15}\ddpssref{solar-system-lightcurves} & B & P2+U2+DR & \centering\cellcolor{yellow!50} T1 & Solar System Object light-curve catalog \\
\cellcolor{black!15}\ddpssref{solar-system-shapes} & B & P2+U3+DR & \centering\cellcolor{yellow!70} T2 & SSO shape catalog from lightcurves \\
\end{tabular}

\vspace{0.4cm}
 \noindent {\bf  \underline{Local Volume (LV) }}\\
\begin{tabular}{ c l l l l l l }
\cellcolor{cb-cats}\ddplvref{lv-catalog-merging} & B & P1+U1+YR & \centering\cellcolor{yellow!50} T1 & Multi-band Merged Point Source catalog \\
\cellcolor{cb-cats}\ddplvref{lv-catalog-variables} & B & P1+U1+YR & \centering\cellcolor{yellow!50} T1 &  \ddplvref{lv-catalog-merging} $+$ Time-Series Information for Periodic Variables\\
\cellcolor{cb-cats}\ddplvref{lv-catalog-list} & B & P1+U1+YR & \centering\cellcolor{yellow!50} T1 &  List-driven Multi-band Photometry\\
\cellcolor{cb-cats}\ddplvref{lv-catalog-force} & B & P2+U1+DR & \centering\cellcolor{yellow!70} T2 &  Joint-pixel Multi-band Forced Photometry\\
\cellcolor{cb-cats}\ddplvref{lv-catalog-force-deblend} & B & P2+U2+DR & \centering\cellcolor{yellow!90} T3 &  Joint-pixel Deblended Multi-band Forced Photometry\\
\cellcolor{black!15}\ddplvref{lv-star-galaxy} & B & P2+U1+DR & \centering\cellcolor{yellow!50} T2 &  Joint-pixel Star/Galaxy Classification\\
\end{tabular}

\vspace{0.4cm}
 \noindent {\bf  \underline{Galactic Plane (GP) }}\\
\begin{tabular}{ c l l l l l l }
\cellcolor{cb-cats}\ddpgpref{mw-catalog-cross-match} & B & Non-baseline & \centering\cellcolor{brown} T4 & Catalogs cross-match \\
\cellcolor{black!15}\ddpgpref{mw-catalog-astrometric solution} & E & Non-baseline & \centering\cellcolor{brown} T4 & Complete Astrometric Solution \\
\cellcolor{black!15}\ddpgpref{mw-stellar-spectra} & R & Non-baseline & \centering\cellcolor{brown} T4 & NISP stellar spectra \\
\cellcolor{cb-stamps}\ddpgpref{mw-euclid-cutouts} & R & Non-baseline & \centering\cellcolor{brown} T4 & Euclid Image cutouts \\
\end{tabular}

\vspace{0.4cm}
 \noindent {\bf  \underline{Local Universe (LU) }}\\
\begin{tabular}{ c l l l l l l }
\cellcolor{cb-stamps}\ddpluref{lu1} & B & P1+U1+YR & \centering\cellcolor{yellow!50} T1 &  Joint pixel processing of large image cutouts\\
\cellcolor{black!15}\ddpluref{lu2} & B & P1+U1+YR & \centering\cellcolor{yellow!50} T1 &  Nearby galaxies structure \& morphological parameters\\
\cellcolor{black!15}\ddpluref{lu3} & B & P1+U1+YR & \centering\cellcolor{yellow!50} T1 &  Dedicated Low Surface Brightness pixel data reductions\\
\cellcolor{cb-cats}\ddpluref{lu4} & B & P1+U1+YR & \centering\cellcolor{yellow!50} T1 &  Multi-band merged catalog with compactness-sensitive measurements\\
\cellcolor{cb-cats}\ddpluref{lu5} & B & P1+U1+DR & \centering\cellcolor{yellow!90} T3 &  Multi-band merged catalog of objects without proper motion\\
\end{tabular}
 
\vspace{0.4cm}
\noindent {\bf  \underline{Transients (TR) }}\\
\begin{tabular}{ c l l l l l l }
\cellcolor{black!15}\ddptrref{transient-lightcurves} & B & P1+U1+RT & \centering\cellcolor{yellow!30} T0 & Transient object light-curves \\
\cellcolor{cb-stamps}\ddptrref{transient-cutouts} & B & P1+U1+RT & \centering\cellcolor{yellow!30} T0 & Transient object cutouts \\
\cellcolor{black!15}\ddptrref{transient-astrometry} & B & P2+U1+RT & \centering\cellcolor{yellow!30} T0 & Transient astrometry \\
\cellcolor{cb-cats}\ddptrref{transient-host-phot} & B & P1+U1+YR & \centering\cellcolor{yellow!50} T1 & Transient host galaxy photometry \\
\cellcolor{cb-phz}\ddptrref{transient-host-z} & B & P1+U1+RT & \centering\cellcolor{yellow!30} T0 & Transient host galaxy redshifts \\
\cellcolor{black!15}\ddptrref{transient-host-spec-params} & B & P2+U2+YR & \centering\cellcolor{yellow!70} T2 & Transient host spectral parameters \\
\cellcolor{black!15}\ddptrref{transient-lightcurve-params} & B & P1+U1+RT & \centering\cellcolor{yellow!30} T0 & Joint transient lightcurve parameters  \\
\cellcolor{black!15}\ddptrref{transient-z} & B & P1+U1+RT & \centering\cellcolor{yellow!30} T0 & Transient redshifts \\
\cellcolor{black!15}\ddptrref{transient-host-params} & B & P1+U1+YR & \centering\cellcolor{yellow!50} T1 & Joint transient host parameters  \\
\cellcolor{black!15}\ddptrref{transient-database} & B & P1+U1+RT & \centering\cellcolor{yellow!30} T0 & Joint transient database and interface \\
\cellcolor{black!15}\ddptrref{transient-detection-efficiencies} & B & P1+U1+DR & \centering\cellcolor{yellow!70} T2 & Transient detection efficiencies \\
\end{tabular}

\begin{center}
\hphantom{na}

\vspace{-1.3cm}
{\bf \Large Recommended Rubin-Euclid Derived Data Products summary table [2/2]}
\vspace{-0.3cm}
\par Acronyms/Codes per column: 1) DDP code name; 2) Community served: B(oth), E(uclid), R(ubin); 3) Priority\linebreak (P1 to P2) + Urgency (U1 to U3) + Timescale (Real Time, Yearly, Data Releases); 4) Production tier (T0 to T3) 
\end{center}
\vspace{-0.2cm}

\noindent {\bf  \underline{Galaxy Evolution (GE) }}\\
\begin{tabular}{ c l l l l l l }
\cellcolor{cb-cats}\ddpgeref{gal1} & B & P1+U1+YR & \centering\cellcolor{yellow!50} T1 & Multi-band catalogs \\
\cellcolor{cb-stamps}\ddpgeref{gal2} & B & P1+U1+YR & \centering\cellcolor{yellow!50} T1 & Postage Stamp Cutouts for Representative Samples \\
\cellcolor{cb-phz}\ddpgeref{gal3} & B & P1+U2+YR & \centering\cellcolor{yellow!70} T2 & Galaxy Photometric Redshifts \\
\cellcolor{black!15}\ddpgeref{gal4} & B & P2+U2+YR & \centering\cellcolor{yellow!70} T2 & Galaxy Physical Properties \\
\cellcolor{black!15}\ddpgeref{gal5} & B & P2+U2+DR & \centering\cellcolor{yellow!90} T3 & Galaxy Morphology \& Structure \\
\cellcolor{black!15}\ddpgeref{gal6} & B & P2+U2+DR & \centering\cellcolor{yellow!90} T3 & Pixel-level decomposition of quasars and their host galaxies \\
\end{tabular}
 
\vspace{0.4cm}
 \noindent {\bf  \underline{Active Galactic Nuclei (AN) }}\\
\begin{tabular}{ c l l l l l l }
\cellcolor{cb-cats}\ddpanref{agn-joint-phot-cat} & B & P1+U1+YR & \centering\cellcolor{yellow!50} T1 & Multi-band Photometric catalogs \\
\cellcolor{cb-cats}\ddpanref{agn-joint-forced-deblend-phot-cat} & B & P1+U1+DR & \centering\cellcolor{yellow!90} T3 & Deblended Multi-band, Multi-epoch and Time-averaged Photometric catalogs \\
\cellcolor{black!15}\ddpanref{agn-joint-class-cat} & B & P1+U1+YR & \centering\cellcolor{yellow!50} T1 & AGN Candidate catalog \\
\cellcolor{cb-phz}\ddpanref{agn-joint-photoz-cat} & B & P1+U1+YR & \centering\cellcolor{yellow!50} T2 & AGN and Host Photometric Redshifts \\
\cellcolor{black!15}\ddpanref{agn-joint-prop-cat} & B & P1+U1+YR & \centering\cellcolor{yellow!50} T2 & AGN Physical Properties \\
\cellcolor{black!15}\ddpanref{agnvar-joint-LCs} & B & P1+U1+RT & \centering\cellcolor{yellow!30} T0 & Prompt Light Curves of Extreme AGN Variability Events \\
\end{tabular}
 
\vspace{0.4cm}
 \noindent {\bf  \underline{Static Cosmology (SC: weak lensing, clustering, clusters)) }}\\
\begin{tabular}{ c l l l l l l }
\cellcolor{cb-cats}\ddpscref{static-cosmo-euclid-source} & E & P1+U1+YR & \centering\cellcolor{yellow!50} T1 & Baseline `list-driven' Y1 ugrizy Rubin photometry for Euclid VIS sources \\
\cellcolor{cb-cats}\ddpscref{static-cosmo-rubin-source} & R & P2+U2+DR & \centering\cellcolor{yellow!70} T2 & Baseline `list-driven' Euclid NISP YJH photometry for Rubin sources \\
\cellcolor{cb-phz}\ddpscref{static-cosmo-spectroscopy} & R & P1+U1+YR & \centering\cellcolor{yellow!50} T1 & Rubin photometric redshift distributions calibrated with Euclid spectroscopy \\
\cellcolor{cb-cats}\ddpscref{static-cosmo-full-depth-rubin} & E & P2+U2+DR & \centering\cellcolor{yellow!90} T3 & Deeper ugrizy Rubin photometry of Euclid VIS sources \\
\cellcolor{cb-cats}\ddpscref{static-cosmo-ambitious-photom-wide} & B & P1+U1+DR & \centering\cellcolor{yellow!70} T2 & Joint-pixel Rubin ugrizy, Euclid VIS \& YJH photometry in the Wide Surveys \\
\cellcolor{black!15}\ddpscref{static-cosmo-ambitious-Shapewide} & B & P2+U2+DR & \centering\cellcolor{yellow!90} T3 & Joint-pixel Rubin-Euclid galaxy shape analysis \\
\cellcolor{black!15}\ddpscref{static-cosmo-ambitious-photom-deep} & B & P1+U1+YR & \centering\cellcolor{yellow!50} T1 & Deep Survey joint-pixel photometry and shear analysis of all Rubin+Euclid bands\\
\end{tabular}

\vspace{0.4cm}
\noindent {\bf  \underline{Strong Lensing (SL) }}\\
\begin{tabular}{c l l l l l l }
\cellcolor{cb-stamps}\ddpslref{sl-1} & B & P1+U1+YR & \centering\cellcolor{yellow!30} T0 & Pansharpenned images of all strong lens candidates \\
\cellcolor{black!15}\ddpslref{sl-2} & B & P1+U1+YR & \centering\cellcolor{yellow!50} T1 & Deblended foreground lens \& background source photometry for lens candidates \\
\cellcolor{black!15}\ddpslref{sl-3} & E & P1+U1+YR & \centering\cellcolor{yellow!50} T1 & A joint colour and morphology catalog for strong lens searches \\
\cellcolor{black!15}\ddpslref{sl-4} & B & P2+U3+DR & \centering\cellcolor{yellow!70} T2 & A strong lens probability for every early type galaxy \\
\cellcolor{cb-stamps}\ddpslref{sl-5} & B & P2+U2+YR & \centering\cellcolor{yellow!50} T1 &  ugrizy,VIS,YJH postage stamps of strong lens candidates \\
\end{tabular}
 
\vspace{0.4cm}
 \noindent {\bf  \underline{Primaeval Universe (PU) }}\\
\begin{tabular}{ c l l l l l l }
\cellcolor{black!15}\ddppuref{Optdetcat} & R & P1+U1+YR & \centering\cellcolor{yellow!50} T1 & Euclid near-infrared photometry for high-redshift galaxies detected in LSST \\
\cellcolor{black!15}\ddppuref{IRdetcat} & E & P1+U1+YR & \centering\cellcolor{yellow!50} T1 & Joint ugrizy,VIS photometric catalogs of Euclid high-${\bf z}$ candidates \\
\cellcolor{black!15}\ddppuref{photostack} & E & P1+U1+YR & \centering\cellcolor{yellow!50} T1 & Photometric measurements on multi-band stacks of Euclid high-${\bf z}$ candidates\\
\cellcolor{cb-stamps}\ddppuref{Stamps} & B & P1+U1+YR & \centering\cellcolor{yellow!50} T1 & Stamps of high-${\bf z}$ candidates \\
\end{tabular}
 
\vspace{0.4cm}

\newpage

\section{Survey strategy optimisation for DDPs} \label{sec:observing}

\noindent\contributors{\hyperref[author:jcuillandre]{Jean-Charles Cuillandre (WG)}, \hyperref[author:lguy]{Leanne Guy (WG)}, Phil Marshall, Peter Yoachim}

\subsection*{Surveys parameters} \label{sec:survey-parameters}

The Vera C. Rubin Observatory Legacy Survey of Space and Time (LSST, \citealp{Ivezi__2019}) will execute the Wide-Fast-Deep (WFD) survey as the bulk of its 10-year program ($\approx$\,80\%). During the remaining time, Rubin will execute a set of special programs, which include observing four deep drilling fields (DDFs): ELAIS-S1, XMM-LSS, Extended Chandra Deep Field-South, COSMOS. The proposed extended DDF over the Euclid Deep Field South would be executed at half the speed as that of the four other Rubin DDFs. 
LSST will cover the sky to unprecedented depths in the u,g,r,i,z,y photometric bands (referred to as ugriz(y) in the science Section of this report), covering an area from the south celestial pole up to a declination of +30 degrees. The Rubin field of view is 9.6 deg$^2$ and the expected median image quality is 0.8 arcsecond in the r-band. The LSST strategy is one of coverage first and depth second.

\begin{figure*}[ht]
\centering
\includegraphics[width=0.85\textwidth]{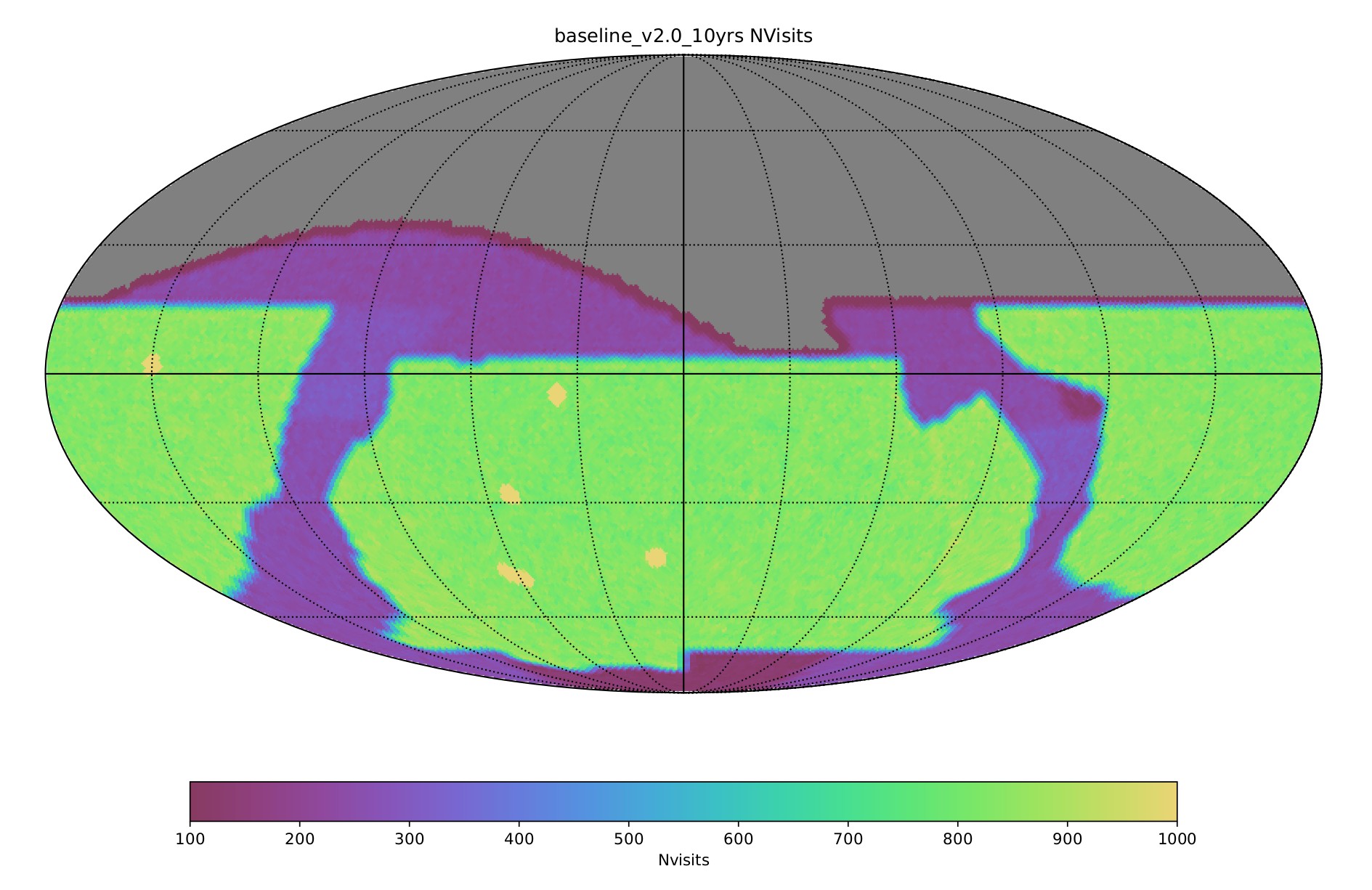} 
\caption{Rubin Observatory’s Legacy Survey of Space and Time (LSST) proposed baseline v2.0 \citep{PSTN-053}, showing an extended Wide-Fast-Deep survey (20,000 deg$^2$, green) reaching all the way south and far north while observing less frequently those areas of the Galactic plane that are most affected by dust (purple). The 4 Deep Drilling Fields, and the proposed DDF over the  Euclid Deep Field South appear in yellow. A northern extension mini-survey could fill the grayed area nearly up to the upper declination limit of the Ecliptic Plane Survey (smoothly curved purple area, Dec\,$\sim$\,$+$30\,deg).}
\label{fig:rubin-survey}
\end{figure*}

The Euclid Survey \citep{Scaramella_2021} is composed of the Wide Survey (15,000 deg$^2$), the bulk of the 6-year effort (65\%), along with a Deep Survey and various deep calibration and auxiliary fields. The three deep fields are EDF-North (North Ecliptic Pole, 10 deg$^2$), EDF-Fornax (Chandra Deep Field-South, 10 deg$^2$) and EDF-South (new extended cosmology field, 23 deg$^2$). The Euclid strategy for the Wide survey is depth first and coverage second. Together these various components will cover the sky above the galactic and the ecliptic planes at unprecedented resolution and spectral range in the broad VIS(ible = r+i+z) band and three near-infrared bands (Y,J,H, referred to as YJH in the science Section of  this report)) as well as a slitless low-resolution near-infrared spectroscopy mode. Expected median image quality is 0.18 arcsecond for the VIS band (0.1$''$/pixel) and 0.45 arcsecond for the near-infrared (0.3$''$/pixel). The Euclid field of view is 0.5 deg$^2$.

\begin{figure*}[t]
\centering\includegraphics[width=15.5cm]
{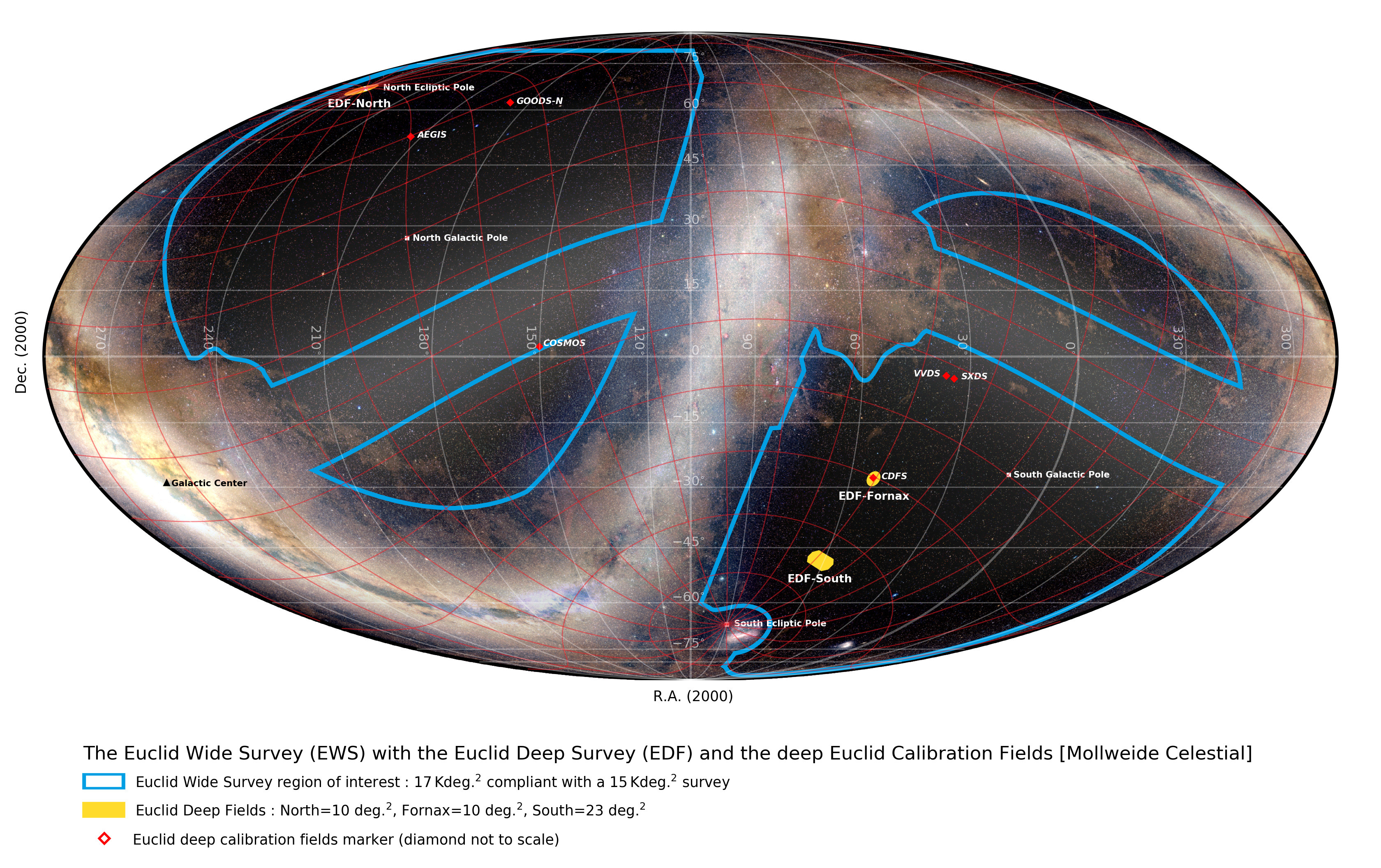} 
\caption{Euclid has identified 17,400\,deg$^2$ of sky compatible with its science requirements (blue outline avoiding the ecliptic and the galactic planes plus areas that are affected by dust). The 6-year baseline survey however only covers the Euclid requirement of a 15,000\,deg$^2$ survey. An extra year of Euclid operation could fill the full 17,400\,deg$^2$. Euclid will cover the wide survey area from the ecliptic pole areas towards the ecliptic plane. The 3 deep fields (yellow) will be 2 magnitudes deeper than the wide survey.}
\label{fig:euclid-survey}
\end{figure*}

\subsection*{Wide Surveys}
\label{sec:widesurvey-parameters}

The LSST Dark Energy Science Collaboration (DESC) argued for modifications to the Wide Fast Deep (WFD) survey towards an extended footprint driven by dust extinction limits. This was endorsed by the Tri-Agency Survey Coordination Task Force \citep{cuillandre2021} for the Rubin-LSST Survey Cadence Optimization Committee (SCOC) since extending the Rubin WFD to Dec\,$+$12.5 to $-$70 degrees enhances the synergy with Euclid,  creating up  to a 9400\,deg$^2$ of overlap to great depths while also augmenting Rubin's follow-up capabilities with imaging and spectroscopic facilities from the North (e.g. Keck, DESI, PFS, TMT). The SCOC note also endorsed survey options reaching all the way down to the South Celestial Pole as they include high quality (dark) Euclid sky that will be observed, by mission design, within the first two years (\figref{fig:euclid-survey}). Following a wide consultation with the Rubin-LSST community, the SCOC has formulated a draft Phase 1 recommendation for the initial survey strategy \citep{PSTN-053} that fulfills all the above wishes (\figref{fig:rubin-survey}).

The mini-survey of the northern sky as proposed in the 2018 Rubin-Euclid white paper \citep{Capak_2019b} suggested pushing the Rubin coverage up to Dec\,+30. In adddition to enhanced Rubin-Euclid synergies, LSST science cases were presented for Targets of Opportunities (ToOs) templates, Local Group, Galactic RR Lyrae, Nearby Universe, DESI synergies, DESI-2, and High-$z$. This has since materialized in some LSST simulations in the form of a northern stripe mini-survey for ToOs, along the Northern Ecliptic Survey, although only in the g, r, i bands. The addition of the z-band in the Euclid area would enable more LSST science while adding an extra 2600\,deg$^2$ overlap with Euclid outside of the extended WFD. This mini-survey completing the WFD would bring the total Rubin possible overlap with the Euclid sky to 12,500\,deg$^2$.

\rec{surveys-1}{Maximize the overlap area of the wide surveys}{The proposed science requires that both surveys observe the same area of the sky. It will therefore be beneficial to maximize the overlap area of the two wide surveys through the LSST WFD and mini-surveys. Maximizing temporal overlap will also be beneficial for the wide surveys, for instance aligning LSST's rolling cadence on the mini-surveys with Euclid's increasing declination observing strategy.}{SO}

\begin{figure*}[t]
\centering\includegraphics[width=17cm]
{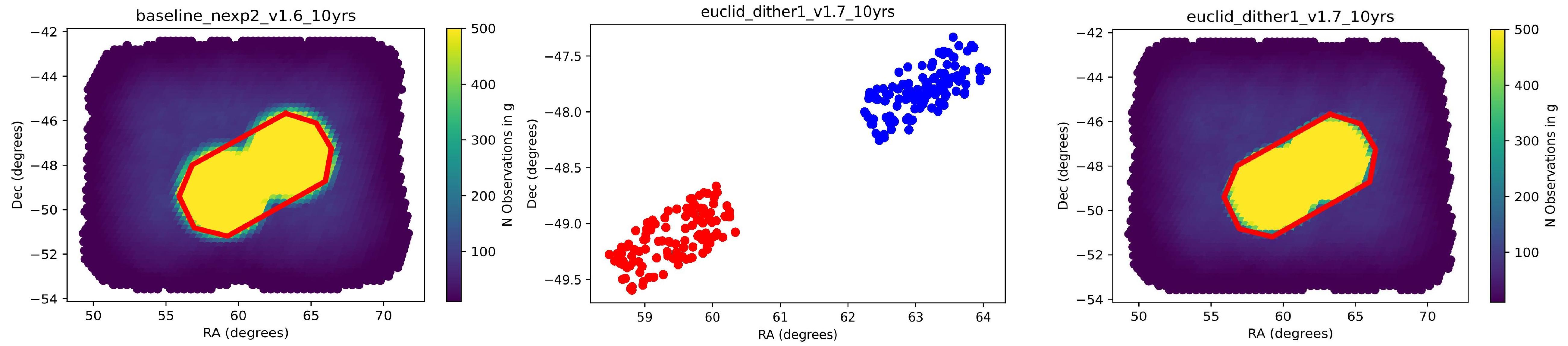} 
\caption{The stadium shape of EDF-South was chosen to minimize the time budget needed by Rubin to cover a contiguous area of at least 20\,deg$^2$, as required by Euclid's science goals. Two independent Rubin DDF positions (left) over EDF-South result in high depth outside the desired footprint. A custom extended Rubin dithering pattern (translation dithers, center) brings better footprint coverage, reaching 21.6\,deg$^2$, out of the stadium's 23\,deg$^2$ footprint, at the required depths (right).}
\label{fig:edfs-survey}
\end{figure*}

\subsection*{Deep Surveys} \label{sec:deepsurvey-parameters}
Euclid selected one of its three deep fields, the CDFS/EDF-Fornax (10\,deg$^2$) field, to match one of the four LSST DDFs. Additionally, there is an informal agreement between Rubin and  Euclid for the fifth LSST DDF to overlap the Euclid Deep Field South (EDF-South, 23\,deg$^2$), which has been shaped into an elongated area to minimize the Rubin time budget needed to reach a uniform depth across the 20 contiguous square degrees required by Euclid's science (\figref{fig:euclid-survey}). As an exercise, the proposed EDF-South has since been integrated in the proposed LSST baseline v2.0 \citep{PSTN-053}in the form of two individual half-speed DDFs, which would lead to a barely contiguous final area of less than 20\,deg$^2$ at the required depth. Interactions between the two projects' survey design teams led to the proposed translation dithers, illustrated in \figref{fig:edfs-survey}, for which Rubin reaches the Euclid depth goal over 21.6\,deg$^2$ within the 23\,deg$^2$ stadium shape area that will be fully covered by Euclid. This dithered approach is favored in the context of the DDPs as it maximizes the science output of the static astrophysics while creating a uniform joint legacy dataset. Transient science is affected but not undermined by such an approach.

Once Euclid is launched early 2023 we will know precisely when the two southern deep fields will be visited over the 6-year period. EDF-Fornax and EDF-South will start getting observed regularly on the second year of the space survey, each visit lasting a fixed 5 to 6 days, every 6 months. To optimize transient science with the LSST and Euclid, cadence should be optimized as much as possible to enable cutting DDPs.

Community exchanges indicated that the development of the most advanced and complex DDPs (e.g. algorithms developments for joint pixel processing) would benefit from an openly shared photometric dataset (meaning sharing the pixel images) across the two projects over a small sky area to full depth. Considering the effort produced by both Rubin and Euclid to accommodate an optimized solution on EDF-South for both telescopes, the EDF-South 23\,deg$^2$ area would be an ideal fit for an openly shared pixel dataset across the two projects.

\rec{surveys-2}{Maximize the Euclid Deep Field South temporal overlap}{The proposed science requires that the both surveys observe the same area of the sky. It is therefore highly beneficial to optimize the observation of the EDF-South surveys in terms of the dithering strategy (translation versus circular) and temporal overlap.}{SO}

\rec{surveys-3}{Enable advances in DDP development by sharing the Euclid Deep Field South pixel data}{The long-term production of the DDPs will benefit from an openly shared photometric dataset across the two projects over a small sky area, EDF-South being an ideal candidate considering the mutual engagement.}{SO}

\begin{sidewaysfigure}
\centering
\includegraphics[width=22.5cm]
{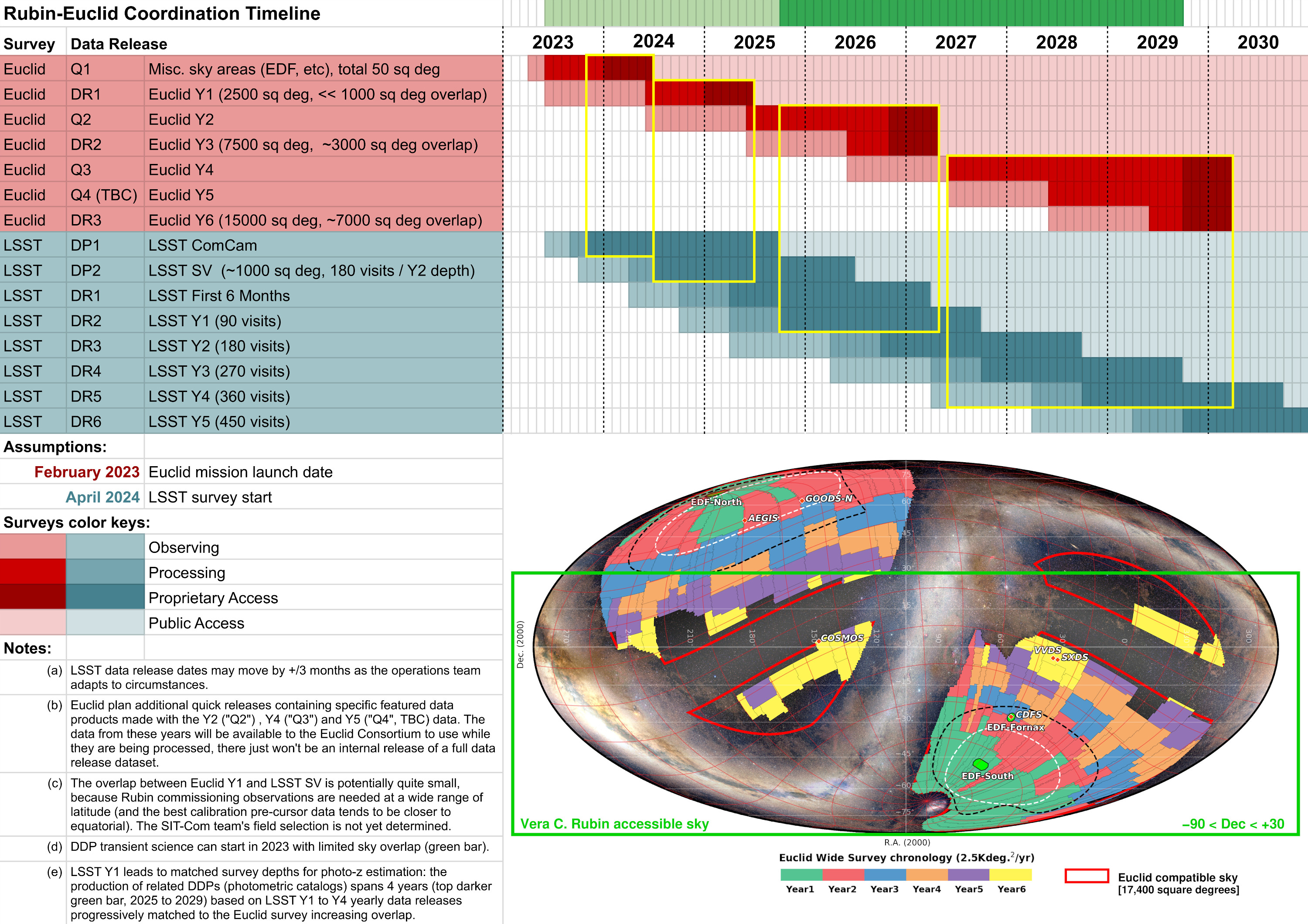} 
\caption{Matched data release schedules for Euclid and Rubin LSST (c.f. included Surveys color keys and Notes on the bottom left). The top green bar shows the proposed time span of the DDPs generation activity while yellow boxes illustrate a scheme for when major DDP generation could take place:  a) the two yellow boxes on the left indicate time window opportunities for investigating the production of a photometric catalog DDP, b) the two yellow boxes on the right indicate that only once LSST Y1 (depth match with Euclid) and subsequent yearly data releases are available, DDPs can then be used for the following LSST releases and the Euclid Data releases 2\&3. The sky map shows a realization of the Euclid Wide Survey per year over 6 years, filling 15,000\,deg$^2$ out of its region of interest (red contours) by starting from the darkest sky from L2 (dashed areas near the ecliptic poles). The green box shows the declination reach of the Rubin Observatory from Chile.}
\label{fig:data-release-schedule}
\end{sidewaysfigure}

\subsection*{Timescales and Data Releases}
LSST will release \textit{Prompt} data products and \textit{Data Release} data products \citep{LSE-231}.  Prompt data products will be released on a 60\,s and 24\,hr timescale. They will include a world public stream of alerts from transient, variable and moving sources, and a prompt products database (PPDB) that will contain catalogs of detections in difference images. Data Release data products will be made available in a series of 11 approximately annual data releases over the duration of the 10-year survey. These will include calibrated images, measurements of positions, fluxes, and shapes, variability information, and a uniform reprocessing of the difference imaging-based Prompt data products. Each subsequent release will comprise a reprocessing of all data taken to date (i.e. not just new data taken since the previous release). The first LSST data release, DR1 will be based on the first six months of data taken and released to LSST data rights holders one year following the start of operations, i.e. mid-2025 if science operations begins in  April 2024 as currently foreseen (\figref{fig:data-release-schedule}).

Euclid also opted for a calendar-driven release scenario     through three data releases (DR) over 6 years, leading to an incremental sky coverage of 2500, 7500, and 15,000\,deg$^2$ at the end of year 1, 3, and 6 respectively ( \figref{fig:data-release-schedule}). All Euclid Consortium scientists will however have access to the processed images and catalogs as soon as the observations are acquired and processed by the Euclid Consortium Science Ground Segment (EC-SGS), and pushed rapidly to the Euclid data repository. A Euclid Date Release integrates a larger effort over a longer timescale, and it is expected that a given Data Release will be accessible to the Euclid Consortium 6 months before it is released to the world, considering all input data were available to scientists since acquisition and that preliminary versions of that given Data Release had already been made available internally in the intervening months.

\begin{figure*}[t]
\centering\includegraphics[width=13cm]
{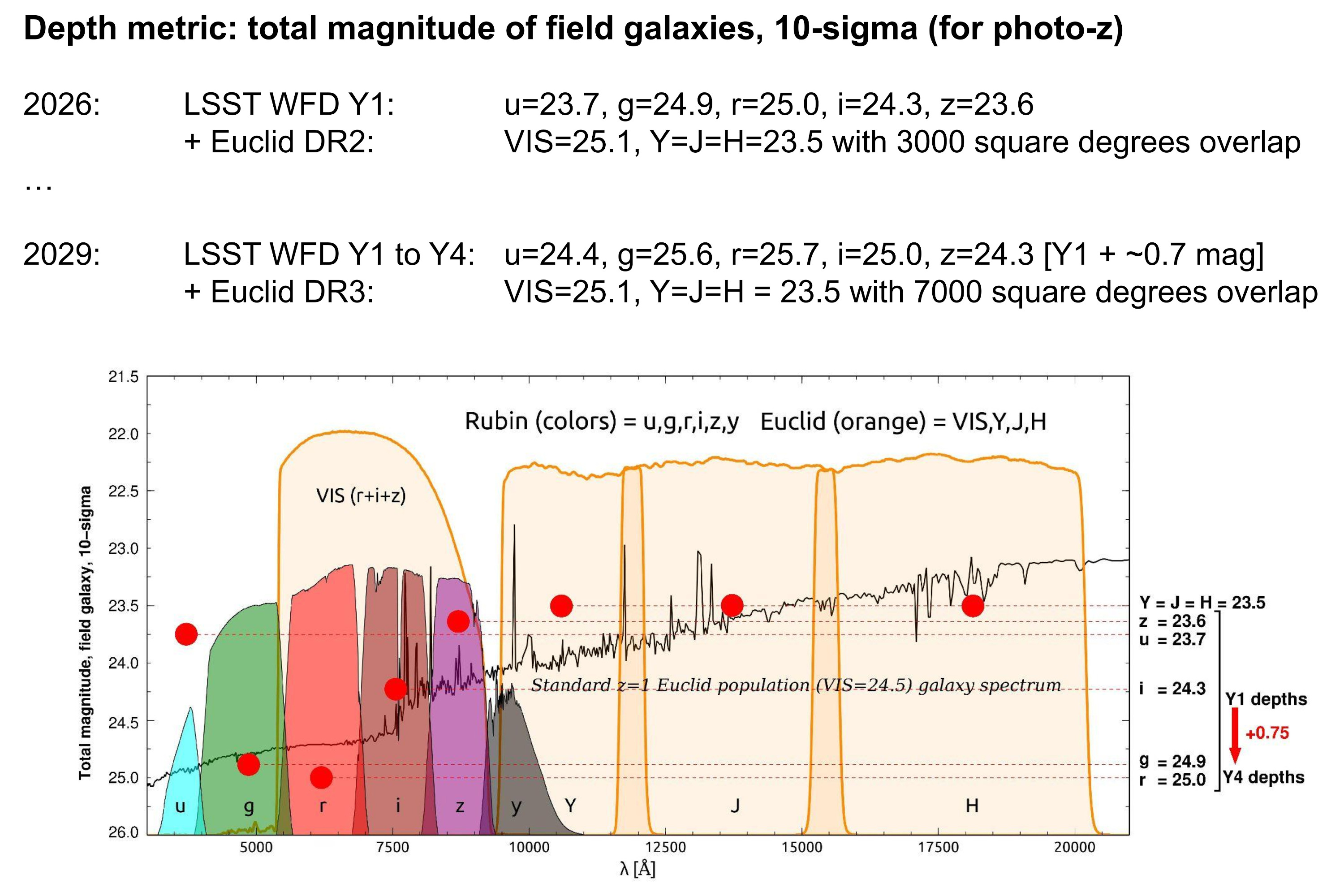}
\caption{\underline{Top:} depth evolution and overlap between the Euclid data releases 2\&3 and the LSST WFD yearly releases, note the photometric redshifts oriented depth metric. \underline {Bottom:} the Rubin+Euclid photometric bands, the full range of the Y-axis representing the 0.0--0.8 sensitivity range in electron per incident photon (e.g. VIS detects $\sim$70\% of all photons entering the telescope's pupil, and Rubin $\sim$45\% in the i-band). The $z$=1 spectrum is a simulation for a typical faint galaxy from the Euclid main lensing population (i $\sim$\,VIS=24.5), showing that the relative depths from the g to the H-band match the general spectral energy distribution (colors from g--r to J--H).}
\label{fig:data-release-depth}
\end{figure*}

\figref{fig:data-release-schedule} shows the nominal data release schedules for both Euclid and Rubin’s LSST at the time of writing. Transient science DDPs can start as early as 2023 as both Euclid and Rubin (with the Commissioning Camera) observe a limited common part of the sky. The more complex DDPs requiring a matched depth of both surveys (photometric catalogs) can be investigated using LSST DP2 in 2024 in preparation for the LSST Y1 (DR2) in 2025, an effort to be conducted on a yearly basis as the LSST depth and overlap with Euclid increase. 
Relative shifting of the two surveys' starting dates could lead to a different matching between data releases. The intent would be to create most DDPs over the widest possible Euclid area with the deepest possible LSST data (except where otherwise noted).
The LSST WFD 2018 baseline adopted for the overlaps in this table covers the declination range from $-$62 to +2 degrees (the LSST proposed baseline v2.0 will bring greater overlap and the coverage numbers mentioned here can be considered a minimum, see \figref{fig:rubin-survey}). \figref{fig:data-release-depth} shows the expected depth of the joint dataset (ugriz,VIS,YJH) with respect to Euclid 2025 DR2 (with LSST Y1, 3000\,deg$^2$ overlap with the proposed baseline v2.0) and 2029 DR3 (with LSST Y4, 7000\,deg$^2$ overlap with the proposed baseline v2.0) areas.

\rec{surveys-5}{Start DDP production as soon as Rubin and Euclid observe a common part of the sky}{In order to meet the scientific expectations of the Rubin and Euclid communities, the production of DDPs should begin as soon as Rubin and Euclid observe a common part of the sky, i.e. as early as 2023.}{SO}

\rec{surveys-4}{Adopt a range of timescales for releasing DDPs}{DDPs that will enable transient science should materialize on a short timescale, e.g. 24\,hr, and should be based on a fast joint processing of the data, whereas more complex DDPs would fit better in the context of the annual Rubin data release scenario
}{SO}

\section{Scientific Motivation}
\label{sec:science}
\subsection{Solar System} \label{sec:solar-system}
 
\contributors{
\hyperref[author:seggl]{Siegfried Eggl (WG)}, \hyperref[author:bcarry]{Benoit Carry}, Matthew M. Knight, Hayden Smotherman, Colin Snodgrass}

\begin{figure}[!ht]
  \includegraphics[width=0.55\textwidth]{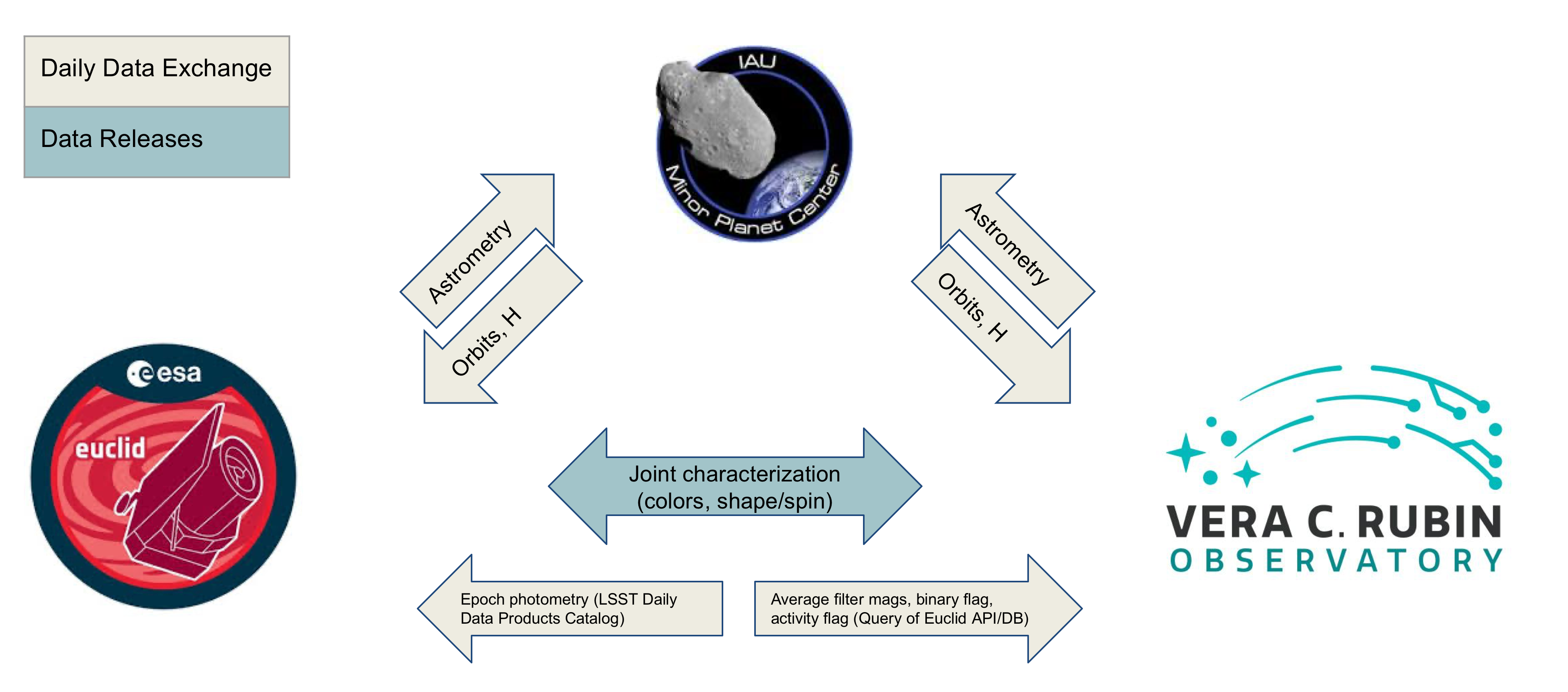}
    \includegraphics[width=0.4\textwidth]{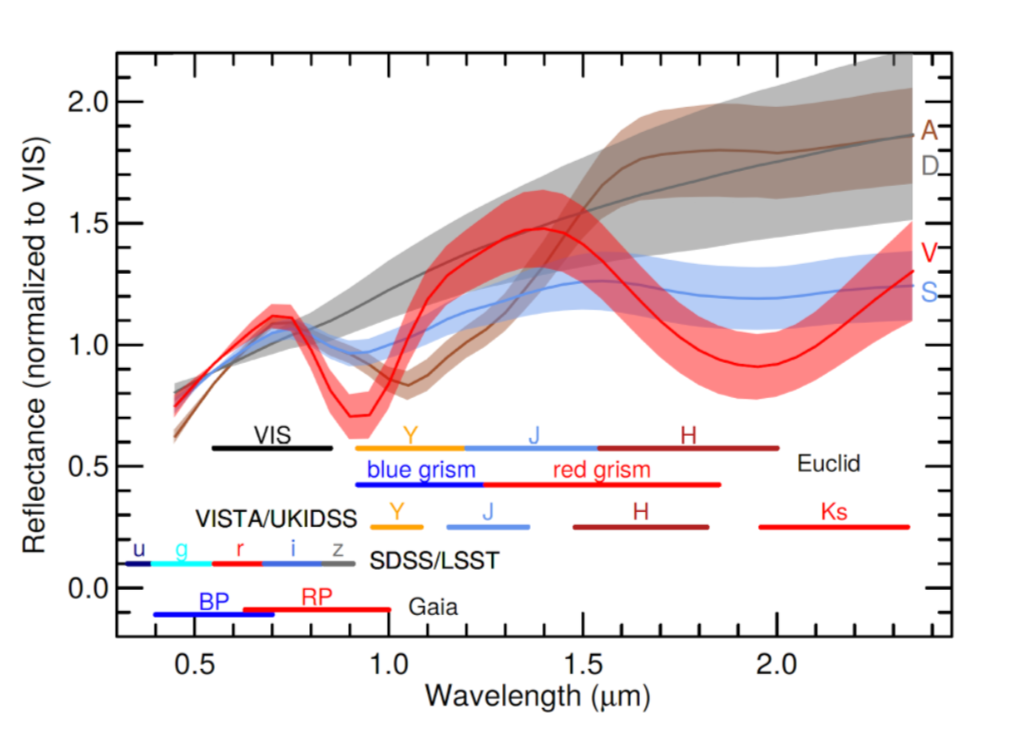}
  
  \caption{Left: Envisaged Derived Data Products exchange between Rubin, Euclid and the IAU Minor Planet Center. SSO astrometry and derived quantities are exchanged on a roughly daily timescale. Joint characterization of SSOs could happen around data releases. Right: Reflectance spectra for various taxonomic classes of asteroids. Euclid and Rubin filters are shown as well. Euclid and Rubin provide complementary color information that can break the degeneracy between different taxomomic types. See \citet{carry2018solar} for details. \label{fig:solar-system-mpc}}
\end{figure}

The Vera C. Rubin Observatory Legacy Survey of Space and Time (LSST) has a cadence that is particularly well suited for the discovery of Solar System Objects (SSOs). Current best estimates suggest that LSST will discover roughly 5 million SSOs, the majority of which are expected to have multiple observations in each of the ugriz filters.
In contrast, the Euclid cadence avoids ecliptic latitudes within 10 degrees of the ecliptic plane.
However, it will still detect roughly 150,000 Solar System objects, only about 1\% of which are known today \citep{carry2018solar}, thanks to 
its sensitivity to faint objects [detection limit in VIS (r+i+z) of m$_{\rm AB} = $26.2 for 5\,$\sigma$ on a point source \citep{Scaramella_2021} and wide field of view of 0.7$\times$0.7\,deg$^2$].
The number of SSO benefiting from combined Euclid and Rubin observations could be increased to encompass the majority of the roughly 5 million SSOs likely to be discovered by LSST, if Euclid were to observe along the ecliptic in, e.g., an extended mission. Euclid and Rubin/LSST data is highly complementary: Euclid provides near-IR colours of SSOs that can differentiate mineral features and can break the degeneracy between spectral classes of asteroids that exist, for instance, at visible wavelengths (\figref{fig:solar-system-mpc}). 
Also, probing potential cometary activity of Centaurs will be a particular strength of Euclid.
The high angular resolution of Euclid (0.1" pixels in the VIS band) is going to provide excellent constraints on the astrometric positions of SSOs. The imaging sequence of Euclid will also provide hour-long lightcurves.
Rubin observatory, on the other hand, is expected to discover roughly 5 million objects as it covers the ecliptic plane and returns to the same fields multiple times. The LSST cadence is designed so as to facilitate orbit determination for SSOs. In contrast, orbits derived from the short arced Euclid data alone would have larger uncertainties associated with them. LSST observations will also allow sparse photometry, i.e. observations with a typical separation in time larger than the rotation period. Sparse photometry lightcurves can be constructed over the length of the LSST survey, in order to study the rotation and shape of SSOs, and their photometric phase curves. Multiple observations in LSST filters for the same objects will furthermore allow visible colors to be measured. 
Last but not least, contemporaneous and even near simultaneous observations between LSST and Euclid are possible, as Euclid is expected to launch early 2023 performing a 6-year survey whereas LSST is to start its 10-year program in 2024. 

LSST will release data on three timescales. Alerts for transients and moving objects are going to be published in near real time. Complete data products for Solar System objects are provided on a daily basis. On a yearly basis LSST data releases will provide solar system data products based on LSST data only.  
Euclid has planned three data releases, the first after year two of operations, the second after year four and data release three is planned at the end of the mission (\figref{fig:data-release-schedule}). 
However, data can be shared between Rubin and Euclid on a much shorter timescale as discussed in \secref{sec:solar-system-products}.

\subsection*{Science Drivers} \label{sec:solar-system-drivers}
Drawing on both the complementary and contemporary nature of Euclid and Rubin SSO data sets, we have identified key areas where combining Euclid and Rubin data would add substantial scientific value. Those key areas are the timely exchange of astrometric measurements and the joint physical characterization of SSOs. 

\subsubsection*{Timely Astrometry and Activity Reporting}
Accurate orbits lie at the heart of SSO research. Size estimates, for instance, rely on absolute magnitudes which are derived from measurements at various positions of the asteroid with respect to the Sun and the observer \citep[e.g.][]{masiero2021uncertainties}. Light curve inversions that can produce shape models for SSOs from sets of photometric observations also rely on accurate information about varying Sun-asteroid-observer geometries \citep{muinonen2020asteroid}. Due to the limited range of radar observatories and the small number of visiting spacecrafts, SSO orbits are almost exclusively based on astrometric measurements. A rapid publication/turnaround of astrometric measurements allows for 1) confirmation of potential new discoveries through follow up observations from other facilities, 2) a more precise targeting of follow-up observations aimed at physical characterization of objects of interest and 3) a better risk assessment for near-Earth objects. The combination of high astrometric precision, parallax and the faint magnitudes at which SSOs can be observed by both Euclid and the Rubin observatory would make a timely publication of astrometric data highly beneficial not only to the projects themselves, but to the entire international scientific community. 
LSST astrometry could be used to improve orbits of faint objects, which would in turn improve attribution of Euclid SSO observations and vice versa. A timely exchange/publication of astrometric data would increase the yield of SSO discoveries, observations in multiple color bands and enhance the quality of derived properties such as absolute magnitudes in individually published Euclid and Rubin data releases. 

One unique possibility that is offered by combined Euclid and Rubin observation is the direct distance measurement to newly discovered SSOs via parallax, in the case of (quasi-)simultaneous observations by both observatories \citep{snodgrass2018simultaneous}. This can substantially improve orbits for newly discovered objects, although similar precision will be reached without simultaneous observations for objects that are independently observed by Rubin at many epochs. The value of near-simultaneous astrometry is therefore greatest for fleeting objects seen only a few times by the LSST cadence pattern, such as very small near-Earth objects, Earth ``mini-moons'' whose orbits bring them to visibility only rarely, 
and possibly interstellar objects passing through our system. All of these are seldom observed and poorly understood, and therefore high value targets. Only a small fraction of Euclid pointings can be observed simultaneously from Rubin (8\%, or $\sim\,$3500 observations). A cost-benefit analysis of attempting such observations still needs to be performed, but the potential impact on LSST scheduling appears be small, since the fields and Euclid exposure times are known well in advance. However, since most of the joint fields would occur during twilight, the full impact of satellite mega-constellations on Rubin twilight observing needs to be better understood before a recommendation regarding  simultaneous Rubin/Euclid observations can be given.

\subsubsection*{Joint Characterization}
Combining Rubin and Euclid data sets, in particular observations of objects recorded in various filters,  would allow for a joint characterization of SSOs. This can happen at two levels: 1) The catalog level: if catalog data is shared between Euclid and Rubin, joint data products such as improved taxonomic classification, improved phase curves and absolute magnitudes (H magnitude), as well as a densification of sparse light curves would result in better constraints on size, shape and spin, multiplicity/binarity as well as composition of observed SSOs.
2) The joint pixel level: combining the high spatial resolution of Euclid with the information of multiple filter observations of LSST would benefit comet activity research, especially for characterising very weakly active bodies in the overlap region between comets and asteroids. The resolution of Euclid will reveal ``fuzziness'' that shows weak activity that would be hidden within the seeing disc (typically 0.8" FWHM) in LSST ground-based imaging, while combined photometry can reveal whether the composition of the object is more alike to comets or asteroids of various classes.
If spectral and photometric information is shared in a timely manner it can also play a vital role in the early detection of comets, outbursts of cometary activity, and newly arising asteroid activity (e.g., from impacts, disintegration, etc.). The latter can be a short-lived phenomena and would be valuable to follow up elsewhere following discovery, on a days-to-weeks timescale.
More reliable constraints on physical parameters of SSOs will improve our understanding the origin and evolution of the Solar System. 

\fnd{solar-system-1}{Combining Rubin and Euclid data enhances our knowledge of the physical properties of SSOs}
{Combining Rubin and Euclid data sets would allow for improved characterization of at least 150,000 Solar System Objects. This includes better orbits and enhanced understanding of physical properties such as spin, shape and taxonomic type.}{SS}

\fnd{solar-system-2}{Near-simultaneous observations of Rubin and Euclid fields are of greatest value}
{Near-simultaneous observations of Rubin and Euclid fields allow for enhanced characterization as asteroids are observed at similar phase angles and rotation states. Moreover, orbits of newly discovered SSOs observed at the same time by Rubin and Euclid can be greatly improved due to the parallax between the observatories.}{SS}

\fnd{solar-system-3}{Timely reporting of astrometry and activity of SSOs enables rapid follow up}
{Timely reporting of astrometry and suspected activity of SSOs will enable a more precise targeting of follow-up observations aimed at physical characterization of objects of interest and improve the risk assessment for near-Earth objects.}{SS}

\fnd{solar-system-4}{Euclid observations of the ecliptic can boost by $\times$30 the number of jointly characterized SSOs}
{If Euclid observes in or near the ecliptic in the framework of an extended mission or during time that has not already been allocated to the main survey, the number of Solar System relevant targets observed could rise from 150,000 to roughly 5\,000\,000. A combined Rubin/Euclid characterization of such a large number of objects would have an enormous impact on Solar System and planetary science as well as planetary defense.}{SS}

\rec{solar-system-1}{Timely publication and exchange of astrometry of Solar System Objects}{A timely publication of astrometric observations and activity flags of Solar System objects from both Rubin/LSST and Euclid through the IAU Minor Planet Center is crucial for cross identification and follow-up observations of high value scientific targets. Rubin SSP is planning on submitting potential discoveries and astrometric observations on a daily basis. The Euclid team is currently assessing the quality of astrometric measurements that could be submitted in a timely manner, but submission is considered.}{SS}

\rec{solar-system-2}{Maximise the number of near-simultaneous Rubin-Euclid observations}{Tweak the Rubin/LSST scheduler to maximise near-simultaneous observations of Euclid fields, where possible. Maximizing near-simultaneous observations will greatly improve orbits of objects in the Inner Solar System and allow for a better characterization of moving objects in the entire Solar System.}{SS}

\rec{solar-system-4}{Combine Rubin and Euclid photometric catalogs of Solar System Objects}{This will improve taxonomic classification, phase curves and absolute magnitudes. Combining photometric catalogs will also lead to a densification of sparse light curves that better constrain the size, shape and spin, and multiplicity/binarity as well as composition of observed SSOs.}{SS} 

\rec{solar-system-5}{Observe the ecliptic plane with Euclid when possible}{Whenever possible, having Euclid observe in or near the ecliptic plane will improve physical characterization of up to millions of SSOs and increase the chances of discovery for small and ``dark'' objects. Both of those contributions would have an enormous impact on Solar System science, planetary science and planetary defense.}{SS}

\subsection*{Specific Derived Data Products} \label{sec:solar-system-products}

\begin{figure*}[ht!]
\centering\includegraphics[width=12cm]{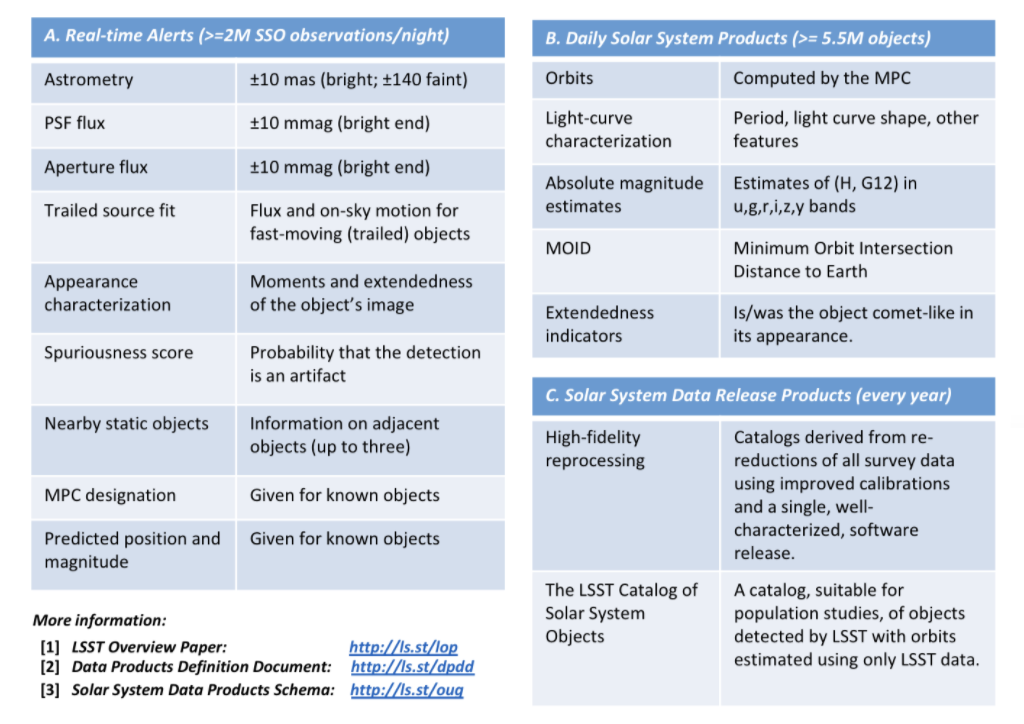}
\caption{Rubin Solar System object data products to be shared with Euclid (\url{https://lse-163.lsst.io/}).  \label{fig:solar-system-rubin_dp}}
\end{figure*}

\begin{figure*}[b]
\centering\includegraphics[width=15cm]{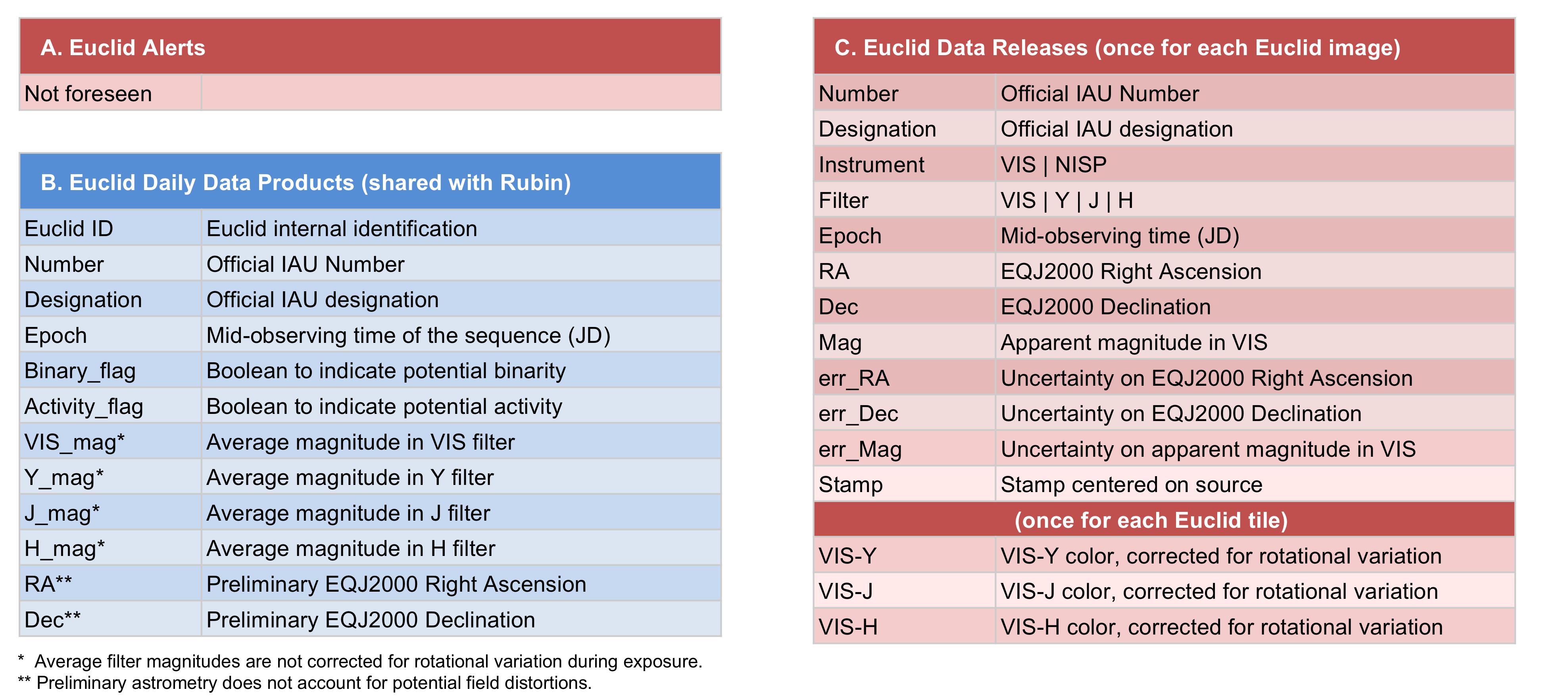}
\caption{Euclid Solar System object data products.  \label{fig:solar-system-euclid_dp}}
\end{figure*}
\subsubsection*{Timely Astrometry and Activity Reporting}
Combining Rubin and Euclid astrometric measurements can be done on a catalog level and does not require a joint pixel level analysis.
LSST will publish astrometric and photometric measurements, appearance related quantities of known Solar System  Objects, and SSO candidate observations in near real-time during the night of observation. These so-called ``alerts'' are publicly available and will be distributed through alert brokers. Observations of known SSOs and SSO candidate observations with a high likelihood of being new discoveries are packaged and sent to the IAU Minor Planet Center (MPC) on a daily basis. Improved orbits for known objects as well as preliminary orbits for new discoveries are then retrieved from the MPC and included in the observation attribution pipeline of the following night. The reason for including the MPC in the loop is to take advantage of the latest data submitted by other observers, such as Euclid. Euclid offers an API to query for preliminary astrometric measurements that is updated on a daily basis. Preliminay astrometry can either be accessed by Rubin Solar System Processing or be forwarded to the MPC. Euclid's high accuracy astrometry will be submitted to the MPC when it becomes available. 
The MPC, thus, constitutes the perfect platform for the two projects to exchange astrometric and orbital data at essentially zero additional cost.


\subsubsection*{Joint Characterization}
Combining Rubin and Euclid color photometry on a catalog level requires the exchange of color information. Astrometry and filter magnitude information from LSST will be submitted to the MPC on a daily basis, where it becomes open-access. Daily Solar System Products such as estimates of absolute magnitudes in ugrizy filters as well as G12 color are accessible to data-rights owners but can be shared with Euclid. In particular, this would encompass the Solar System Object and Solar System Source tables and the corresponding DIA source table entries as outlined in the Rubin Data Products Definition Document \citep{LSE-163}. 
Euclid activity flags and color information can be made available to Rubin exclusively with the express understanding that this data or derived data products would not be published before Euclid data releases. See also \figref{fig:solar-system-rubin_dp} and \figref{fig:solar-system-euclid_dp}.
Color photometry and/or general image analysis of combined stamps would require the exchange of stamps. Stamps for LSST alerts are accessible via \href{https://www.lsst.org/scientists/alert-brokers}{community alert brokers}. Stamp generation is not foreseen for each Euclid observation, but stamps could be generated on the fly upon request. A combined stamp analysis would be relevant for near-simultaneous observations of asteroids and comets, and observations of extremely slow moving objects in the outer Solar System that are observed by both Rubin and Euclid.

\ddp{solar-system-astrometry}{Timely Solar System Object astrometry}
{A catalog of astrometric measurements of Solar System Objects individually submitted to the IAU Minor Planet Center by Rubin and Euclid in a timely manner (within days).}{From the start.}{SS}

\ddp{solar-system-stamps}{Stamps for Solar System Objects}
{Stamps of Solar System Objects jointly observed by Euclid and Rubin and derived pixel values. Stamps will be most useful for candidate SSOs that were observed at the same time by both surveys. Stamps sizes analogous to LSST alert cut-outs are recommended. Observations of active/extended objects such as comets could merit the occasional, mutual exchange of larger stamps.}{From the start.}{SS}

\ddp{solar-system-lightcurves}{Solar System Object light-curve catalog}{A multi-timescale catalog of Solar System Object light curves combining both Rubin (sparse) and Euclid (dense) light curve data.}{Post Euclid DR1, repeated after every Euclid DR.}{SS}

\ddp{solar-system-shapes}{SSO shape catalog from lightcurves}{A catalog of approximate shapes of asteroids from combined Rubin (sparse) and Euclid (dense) light curves.}{At LSST DR5 = Euclid DR3.}{SS}



\subsection*{Algorithms} \label{sec:solar-system-algos}

\subsubsection*{Timely Astrometry and Activity Reporting}
Algorithms for combining astrometry and orbit fitting exist at the Minor Planet Center. Reprocessing of data based on updated orbits is done individually by the Rubin and Euclid projects.
Linking algorithms to identify observations of known SSO and potential SSO discoveries exist \cite[e.g.][]{eggl2020heliolinc3d,moeyens2021thor} and have been developed by Rubin and Euclid independently due to different modes of observation. 
\subsubsection*{Joint Characterization}
Algorithms for combining color photometry on a catalog level exist and have been demonstrated in e.g. \citet{carry2016spectral} to derive taxonomic classifications and \citet{moeyens2020atm} and \citet{ivezic2021predicting} to provide constraints on SSO sizes. There is little doubt that those algorithms can be successfully applied to Rubin and Euclid data. 
Algorithms for shape modeling exist \citep[e.g.][]{muinonen2020asteroid}, but their integration with Rubin and Euclid colors has not been demonstrated. However, those algorithms have been successfully demonstrated for a combination of ground based and sparse Gaia photometry.
Algorithms for image analysis of combined stamps do not exist because they will depend on individual users' science goals. 
Basic algorithms for detection of cometary activity (quantifying ``fuzziness'' or extendedness) exist for both Rubin and Euclid and results from them are included as activity detection flags in instant/daily releases. More advanced algorithms (e.g. tail detection) are being developed within the Rubin SSSC, these could also be applied to stamp images from Euclid. Some adaption may be required to deal with the fact that Euclid SSO images will be trailed due to object motion and long integration time.

\subsection*{Technical Considerations} \label{sec:solar-system-resources}

\begin{table*}[t]
\centering
\begin{tabular}{ p{3cm}|p{3cm} |p{5cm} |p{5cm}  }
 \textbf{Name} & \textbf{Approx. Size} & \textbf{Details} &\textbf{Description}\\ [0.5ex] 
 \hline
shapes &  2GB &  150k * 1k light curve points * vertex xyz * 32bit float & Shapes can be stored in text files containing vertex coordinate information.  \\
light curves & 1GB & 150k * (1k light curve points + 10 derived parameters) * 32bit float  & Joint Euclid - Rubin light curve data files with filters, measured filter magnitudes, time stamps and derived color information such as spectral slopes and absolute filter magnitudes.\\
stamps & 13GB & 150k * 10 filters * 30x30pixel * (32 + 32 + 16)bit float (flux + variance + mask)  & Stamps of objects jointly observed by Euclid and Rubin and derived pixel values in all filters. \\ [1ex] 
 \hline
\end{tabular}
\caption{A preliminary list of proposed Rubin-Euclid derived data products (DDP) for Solar System science. DDP size estimates are approximate, uncompressed file sizes for a population of 150k jointly observed Solar System Objects.}
\label{tab:ddp}
\end{table*}

\subsubsection*{Timely Astrometry and Activity Reporting}
Due to the existing interface with the MPC no extra processing power is required on the Rubin side. 
Similarly, the process of generating and submitting astrometric data to the MPC is being performed by Euclid independently.
As for sharing activity information, Euclid will make activity and binarity flags publicly available via API query. LSST data brokers could obtain this information and merge it with the LSST alert stream. However, details of this process have yet to be discussed with brokers. 

\subsubsection*{Joint Characterization}
When combining color photometry on a catalog level, processing requirements are negligible. However, if this is to be done on a timely manner, an exchange server/protocol is needed. Rubin could grant Euclid access to daily snapshots of the Solar System Object tables. The exchange of Euclid Solar System Object data would likely require a small database to be set up and accessible via an API, which is planned on the Euclid side.
Combined shape modeling from light curves can be more resource intensive. We estimate that a full inversion of all jointly observed asteroids (roughly 25k per year) would require on the order of 100k CPU hours per year, based on Asteroid@home statistics and \citep{muinonen2020asteroid}. RAM requirements are negligible. GPU implementations may be available. The time scale for a joint processing would be once per Data Release. 
Detailed estimates for joint image analysis or activity detection in combined stamps are difficult to assess as they depend on the use case. 
Table \ref{tab:ddp} contains a list of potential derived data products.

\subsection*{Timescale}  \label{sec:solar-system-timescale}

\subsubsection*{Timely Astrometry and Activity Reporting}
Depending on the agreed submission rate of Euclid data to the MPC, the exchange of astrometry and derived data products via the MPC is expected to happen roughly on a daily basis.
Similarly, updates on activity/binarity of objects on the Euclid side that can be shared through LSST data brokers are expected roughly on a daily basis.

\subsubsection*{Joint Characterization}
Joint analysis of color photometry on a catalog level as well as combined shape modeling will likely happen around Data Releases. 
Time frames for the analysis of combined stamps are unclear at this stage, but for activity detection there is a scientific motivation for alert generation on short (days) timescales.
\subsection{Milky Way halo \& Local Volume}\label{MW_LV}

\contributors{\hyperref[author:aferguson]{Annette Ferguson}, Keith Bechtol, Jeff Carlin, Roelof de Jong, Ariane Lan\c{c}on, Søren Larsen, Marina Rejkuba}

The high spatial resolution near-IR imagery of Euclid, and the deep optical time-sampled photometry of Rubin, will each offer unprecedented opportunities to study resolved stellar populations in the Milky Way, the Local Group and Local Volume galaxies out to $\approx\,$5-7\,Mpc.  While each survey on its own is poised for breakthrough science,  their combination is likely to be truly transformative. In this Section we highlight the high-level science drivers, and detail the kinds of DDPs and algorithms that this community requires in order to realise the full potential of these unique and complementary datasets. 

\subsection*{Science Drivers} \label{sec:MW_LV-drivers}

Our understanding of the assembly of the Milky Way has undergone a dramatic revolution in the last few years thanks to the spectacular dataset provided by ESA's Gaia mission.  The delivery of precise distances and proper motions for over a billion stars has led to a variety of exciting results, such as the discovery of at least one significant early accretion event in the history of our Galaxy, a large number of new stellar streams in the Galactic halo and the confirmation of huge low surface brightness tidal extensions around many globular star clusters \citep[e.g.][]{Helmi2018, Ibata2021, Kuzma2021}.  While the remaining Gaia data releases promise much more in the coming years, especially when coupled with new and forthcoming  high-multiplex spectroscopic facilities such as WEAVE, DESI and 4MOST, some of the most fascinating and poorly-understood components of the Milky Way remain beyond its reach.  For example, at one extreme,  main sequence stars lying at radii of $\gtrsim\,$30\,kpc are too faint for Gaia, making the outer halo of the Milky Way, with its repository of diverse dwarf satellites, ancient globular star clusters and predicted copious quantity of low surface brightness streams, essentially inaccessible for detailed study.  This limits our ability to decipher fully the accretion history of the Galaxy, because the clues gleaned from the inner halo are extremely challenging to interpret due to the rapid phase-mixing in these parts.  On the contrary,  the long dynamical timescales which characterise the outer halo allow it to preserve a rich source of information about the accretion history. Moreover, the outer halo provides an excellent laboratory for hunting for clues about the nature of dark matter. For example, the existence of gaps, spurs and peaks in cold tidal streams may signify impacts with dark matter sub-halos \citep[e.g.][]{Bonaca2019}. At the other extreme, crowding and extinction in the disc plane and central regions of the Milky Way have rendered these regions also very challenging for Gaia. Yet, these are the locations where the bulk of the Milky Way’s stellar mass resides. These components of the Milky Way are discussed further in \secref{MW_GP} of this report.

The success of the Gaia mission has also brought to the fore a more fundamental question -- how should we view the Milky Way in terms of the disc galaxy population at large?
Indeed, it is becoming clear that the Milky Way cannot even be crudely described as evolving in isolation, with the influence of the Magellanic Clouds in shaping the structure of the outer disc and stellar halo becoming increasingly recognised  \citep[e.g.][]{Garavito2021}.
While steady progress is being made on understanding the formation history of our nearest large neighbour, M\,31 \citep[see the review by][]{Ferguson2016}, a considerably larger sample is required if we are to definitively address the question of whether the Milky Way is representative. The Local Volume, loosely defined as within 5--7 Mpc of the Milky Way,  contains many excellent targets for such galactic archaeological studies, including searches for tidal streams, stellar halos and faint satellites.  

The Euclid and Rubin surveys are each positioned to make major breakthroughs on many of the questions outlined here. Covering $\approx$\,15,000\,deg$^2$ of sky, Euclid's deep near-IR photometry  will open windows on luminous evolved stellar populations (e.g. old red giant branch (RGB), intermediate-age asymptotic giant branch (AGB), young red and blue supergiants) in the Galactic halo and in galaxies out to $\sim$\,5\,Mpc distances.  Furthermore, its high spatial resolution imagery will facilitate unprecedented star/galaxy separation over very large scales, currently the most prohibitive obstacle for studies of extremely diffuse structures in the Milky Way halo and in the peripheral regions of nearby galaxies \citep[e.g.][]{Bernard2016,Pucha2019}. Euclid will also enable the identification of globular star clusters in the Local Volume directly through their partially-resolved morphologies, opening up the possibility to hunt for such objects in the field as well as in the vicinity of galaxies.   On the other hand, Rubin provides great depth in the optical pass-band, where the more numerous but fainter main sequence stars dominate, and  also time-sampled photometry, enabling the identification of periodic variable stars, which can be used for distance derivation and population dating.   Rubin's multi-band optical photometry is also critical for characterising the nature and properties of different stellar populations (e.g. spectral types, photometric metallicities). 

The DDPs proposed here will allow the Euclid and Rubin communities to:

\begin{itemize}

\item Determine the accretion history of the Milky Way through studying the properties of the Galactic outer halo, including the detection and characterization of stellar debris within it;

\item Search for and study new dwarf galaxies in the Milky Way outer halo, which constrain the satellite luminosity function, the nature of dark matter and the formation of galaxies at the extreme end of the galaxy mass scale;

\item Analyse and model the detailed morphology of thin stellar streams in the outer halo, yielding insight on the lumpiness of the Milky Way's gravitational potential;

\item Search for and characterise extended stellar halos, tidal debris streams, new dwarf satellites and globular clusters in Local Volume galaxies out to $\sim$\,5-7\,Mpc, enabling a first detailed measurement of the ensemble (recent) accretion histories of Milky Way analogues and enabling our understanding of the Milky Way's history to be placed in context;

\item Quantify the recent star formation and chemical enrichment histories of galaxies out to $\sim$\,5-7\,Mpc through analysis of bright evolved stellar tracers, such as Mira Long Period Variable stars, and multi-band photometry. 
 
 \end{itemize}

\subsection*{Specific Derived Data products} \label{sec:LV-products}

Two common themes that emerged from the submitted cases are multi-band photometry and star/galaxy separation; these are both reflected in DDPs suggested below.  

\ddp{lv-catalog-merging}{Multi-band Merged Point Source Catalog}{A minimal desire is a multi-band merged catalog containing ugrizyYJH photometry for point sources that have been detected above a fixed threshold (e.g. 5 sigma) in Euclid NISP and in Rubin.  This straightforward joint data product would provide static point source photometry across a broad photometric baseline, allowing crude spectral-typing (and hence distance estimation) and photometric metallicity determination.   Stellar sources would be defined on the basis of NISP photometric measures (e.g. PSF versus fixed aperture).}{From the start.}{LV} 

\ddp{lv-catalog-variables}{Multi-band Merged Point Source Catalog with Time-Series Information for Periodic Variables}{Pulsating variable stars, such as classical Cepheids, Mira Long Period Variable stars and RR Lyrae, are  valuable tools for resolved stellar population studies. The Cepheids are young stellar populations, tracing recent star formation events, and are well-known standard candles. The Miras are located on the AGB phase and 
trace intermediate-age and old stellar populations, while RR Lyrae trace old Population II stellar populations that formed $\sim$\,10\,Gyr ago. The distributions of periods and amplitudes are relatively easy to obtain from Rubin multi-epoch photometry, while Euclid will provide the near-IR magnitudes that are fundamental for accurate distance measurements through well-defined period-luminosity relations in near-IR. The DDP would consist of merged catalog point source photometry and period/amplitude information for pulsating variable stars identified by Rubin.}{From the start.}{LV} 

\ddp{lv-catalog-list}{List-driven Multi-band Photometry}{The first improvement on simple catalog merging would be a DDP consisting of list-driven ugrizy and YJH photometry for sources detected in Euclid NISP, and in Rubin, above some detection threshold. This is likely to lead to more multi-band detections, bringing  in sources with extreme colours and/or near the detection limits of the surveys.  Star/galaxy information would be provided by NISP photometric measures. }{From the start.}{LV} 

\ddp{lv-catalog-force}{Joint-pixel Multi-band Forced Photometry}{Joint pixel ugrizyYJH photometry is likely to substantially increase the size and quality of the catalog, bringing in further faint sources and improving photometric and astrometric precision.  This will in turn yield better statistics on the faintest stellar populations (the bulk of which will reside in Local Group and Local Volume galaxies) and enable tighter constraints on stellar properties. Star/galaxy information would be provided by NISP photometric measures. This DDP is dependent on algorithm development.}{Post Euclid and Rubin DR1}{LV} 

\ddp{lv-catalog-force-deblend}{Joint-pixel Deblended Multi-band Forced Photometry}{Joint pixel ugrizyYJH deblended photometry using VIS will yield gains for stellar populations where there is a moderate-to-high degree of stellar crowding. This includes the peripheral regions of Milky Way globular clusters, the main bodies of bright Milky Way dwarf satellites and galaxies in the Local Group and Local Volume. This DDP is dependent on algorithm development and represents the most complex approach.}{Post Euclid and Rubin DR1 }{LV} 

\ddp{lv-star-galaxy}{Joint-pixel Star/Galaxy Classification}{A desired eventual DDP is star/galaxy classification based on the joint analysis of Euclid VIS morphology and Euclid+Rubin ugrizyYJH photometry of all sources  common to Euclid and Rubin.  Incorporating both morphology and colour information will likely  improve the classification for faintest  sources, impacting analyses of the most diffuse structures in the Milky Way halo and those around galaxies out to several Mpc distances. This DDP is dependent on algorithms development and is an addition to \ddplvref{lv-catalog-force-deblend}.}
{Post Euclid and Rubin DR1}{LV} 

\subsection*{Algorithms} \label{sec:LV-algos}

The DDPs outlined here are for the most part straightforward and have many commonalities with those emerging from other science areas; as a result, they should not require algorithmic development beyond that outlined elsewhere in this report (e.g. techniques for forced photometric extraction and deblending). 

The only exception to this is the desire for eventual joint star/galaxy separation using both Euclid VIS shape information and multi-band photometry.  Previous efforts provide a good starting point \citep[e.g.][]{Fadely2012, Kim2017, Slater2020} but more effort will be needed to tailor the approach to the specific Euclid and Rubin datasets.

\rec{local-volume-1}{An Optimal Star/Galaxy Classification}
{Explore how to optimise star/galaxy separation using a combination of Euclid VIS morphology and multi-band ugrizyYJH photometry. }{LV}

\subsection*{Technical Considerations} \label{sec:LV-resources}

There are no technical considerations beyond those outlined elsewhere in this report. 

\subsection*{Timescale}  \label{sec:LV-timescale}
Desired timescales have been indicated against the individual DDPs outlined above. The joint-pixel processing efforts are likely to be the most challenging and our needs are not likely to drive the timescales for the production of these. 

\subsection{The Galactic Plane}\label{MW_GP}

\contributors{
\hyperref[author:ebachelet]{Etienne Bachelet (WG)}, Robert Blum, Herv\'{e} Bouy, Leo Girardi, Rodrigo Ibata, Eduardo L. Mart\`in}

\secref{sec:observing}: Survey strategy optimisation for DDPs, describes the nominal Rubin and Euclid surveys that are the basis for the general DDP discussion. Rubin/LSST aims to cover about 18,000\,deg$^2$, mostly in the Southern sky through its Wide-Fast-Deep component, including the Galactic plane. On the other hand, the Euclid Wide survey focuses only on high galactic latitudes regions ($|\bf b|$\,$>$\,23), and will cover 15,000\,deg$^2$ over the whole sky. However, starting year 3 of its 6-year long survey mission, Euclid will exhaust the observable area within its Region of Interest (\figref{fig:euclid-survey}) when it crosses the plane of the Milky Way. This will result in available observing time (nearly 9 months total from year 3 to year 6) for use by scientists outside of the Euclid Consortium. This time will be allocated by ESA approximately two years after launch to specific scientific investigations in areas outside the Euclid Region of Interest or revisit areas previously observed. This Section derives from this consideration and focuses on a few non-binding illustrative science cases exploiting the complementarity both in wavelength and time of the LSST and Euclid surveys as unique opportunities to study the Milky Way plane. 

This Section was developed as an exploratory exercise to illustrate the scientific potential of new observations tailored to maximize the Rubin-Euclid scientific synergies. DDPs are suggested but are conditional to decisions that have yet to be made. As such they are on a different level from all the other DDPs presented in this report and are assigned to a dedicated conditional Tier 4 effort.

\subsection*{Science Drivers} \label{sec:MW_GP-drivers}

{\bf Microlensing}

Thanks to its large field of view and faint detection limit, it is expected that LSST will detect thousands of microlensing events in the entire Milky Way every year \citep{Sajadian2019}, and potentially towards the Magellanic Clouds \citep{Sajadian2021}. This is a formidable opportunity to study the population of faint objects of the galaxy, including (free-floating) planets, brown dwarfs and stellar remnants. The multi-band observations from LSST will provide constraints on all microlensing sources as well as helping on the classification of microlensing alerts (because the microlensing magnification is achromatic). Twice a year, the Galactic plane will dominate the accessible sky that Euclid could possibly observe due to observing constraints. Euclid could provide complementary near-infrared bands and additional epochs that offer unique constraints on the microlensing models. We note that these complementary epochs could also provide unique constraints on the longest events that will be detected by the Roman microlensing survey \citep{Spergel2015}. Moreover, the distance between Euclid (at the Lagrangian point L2) and the Earth leads to a small difference of the line of sight towards microlensing events which offers a unique opportunity to systematically measure the microlensing parallax. This ultimately leads to the estimation of the mass and distance of lenses, see for example \citep{Wyrzykowski2020}. Finally, the superior resolution of Euclid would provide information on the lens and source proper motions and brightnesses, providing unique constraints of the mass and distance of lenses (Bachelet et al. 2021, under review).

{\bf Lucky astrometry with Rubin+Euclid}

Most of the baryonic mass of our Galaxy is concentrated near the Galactic plane, where the stellar populations overlap in a complex mixture of components and sub-structures. Disentangling these features in order to understand their current structure and evolutionary history will require excellent phase-space measurements and excellent photometry over a large fraction (half) of the Galactic plane. The necessary photometric and astrometric data could be provided by Rubin, but this survey will be hampered by the fact that the relevant areas of sky correspond to the most crowded regions of the Milky Way. The broad optical Euclid band could provide the key information to render the Rubin observations most useful for Galactic dynamics: it will be possible to use the high-resolution Euclid imaging to identify those stars that are (by chance) unaffected by nearby neighbors at Rubin resolution, and thus obtain clean astrometric and photometric measurements that can be used for dynamical analyses.
Note that similar surveys in the Galactic disk conducted with the Nancy Grace Roman Space Telescope will be able to reach much deeper magnitudes with better spatial resolution, but they will not be competitive with Rubin+Euclid in terms of their spatial coverage. Galaxies are complex and spatially non-homogeneous structures, and therefore require panoramic studies to unveil their full dynamical information.
Here we envisage skirting the Galactic disk approximately 180 degrees along the plane, with a continuous band that is approximately 5 degrees wide in Galactic latitude. This will allow us to obtain the best large-scale kinematic portrait of our Galaxy’s disk using the Rubin astrometry and photometry at the location of stars shown to be uncontaminated by neighbors from the Euclid images (or for instance, the $\sim$10\% least contaminated stars in a region could be taken).

{\bf Star Forming Regions}

The study of star forming regions has brought rich information on stellar populations, and in particular for low mass objects \citep{Zapatero2006, Burgess2009, Quanz2010, Marsh2010}. This includes free-floating planets, celestial objects of planetary mass that are not in orbit of a known host. The detection of such bodies is difficult due to their extreme faintness. Generally, candidates are selected using a combination of multi-wavelength photometry (optical and near-infrared) as well as proper motions. This has be done in the past, but only for relative small samples, using the Hubble Space Telescope and Adaptive Optics observations \citep{Kirkpatrick2012, Esplin2017, Muzic2017}, or deep multi-wavelength surveys \citep{Esplin2019, Miret-Roig2020}. The large field of view, complementary bands and spatial resolution of LSST and Euclid are ideal to conduct an unique homogeneous census of these regions and to provide unique constraints on the end of the initial mass function and the still elusive fragmentation limit. In particular, conducting a joint survey of $\sim$\,150 square degrees of the star forming regions located in Ophiuchus, Scorpius and Orion constellation would allow a complete census of the stellar population down to the planetary mass regime, i.e. 1 to 2\,$M_{\rm Jup}$. The detection of hundreds of ultra-faint free-floating planets and thousands of brown dwarfs are expected, and these objects will be privileged targets for the future generation of great observatories such as the James Webb Space Telescope, the Extremely Large Telescope and the Thirty Meter Telescope. Euclid near-IR spectra will be a source of information of the brightest young planets' atmosphere, otherwise extremely challenging and difficult to obtain because of the dazzling light of a host star.  The unique combination of deep, homogeneous and multi-band photometry of the entire area from both surveys is critical for the objects characterization and the completeness of the census, as well as the elimination of contaminants. By combining the datasets, the proper motions can also be measured to a precision well below $\sim$ 1 mas/yr (or 5 to 20\,$\sigma$ given the average motions of these groups), providing unique constraints on the different populations. Moreover, the superior Euclid VIS resolution will provide a view of multiple systems to unprecedented details (down to 20\,au in USco  and 60\,au in Orion). This will permit systematic study of the multiplicity of the entire regions, and ultimately place constraints on the formation process of brown dwarfs and free-floating planets.

\fnd{galactic-plane-1}{Euclid-Rubin joint observation enhances the characterization of microlensing events}{Combining datasets of the two surveys provides a better coverage of the lightcurves as well as color-information that evolves during the course of the event and also provides constraints on the microlensing parallax. The superior resolution of Euclid also allows measurements of the relative proper motion and the fluxes of sources and lenses.}{GP}

\fnd{galactic-plane-3}{Euclid Image cutouts}
{The superior Euclid resolution offers the unique opportunity to study the motions of the lenses and sources prior / posterior to the microlensing magnification peaks. This can provide unique constraints on the relative proper motions and star fluxes that ultimately constrain the masses and distances of microlenses. The needed angular size of cutouts is relatively small, a few arcseconds around the location of events are sufficient.}{GP}

\fnd{galactic-plane-4}{Catalog cross-match}
{The combination of the photometric catalogs will improve the classification, characterization and modeling of the previously described objects of interest.
The combination of astrometry is much more involved but individual measurements from both missions will be useful for initial reconstruction of the proper motions.}{GP}

\rec{galactic-plane-1}{Maximize the survey overlap area}
{The proposed microlensing science requires that the surveys observe the same area of the sky and therefore it will be beneficial to maximize both surveys' overlap region, as well as synchronizing their cadence as much as possible.}{GP}

\subsection*{Specific Derived Data products} \label{sec:GP-products}

\label{sec:GP1-products}
\ddp{mw-catalog-cross-match}{Catalog cross-match}{Both photometric and astrometric measurements can be matched at a catalog level. LSST will publicly release alerts on transient events via brokers, which will carry out classification. The possibility to access all information collected by both missions for any sky location is vital for all the science cases discussed above.}{Pending definition of Euclid's non-allocated time.}{GP}

\ddp{mw-catalog-astrometric solution}{Complete Astrometric Solution}{The complete astrometric solution at each individual epoch, coupled with the combined grizYJH photometry, enables the study of FFP and low mass brown dwarfs in star forming regions.}{Pending definition of Euclid's non-allocated time.}{GP}

\ddp{mw-stellar-spectra}{NISP stellar spectra}{Although the field will be rather crowded, a fair number of spectra (or portions of) will be available and will allow us to refine the selection and derive accurate effective temperatures.}{Pending definition of Euclid's non-allocated time.}{GP}

\ddp{mw-euclid-cutouts}{Euclid Image cutouts}{Analyzing Euclid image cutouts at the location of microlensing events will allow the estimation of the source and lens fluxes, as well as their relative proper motions, via double-PSF fitting.}{Pending definition of Euclid's non-allocated time.}{GP}

\subsection*{Algorithms} \label{sec:GP-algos}
\begin{itemize}

\item {\bf Astrometric and photometric catalog cross-match}
The various science cases previously discussed require joint photometry, including difference image and forced photometry. The algorithms exist but some work is needed to adjust to Rubin and Euclid specifications.

\item {\bf Photometric classification and modeling}
The classification of catalog objects and lightcurves is desirable for most of the science cases. Many algorithms have been developed in the past for this purpose, especially using machine learning techniques. But the implementation on multiple pass-bands, especially with near-infrared data from Euclid, remains an open question. The modeling of microlensing events can be done in real time using the pyLIMA package \citep{Bachelet2017}. This has be done already by the FINK broker \citep{Moller2021}.

\item {\bf Euclid cutouts}
Algorithms to collect cutouts of Euclid images based on sky location requests are well established. The Aladin Service at CDS is a good example. However, the scalability of such algorithms to the data volume considered needs to be assessed.

\item {\bf Complete astrometric solution}
The complete astrometric solution algorithms exist but are resources intensive. Therefore, it is necessary to study how they can be implemented for the large data volume considered.

\end{itemize}

\subsection*{Technical Considerations} \label{sec:GP-resources}

Additional computing should be minimal, unless the Euclid data were to be used as an extra epoch for the astrometric solutions.

\subsection*{Timescale}  \label{sec:GP-timescale}
The microlensing science case needs almost real-time DDPs for classification/modeling and additional follow-up observations. A daily update on the lightcurve DDP seems a minimum. Rapid access to Euclid images cutouts is desirable. For the other science cases, yearly updates are sufficient.

\subsection{Local Universe} \label{sec:LU}

\contributors{\hyperref[author:cconselice]{Christopher Conselice (WG)}, \hyperref[author:jcuillandre]{Jean-Charles Cuillandre (WG)}, Ivan Baldry, Sarah Brough, Michele Cantiello, Jeff Carlin, Chris Collins, Pierre-Alain Duc, Annette Ferguson, Leslie Hunt, Sugata Kaviraj, Johan Knapen, Ariane Lançon, Søren Larsen, Mireia Montes, Polis Papaderos, Reynier Peletier, Javier Roman, Crescenzo Tortora, Chris Usher, Karina Voggel, Aaron Watkins}

\subsection*{Science Drivers} \label{sec:LU-drivers}

The local universe includes galaxies that are nearby and thus we can study them in greater detail than more distant galaxies. This gives us an opportunity to study the properties and evolution of galaxies in a way that is simply not possible to do for more distant galaxies. A joint multi-wavelength Rubin--Euclid analysis of these galaxies from the optical to the near-infrared is critical for better understanding their properties, including their stellar populations and structure. Nearby galaxies also allow us to study the structures of galaxies in a way that has not been possible before, namely the ability to study the distribution of diffuse outer light of galaxies. This outer light can include halo and debris from long past galaxy interactions. This also allows us to probe low surface brightness galaxies that have yet to be be discovered. 

There are therefore a few major issues that need to be addressed in the local universe with a combined Rubin--Euclid dataset. Many of these are outlined in \citet{Rhodes_2017}. We give an overview of the topics and issues here that are relevant to the local universe, as reported during the DDP discussion.\\

{\bf Pixel/Resolution Element SEDs}

Matching the pixels of a combined dataset between Euclid and Rubin leads to spectral energy distributions (SEDs) for nearby galaxies. Within this pixel matching, where the pixel sizes and astrometric alignment are done on Rubin and Euclid data, a unique science unfolds beyond what we can currently do for all but a handful of nearby galaxies and for each dataset alone. Euclid and Rubin probe different wavelength domains which means that there is spectral energy distribution for all objects for which both surveys overlap.   This can lead to, for galaxies, the determination of galaxy star formation histories, metallicities, dust contents, amongst other properties. The retrieval of these parameters and features is now commonly done using the so called ``SED fitting''.

With the high resolution data obtained from Rubin and Euclid for nearby galaxies this methodology can be done on a pixel by pixel basis for individual galaxies. This is a very powerful methodology as it allows us to determine the properties of not only individual galaxies, but sub-regions of galaxies. This also goes beyond a simple structural decomposition whereby we might be able to determine the properties of spiral arms/bulges/bars to one where we are determining the star formation and metal enrichment history of small aspects of galaxies.  The resolution element in which we can carry out this experiment depends upon the distance to the galaxy and is ultimately limited by the resolution of the instrument with poorer seeing, which will in this instance be from Rubin.

This requires that the actual data, that is, the imaging itself is aligned between the two data sets to a high accuracy. This will allow studies of the internal structures of tens of millions of galaxies in what can be considered a photometric Integral Field Unit (IFU) analogy. This will give us unique information about the formation history of galaxies that goes well beyond the thousands of actual IFU measures now ongoing or the total galaxy SEDs we can now measure for galaxies. This however requires that the data from Euclid and Rubin be processed together to assure this alignment and PSF/seeing matching is successful.\\

{\bf Low Surface Brightness Features}

A major advance that Rubin and Euclid will provide is the systematic search for low surface brightness galaxies and an in-depth exploration of the low surface brightness portions of galaxies. Both of these are relatively unexplored aspects of galaxies, and we still know very little about both of them. LSB galaxies have been known to exist in abundance since the 1980s, yet we are still discovering these systems. The great depth of Rubin and Euclid will provide us with a opportunity to study these systems and discover ones that go beyond the limits of what we can currently do.

LSB light in the outer portions of galaxies often contain diffuse structures that typically can only be seen through deep imaging. This includes light halos whose properties and features are still being characterized and discovered. These halos however probably can tell us important information about the formation history of galaxies.  Within the outer portions of galaxies we also find considerable amount of tidal debris that originate most likely from merging events.  Typically these are seen as either diffuse ``fan'' like structures or more thin diffuse ``lines'' that likely trace the previous orbit of a merging satellite galaxy with the primary system. The reason these outer diffuse structures are interesting for galaxy formation is that they are long lived, and thus can survive in the outer parts of massive galaxies for many Gyr.  They thus are a remnant of the earlier formation of the galaxy with structural and kinematic information that has in principle  been removed from the inner light due to dynamical processes such as relaxation.

Euclid and Rubin will both be able to probe this outer light in unique and complementary ways from the optical to the near-infrared. First, the vast number of galaxies sampled by both telescopes (nearby ``large'' galaxies in the local universe cover nearly 2\% of the sky) will allow us to study these properties in great detail for the first time and go beyond the tens-hundreds of galaxies studied in this way at present.  Together Euclid and Rubin will be able to examine the overlapping systems to study the nature of this diffuse light and discover together new LSB systems, Euclid opening in particular an unexplored LSB window in the near-infrared.  The combination of both telescopes will allow the full SED of these galaxies to be characterized and fit to stellar population models.  Furthermore, the discovery space of LSB galaxies will greatly expand by having deep data from both telescopes. It is perhaps possible to combine and confirm detections from one telescope with the other, in particular with VIS detections on Euclid. The combination of Rubin depth and Euclid resolution will be very complementary for the study of these systems and their discovery.

Diffuse galactic light is present over the whole sky and will appear in its most structured form to various degrees across the high galactic latitude sky common to Rubin and Euclid \citep{Scaramella_2021}. Recent studies \citep{Miville-Deschenes-2016,Roman2020,Marchuk2021} characterize the properties of these galactic cirrus and illustrate how they contaminate deep extragalactic observations. The color and morphological information offered by the joint Rubin-Euclid dataset will be key to disentangle the structured galactic cirrus from extragalactic features.\\

{\bf Globular clusters outside the Local Group}

Globular clusters populate the high density end of the range of galaxy substructures. Extragalactic globular clusters (EGCs) are interesting in many respects: they are tracers of galaxy halo masses as well as tracers of galaxies merging histories; their own stellar populations display specific chemical patterns that point to peculiar formation histories; they may be the origin of significant fractions of galaxy halo and field stars; they relate to ultra compact dwarf galaxies (UCDs) and to nuclear clusters in ways that are still unclear; free-floating clusters may remain to be discovered. Our current knowledge of EGC populations relies mostly on pointed observations or on surveys of the nearest galaxy clusters. With Rubin and Euclid, it will be possible to construct volume-limited catalogs of cluster candidates and UCDs over vast sky areas, without the selection biases associated with pointed observations. It will also be possible to overcome the systematic photometric errors that have plagued comparisons between the photometric properties of samples from different ground-based surveys, in particular in the near-infrared.

\subsection*{Specific Derived Data Products} \label{sec:LU-products}

The data products that are needed for the above science cases are the imaging data for both Rubin and Euclid. The pixelated data are required to carry out the analysis for the nearby galaxies where this science can be carried out, which are well resolved. However, this does not require the entire surveys' pixelated dataset, just those areas that cover nearby galaxies and their extended low surface brightness features (halo, streams, etc). This dataset implies developments of analysis tools and procedures that are unique to these large imaging experiments.

One issue that needs to be addressed is the systematic detection of nearby galaxies across both surveys. This basic process is fraught with issues, as has been shown when trying to detect large galaxies with standard procedures through e.g., the Sloan Digital Sky Survey. What occurs when using standard extraction is that a large galaxy gets shredded into many different detections as multiple objects. This is a real problem when trying to analyze these systems.  

We can get some idea of the influence of this issue within Euclid and Rubin by considering what fraction of the sky has galaxies such as these. A simple test for this is to determine how many galaxies across the sky will need special handling that is not provided by the typical routines for detection and deblending of images. This is done by taking the mass function for nearby galaxies from \cite{Baldry2012} and determining how many massive galaxies there are in the nearby universe that have a mass $>$\,10$^{\bf 10.75}$\,M$_\odot$.  Using the number density of these objects and including the size distribution from \cite{Lange2015} we calculate that roughly 2\% of the sky will be covered by galaxies larger than $\sim$\,10 \,arcminutes.  These galaxies will need special care to source extract.  The benefit of Rubin and Euclid for these is that the two surveys can be used together to optimize the source extraction for these systems from the optical to the near-infrared.  The use of nearby galaxy catalogs may also help in this process, or a data product of the detection for both telescopes such that they are cross correlated to ensure that the correct amount of light from given galaxies is equally measured from both surveys.

\begin{figure*}[t]
\centering\includegraphics[angle=90,width=9cm]{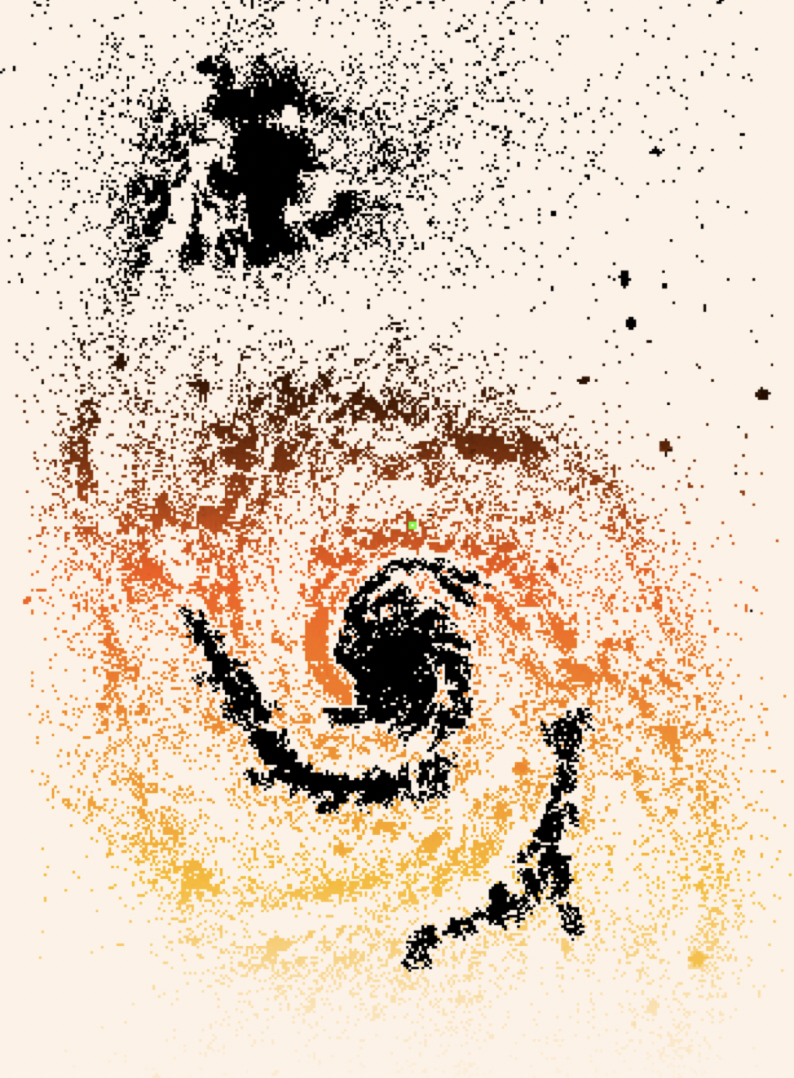} 
\caption{Example of how a nearby galaxy, in this case Messier\,51, will become shredded during the process of source extraction using parameters optimized to detected small distant galaxies, which are the primary targets for the deep Euclid imaging.  Each colour represents a different `detection' which is cataloged as a single unique object.  In a blind catalog using the same source extraction this galaxies, and others like it, would not be properly accounted for and any measurements of its properties, including photometry, would be entirely incorrect.    
\label{fig:LU_Figure1}}
\end{figure*}

For clusters of galaxies we will have the opposite problem in that nearby galaxies can become segmented into the same object.  This is a real problem when trying to determine the properties of galaxies as a function of their environment.  Using both Euclid and Rubin we should be able to determine the success of the source extraction methodology by examining how unique and successful this extraction has been done both both surveys. This would require the data product of catalogs that list positions as well as magnitudes for galaxies in dense regions. This data product however may need to be interactively done such that the optimization of the detection of galaxies in these dense fields is done consistently and correctly for both telescopes.

\rec{lusim}{Start a simulations study}{Starting a simulations study for issues outline above was identified as a high priority area.}{LU}

\ddp{lu1}{Joint pixel processing of large image cutouts}{Special reductions of both data sets are needed for some local universe science.  While it is unlikely that the pixel data from both Euclid and Rubin will be available for all galaxies, to properly process and analyze these nearby galaxies the individual pixels from both Rubin and Euclid are needed. This is a requirement to produce catalogs of `pixels' or `resolution elements' (however these may be defined) such that the flux or magnitude in each wavelength for both telescopes is produced. This will be crucial for all Rubin$+$Euclid ugrizYJH photometric bands as all of this data will be useful for SED fitting and thus the derivation of parameters for each individual element that can be combined to infer the total history of formation for that particular galaxy.}{From the start, tiered approach over the lifetime of the surveys and synchronised with the data releases.}{LU} 

\ddp{lu2}{Nearby galaxies structure \& morphological parameters}{Measuring structures of nearby galaxies to a high degree is made possible due to their nearby distance such that we are able to resolve the individual components of these galaxies with a high precision.  The data products in which we need to do this is the high resolution imaging of the galaxies themselves as imaged with both Euclid and Rubin.  These high resolution images can then be fitted together and separately to determine the robustness of not only basic morphological parameters such as size and Sersic indices, but also to identify and confirm the existence of features that require a high resolution, such as multiple bars, rings, and star forming complexes within the galaxies themselves.  The data in the form of pixels from both telescopes are needed for this as  Rubin will obtain deeper data useful for the outer portions of galaxies, and Euclid will provide higher resolution imaging with VIS in particular that is essential for fitting the inner portions and substructure.  Algorithms should be developed to make use of both datasets in simultaneous fitting to provide robust structural parameters.}{From the start, tiered approach over the lifetime of the surveys and synchronised with the data releases.}{LU}

\ddp{lu3}{Dedicated Low Surface Brightness pixel data reductions} {To detect and study the low surface brightness light coming from galaxies we will need to reduce the data from both telescopes in unique ways that accounts for the flat fielding and sky background removal.  Imaging for nearby galaxies from both telescopes will also be required to determine if the feature which we see in one telescope are also present in the other, and help disentangling extragalactic features from galactic cirrus.  Therefore for this, the pixel data is uniquely needed for this DDP from both telescopes. This will be required to ensure a full accounting of this diffuse light and to avoid, or at least limit, the amount of false positive detections within these features.}{From the start, tiered approach over the lifetime of the surveys and synchronised with the data releases.}{LU} 

\ddp{lu4}{Multi-band merged catalog with compactness-sensitive measurements}{The first step in any study of extragalactic globular clusters (EGCs) is to separate them from foreground stars or background galaxies. This can be done on the basis of morphology (relatively nearby) and of color. The combination of u, i and H photometry is expected to be particularly efficient in the Rubin-Euclid context. With typical half-light diameters of 5\,pc, EGCs progressively become point-sources beyond $\sim$\,20\,Mpc (depending on their individual luminosity), and UCDs much further out. The Multi-band Merged Catalog requested here (unlike \ddplvref{lv-catalog-merging} for instance) should include any VIS measurements that may be exploited to detect departures from a pure PSF-shape. It should also include comparable measurements of the integrated photometry in all bands, i.e. measurements after the PSFs in the filters have been homogenized. These are needed to interpret spectral energy distributions and estimate parameters such as metallicity, age, extinction. In principle the requested catalog can be ``list-driven'', i.e. based on a pre-selection, but any pre-selection based on either Euclid only, or Rubin only, will be significantly more permissive to contamination and biased than a combined selection.}{From the start (``list-driven'' approach), with a tiered approach over the lifetime of the surveys as pixels become co-located, synchronised with the yearly data releases.}{LU}

\ddp{lu5}{Multi-band merged catalog of objects without proper motion}{A  merged catalog complementing \ddpluref{lu4} with Rubin and Euclid-based proper motion information (relative to Gaia, for instance), will be instrumental to reduce stellar contamination in catalogs at the faint end: any source with a proper motion will not be an extragalactic globular cluster.}{Towards the end of the surveys.}{LU}

\subsection*{Algorithms} \label{sec:LU-algos}

There are several algorithms that can be applied to combining Euclid and Rubin data.  The most obvious is pixel matching, whereby the same coordinate of objects is aligned to sub-pixel scale accuracy.  This can be achieved through a variety of methods, including using stellar positions to calibrate the astrometry to a subarcsecond precision.

For the SED fiting we will need several tools that exist in some format, but can be better adopted to use the unique aspects of Euclid/Rubin data combined.   We also will need to have algorithms that will match the pixels between the two surveys and produce seeing element spectral energy distributions that can be fitted by these codes.  The algorithms for this must be able to handle the spatial information provided by the data and thus must be adapted from the SED fitting methods that are now applied to entire galaxies.

For the LSB galaxies and their features, we will need new algorithms to both reduce the data to find these features, as well as measuring them.  Much of this work is now done by hand as the number of systems is still small, but this will not be possible with Euclid/Rubin when we will have orders of magnitude more galaxies where the outer light can be effectively measured.  Algorithms in the spirit of AutoProf \citep{Stone2021} thus need to be developed that can not only automatically measure this light and its spatial distribution, but also utilise the fact that this light can be measured by both Rubin and Euclid and make use of imaging from both telescopes when creating a data product of LSB features.

\subsection*{Technical Considerations} \label{sec:LU-resources}
See remarks in the previous section.

\subsection*{Timescale} \label{sec:LU-timescale}

The data can be used progressively until the survey areas are complete. The u-band data is particularly precious both for the separation of EGCs from contaminants and for the study of the internal stellar populations of EGCs. Hence we expect the EGC catalogs will continue to improve through the Rubin yearly releases.

\subsection{Transients} \label{sec:transients}

\contributors{\hyperref[author:ihook]{Isobel Hook (WG)}, \hyperref[author:ebachelet]{Etienne Bachelet (WG)}, Pierre Astier, Maria Teresa Botticella, Enrico Cappellaro, Stefano Cavuoti, Jose Diego, Dominique Fouchez, Melissa Graham, Jens Jasche, Rubina Kotak, Guilhem Lavaux, Florent Leclercq, Giuseppe Longo, Seppo Matilla, Gautham Narayan, Stephen Smartt, Charling Tao, Sjoert van Velzen, Benjamin Wandelt}

The study of transients requires a degree of urgency. It is not generally possible to return to the object later to get more data, and we often need to make rapid decisions to trigger follow-up observations. The sharing of data during the proprietary periods of Euclid and Rubin will be particularly beneficial, and the transients science case is an ideal driver for a shared DDP. Transients are also relatively rare sources (the number of pixels involved is not huge) and so could be considered as a special case where some constraints for the general sharing of data could be relaxed. 

Both LSST and Euclid surveys will discover new transients, and both will detect transients discovered by the other survey. The combined time-series data (including information about non-detections) on the transients themselves, and also the combined ``static'' data describing the environments of the transients are both highly valuable.  

\fnd{trans}{Joint Rubin-Euclid analysis of transients and their environments is highly valuable}
{Overall, joint Rubin-Euclid measurement of transient lightcurves and their environments are highly valuable because of the extended colour, spatial and timing information that the combination provides.}{TR}

\fnd{trans-subset}{Transients may be considered a special subset of Rubin and Euclid sources}
{Although numerous, transients still represent a relatively small fraction of the sources that Rubin and/or Euclid will detect. Hence they could be considered a special case for definition and later implementation of DDPs, where some constraints could be relaxed.}{TR}

\subsection*{Science Drivers} \label{sec:transients-drivers}

{\bf Supernova rate}

Measurements of the rates of all types of supernovae as a function of the star formation history for different parent stellar populations can be exploited to constrain the SN progenitor scenarios, the galaxy chemical evolution and the population of compact objects, i.e. neutron stars and black holes \citep{2017A&A...598A..50B, 2019A&A...625A.113G}. The required observational data are:
\vspace{-0.4cm}
\begin{itemize}[noitemsep]
\item  the statistics of SN detections and their (photometric) classification and,
\item the galaxy counts as a function of (photometric) redshift and their spectra energy distribution (SED) from which are derived their stellar population content (mass, star formation rate and history).
\end{itemize}
\vspace{-0.2cm}
The rates of the different SN types are linked to the properties of the host stellar population through the SN progenitor scenarios. The combination of Rubin/Euclid data is required to achieve (1) a much better constraint of the galaxy photometric redshifts and SEDs, because of the extended wavelength range, and (2) a more accurate assessment of the transient detection systematics, in particular for the estimate of the dust extinction bias.

{\bf Supernova cosmology}

Type Ia supernovae (SNe~Ia) are powerful standard candles that can be used to measure cosmological parameters. The next generation of surveys will dramatically increase the number of well-measured SNe~Ia, while also providing superior control of systematic effects. This will enable more accurate and precise measurements of the equation of state of dark energy and of its evolution with redshift. This in turn may help us understand the underlying cause of the accelerating expansion of the universe.

LSST is ideal for finding distant SNe Ia up to $z\sim\,$1. Its multi-colour lightcurves will enable distance measurements (provided they are sufficiently well-sampled, which depends on the survey strategy that is being developed now). Euclid has the potential to improve the results by adding near-infrared photometry that helps constrain the effects of dust extinction on SN brightness. In addition, there is a good opportunity to use Euclid near-infrared imaging in the YJH bands to extend a high-quality SN Hubble diagram to higher redshifts (the y-band in Rubin is less suitable because of observing conditions from the ground). Euclid will provide near-infrared measurements useful to probe higher redshift domains in the two common deep fields and to constrain dust extinction for lower redshifts in the wide surveys. Note that the detection of ``live'' SNe~Ia by Euclid will be rather limited by the fact that it will only make a single pass over most of the sky, and sampling even in the Euclid Deep fields will generally be too sparse to detect transients more than once or twice during a lightcurve spanning of order one month. Coordinated observations of the common deep fields (EDF-Fornax and EDF-South) would improve this situation.

The combination of Euclid and LSST deep imaging observations of the host galaxies of SNe~Ia (see also next section) provides additional constraints on host properties such as stellar mass and star formation rate that are known to correlate with the peak brightness of SNe~Ia even after the usual standardisation by lightcurve width and colour. Measurement of host galaxy properties will therefore help improve the accuracy of the LSST SNe for cosmology. Euclid may also provide spectroscopic host galaxy redshifts of previously discovered LSST SNe, which will increase the target sample for photometric classification. The SNe~Ia we identify by these methods can be incorporated into a cosmological analysis. 

The SN Ia cosmology case has other interesting extensions. For example, transient data and galaxy clustering are correlated since they probe the same underlying structure. In combination they additionally provide information about large-scale peculiar velocity flows. Therefore, their cross-correlation can probe the accelerated expansion of the Universe beyond the classical redshift-distance test applied to individual objects \citep{2018arXiv180806615M, 2021MNRAS.504.3884A}. A first implementation of such an approach has been demonstrated on mock surveys for the analogous case that uses dark gravitational wave sirens as standard candles \citep{2021PhRvD.103d3520M}. Specifically, we will perform a generalization of the Alcock-Paczy\'{n}ski test that pairs Rubin’s high-cadence observations of transients and Euclid’s accurate spectroscopic galaxy redshift survey, in
overlapping regions. As a result, we will test the accelerated expansion by constraining the time-varying equation-of-state of dark energy. The approach can further be applied to scenarios with other models of cosmic acceleration and nonzero curvature. As a natural byproduct, we will provide the community with accurate reconstructions of 3D density and peculiar velocity fields, which can be
used, e.g., to constrain galaxy properties as a function of their environment, or to correct measurements of $H_0$ based on gravitational-wave standard sirens \citep{Mukherjee2021}.

{\bf Environments of extragalactic transients}

Joint LSST-Euclid imaging of even relatively small regions around transients would enable a wide range of science. For example, \vspace{-0.4cm}\begin{itemize}[noitemsep]
\item Morphology of host galaxies, and in some cases unique identification of high-$z$ host galaxies.
\item For nearby galaxies: identification of the progenitor location (star-forming, passive) and in some cases direct identification of the progenitor star or binary system. This has been successfully done with HST, and Euclid’s high resolution images will provide a pre-explosion image for many. The near-infrared is particularly useful, as red supergiant progenitors are bright in YJH.
\item For transients with no host in LSST data: possible identification of a high-$z$ host (in YJH), allowing identification of high-$z$ transients in LSST.
\item For identification of gravitational wave counterparts: LSST will likely map the sky locations of GW sources and will find many faint sources. Euclid postage stamps of these will indicate any which are coincident with faint, near-IR sources, and so are likely foreground stars or background SNe. While LSST will provide host galaxy information, the near-IR information from Euclid will also be powerful for those that do not have an obvious host.
\item Enhanced photo-$z$ measurements for host galaxies of transients - a very powerful way of selecting certain types of transients. If the YJH photometry can enhance the ugrizy LSST photo-$z$ then a more reliable photo-$z$ can be used to estimate the absolute magnitude of the host.
\item Spectroscopic redshift from Euclid of nearest host - would provide confirmation of high-$z$ hosts.
\end{itemize}
	
{\bf Near-infrared photometry for LSST transients}

The near-infrared emission from extragalactic transients reveals unique information about the energy source and the radiative processes in stellar explosions such as kilonovae \citep[e.g.][]{Villar2017} and supernovae \citep[e.g.][]{Graur2020}. Near-infrared photometry for LSST transients would be a scientifically powerful addition to the optical light curves.

{\bf Tidal Disruption Events}

Rubin has the potential to find Tidal Disruption Events (TDEs) but the key challenge is selecting a clean TDE sample again a background of SNe and AGN. Adding extra information from Euclid would allow selection of a clean sample, which enables a whole array of TDE science. 
The Euclid positions and photometry (light curves) will complement the  Rubin-based filtering of the alert stream for TDEs. Specifically, precise astrometry from Euclid and determination of the host-flare offset can help the characterization of these events. Complementary lightcurves in wavelength range and time are valuable for the characterization of the event itself, but probably also for the faintest host galaxy.
For this to work, the Euclid data would have to be available to (some) of the Rubin brokers, in semi real-time ($\sim\,$1\,day). This would likely be a high-level filter, that is applied to a cleaned part of the Rubin stream, so the required compute resources are modest (and the computing would happen in the brokers).

{\bf Star caustic crossings (overlap with Strong Lensing Section)}

LSST’s cadence and depth may be able to detect caustic crossings of luminous $z<$\,1 stars. Euclid’s higher resolution imaging will be useful for a better derivation of the lens model, needed to interpret the results. Stars can be magnified by factors $>$10,000 during a caustic crossing by a galaxy. A bright star with absolute magnitude $-$12, at $z$\,$=$\,1, and magnified by a factor 10,000 would have apparent magnitude $\sim\,$22, well within reach of LSST and Euclid, but since these events are short-lived (few days with sufficient brightness to be observed by LSST), it will be most likely missed by Euclid, but could be observed by LSST.

\subsection*{Specific Derived Data Products} \label{sec:transients-products}

{\bf Photometry of transients} 

A common theme that emerged from several transients science cases is the need for joint, multi-band photometric time series of the transients. Such time series should include detections in all Rubin and Euclid imaging filters and should include upper limits, derived by forced photometry at the transient location, in cases where either Euclid or Rubin observed the location of the transient but did not detect it. Such upper limits provide (1) additional colour information that can be crucial for classification of transients and (2) pre-explosion upper limits provide very valuable information on the time of explosion. Science cases that would benefit from such a DDP include SN cosmology, SN rates, Environments of transients, Tidal Disruption Events, Star caustic crossings and the near-infrared photometry case. Gravitationally lensed transients should be included in the list of sources.

\rec{upperlims}{A joint Rubin-Euclid photometric catalog should include time-series information and upper limits (in cases of non-detection by either facility) for the subset of sources that are transients}{Non detections are important because (1) additional colour information that can be crucial for classification of transients and (2) pre-explosion upper limits provide very valuable information on the time of explosion.}{TR}

\ddp{transient-lightcurves}{Transient object light-curves}
{A joint time-series photometric catalog of all transients (including upper limits) combining both Rubin (dense) and Euclid (sparse) data.}{From the start, updated nightly (provided an observation of the location was made) to enable follow-up observations. Updates of the entire lightcurves should also be made periodically (every month initially, then every season) as the depth of the template (reference) images increases.}{TR}

\vspace{1em}

{\bf Image cutouts} 

The image of a transient and its surrounding area provides a wealth of information about the host galaxy and environment of the object. Therefore a valuable DDP would be regularly-updated postage stamp images in all LSST bands (ugriz) and Euclid VIS, YJH at the full stacked depth available at the time. The science cases that would benefit from such images include Environments of transients, Supernova cosmology, and star caustic crossings. The supernova cosmology case would benefit from pixel values in all filters for joint fitting of the SN and host galaxy. While the pixel values in these images will be used for quantitative scientific analysis, they are also useful as finding charts for follow-up observations on other facilities. The suggested size of the cutouts is 30 arcsec $\times$ 30 arcsec, to match that provided in the public Rubin alert stream. Alternatively, both Euclid and LSST could consider agreeing to share a larger stamp (e.g. 3 arcmin $\times$ 3 arcmin). However, analysis should be done to determine whether the total sky area covered in the latter case becomes infeasible to share as a DDP. Note that the expected rates of transients will depend somewhat on the final survey strategies chosen for both facilities.

\fnd{transients-cutouts}{Image cutouts in multiple bands of a small region around transient sources provide valuable information about the environment of the transient and its host galaxy}{Cutouts would enable creation of finding charts for further follow-up and important scientific input into the classification of transients (which impacts all transient science cases). In addition, specific host galaxy information that is likely to improve the derivation of cosmological parameters from Type Ia supernovae.}{TR}

\rec{transients-sky-area}{Carry out a study of expected transient rates and sky coverage}{A study should be carried out to estimate the rate of all transients and hence the total cumulative area of sky that would need to be shared for different sizes of image cutouts. The expected rates of transients will depend on the final survey strategies chosen for both facilities.}{TR}

\ddp{transient-cutouts}{Transient object cutouts}
{Time series of multi-band (ugrizYJH) image cutouts of size 30 arcsec $\times$ 30 arcsec centered on all transients, and derived pixel values.}{From the start, updated nightly (provided an observation of the location was made). Updates of the entire sequence should also be made periodically (every month initially, then every season) as the depth of the template (reference) images increases.}{TR}

\vspace{1em}

{\bf Astrometry of transients} 

The most accurate position measurement of the transient is valuable when considering its environment. It can also be used as a valuable classification tool. For example, accurate Euclid positions of transients relative to the centres of their host galaxies will help determine whether a transient is a possible Tidal Disruption Event candidate.

\fnd{transient-astrometry}{The most accurate position measurement transients is valuable information}{This is important when considering the local environment of a transient within its host galaxy, and can be used as an additional classification tool, which in turn enables rapid follow-up.}{TR}

\ddp{transient-astrometry}{Transient astrometry}
{Position of the transient as determined jointly from LSST and Euclid imaging. The position relative to the centre of the host galaxy is also requested.}
{From the start, updated nightly (provided an observation of the location was made) to enable classification and follow-up observations.}{TR}

\vspace{1em}

{\bf Photometry of host galaxies} 

Multi-band photometry of transient host galaxies is valuable for many science cases including Environments of transients, SN rates and SN cosmology. For some transients it may not be clear which galaxy is the host, in which case photometry of several surrounding objects would be valuable. Therefore a valuable DDP would be a joint Rubin-Euclid catalog of galaxies near the transient and their multi-color photometry. The catalog should be updated regularly as the stacked depth of the images increases.

\ddp{transient-host-phot}{Transient host galaxy photometry}
{Joint catalog of positions and and multi-colour photometry of galaxies (SEDs) within a 30\,\arcsec $\times$ 30\,\arcsec region centered on each transient (matching the size of the cutout image DDP above.)}{From the start, updated prior to major data releases.}{TR}

\vspace{1em}

{\bf Transient/host Spectroscopy} 

Spectroscopic redshifts of transients and their host galaxies provides crucial information that help constrain the classification of the transient and its cosmological epoch (e.g. for measuring SN rates). It also constrains the properties of the host galaxy (for the transient environment case). For Type Ia SNe, it is a crucial parameter for placing the SN on a Hubble diagram and deriving cosmological parameters. Host galaxy spectroscopic properties (such as star formation rate derived from emission lines) would be valuable for the SN cosmology case. Finally, the expansion case requires Euclid spectroscopic redshifts for all galaxies in order to reconstruct the
density field, as well as the full catalog of LSST supernovae to reconstruct the velocity field and perform the proposed test.

\ddp{transient-host-z}{Transient host galaxy redshifts}
{Joint catalog of redshifts (photometric or spectroscopic if available) of galaxies in a 30\,\arcsec $\times$ 30\,\arcsec box centered on each transient. If the source lies in a galaxy cluster then the cluster redshift should also be provided.}
{From the start, updated nightly (required for real-time classification).}{TR}

\ddp{transient-host-spec-params}{Transient host spectral parameters}
{Spectral parameters of host galaxies (e.g. SFR from emission lines) of the host galaxies of transients}
{From the start, updated prior to major data releases.}{TR}

\vspace{1em}

{\bf Higher-level products} 

Combined light curve parameters such as stretch, colour, date-of-maximum would be valuable for the SN cosmology case. Similarly, combined host galaxy parameters (stellar mass, star formation rate, morphology, 
multicolour light profiles, photometric redshifts) would also be useful for the SN cosmology case and others such as rates and environments. In some cases, knowledge of these parameters may help with real-time classification of transients that is required in order to arrange follow-up observations.

Many transient cases (e.g. environments, Tidal Disruption Event, SN cosmology) would benefit from an easily-accessible transient database. In addition, detection efficiencies as a function of magnitude and sky location are required in order to make quantitative measurements (required for rates, expansion of the universe, SN cosmology cases).

\ddp{transient-lightcurve-params}{Joint transient lightcurve parameters }
{Combined light curve parameters such as stretch, colour, date-of-maximum light, derived from the joint optical\&near-infrared lightcurves. This requires an estimate of the redshift, from the host galaxy or from the transient itself.}
{From the start, with updates as new lightcurve points are added.}{TR}

\ddp{transient-z}{Transient redshifts}
{Joint catalog of redshifts of the transients, derived from all available Rubin and Euclid photometry and Euclid spectroscopy (of the host or the transient itself).}
{From the start, updated every season.}{TR}

\ddp{transient-host-params}{Joint transient host parameters }
{Combined host galaxy parameters including stellar mass, star formation rate, morphology, multicolour light profiles and photometric redshifts (spectroscopic redshifts are covered above), derived from the combined Rubin and Euclid imaging and spectroscopic data. Needed for statistical analysis (e.g. cosmology, rates) and may also help real-time classification of transients.}
{From the start, updated prior to major data releases.}{TR}

\ddp{transient-database}{Joint transient database and interface}
{All public Rubin and Euclid information on transients in easily accessible form.}
{From the start, updated daily.}{TR}

 \ddp{transient-detection-efficiencies}{Transient detection efficiencies}
{Rubin and Euclid detection efficiencies as a function of magnitude corresponding to each transient.}
{Updated at major data releases.}{TR}

{\bf Illustration of a joint analysis based on transients DDPs} 

The following suggestion is considered to be a joint analysis of separate data products from Rubin and Euclid rather than a jointly-derived data product requiring joint pixel-level processing. Therefore we retain the description of the case and its requirements for completeness.

The generalised Alcock-Paczy\'{n}ski test of the expansion requires two overlapping 3D fields, one of galaxies (redshift tracers) and one of SNe (distance tracers, corrected for peculiar velocities). Bayesian algorithms will produce the required field-based reconstructions: (i) BORG \citep{2012MNRAS.425.1042J, 2015JCAP...01..036J, 2019AA...625A..64J, 2019arXiv190906396L} will perform detailed inference of the density field over the entire volume probed by Euclid observations; (ii) ViRBIUS \citep{Lavaux2016, 2019MNRAS.488.5438G, 2021arXiv211115535B} will directly constrain peculiar velocity fields based on transient information, giving access to a luminosity distance space. Finally, combining the BORG and ViRBIUS likelihoods will simultaneously constrain density and velocity fields.

{\bf Other requirements}

Although not a derived data product, joint transient science with Rubin and Euclid is strongly affected by the coordinated observations of the two surveys. In particular the choice of Deep field locations and observation times (within the same observing season) will improve the chance of joint detection of transients. Finally, we note that it would be ideal for Euclid transient alerts to be made public (as planned for LSST).

\rec{rec-transients-1}{Survey overlap and timing}{The Rubin and Euclid projects should remain coordinated in their survey plans, in particular regarding the location and timing of deep field observations.}{TR}

\rec{transients-alerts}{Transient alerts}{The Euclid project should issue public transient alerts in real time.}{TR}

\subsection*{Algorithms} \label{sec:transients-algos}

The transient science cases require difference imaging and forced photometry on the difference images. Algorithms exist but may need adaptation to cope with the different spatial resolution of Rubin and Euclid. Pixel-based SN/host deblending routines will be required. For cosmology using Type Ia SNe, a light-curve fitter suitable for optical\&near-infrared and ground and space-based data is required.
For many science cases (e.g. environments, rates, SN cosmology) transient photometric classification tools are required, for both real-time data streams and final lightcurves. These algorithms are being developed by different groups. 

\subsection*{Technical Considerations} \label{sec:transients-resources}

The expansion of the universe case provided requirements on computing of HPC with 1000 cores and storage of 20TB. Other cases also require a platform that hosts co-located data and computational tools, and infrastructure for catalog access. A suggestion was made for common simulations and we recommend that a computing task force is created to look at requirements, particularly for the SN cosmology case.

\rec{transients-comp}{Set up a computing task force to consider requirements for the joint analysis of transients}{}{TR}

\subsection*{Timescale}  \label{sec:transients-timescale}

Some cases (e.g. SN cosmology, environments and perhaps rates) need approximately daily/real-time DDPs for classification/selection and triggering transient follow-up on other facilities.  Regular, rapid updates to joint lightcurves, image stamps and host galaxy/transient redshifts are the most important real-time data. In addition, real-time, rapid access to the DDP is needed as soon as new transients are identified. Longer timescales are sufficient for late-time lightcurves ($\sim$\,100\,days) as for the near-infrared case.

Some cases (rates, expansion of the universe, SN cosmology) also use “final” transient datasets (large sets of full light curves or derived parameters), these will build up with time. In addition, the quality of lightcurves and image stamps will continue to improve even after the transients have faded because of improved reference (``template'') depth. For this reason, all transient DDPs should continue to be re-derived as the surveys continue. Approximately yearly updates would be sufficient.

\subsection{Galaxy Evolution} \label{sec:galaxy-evolution}

\contributors{\hyperref[author:mbanerji]{Manda Banerji (WG)}, \hyperref[author:mmoresco]{Michele Moresco (WG)}, Viola Allevato, Laura Bisigello, Micol Bolzonella, Jarle Brinchmann, Olga Cucciati, Rapha\"el Gavazzi, Peter Hatfield, Olivier Ilbert, Clotidle Laigle, Guilhem Lavaux, Claudia Maraston, Manuela Magliocchetti, Henry Joy McCracken, Lucia Pozzetti, E. Sarpa, M. Shuntov, Margherita Talia, Niraj Welikala, Ilsang Yoon, Elena Zucca}

Rubin and Euclid will both result in the discovery and characterisation of a vast and diverse galaxy population over an extensive redshift range, enabling the evolution of galaxy properties to be traced over much of cosmic time and in a variety of environments. A key motivation in the production of the DDPs is to allow robust galaxy properties to be derived as well as identification of rare and unusual galaxy populations where multi-colour data is often essential to distinguish them from contaminant populations. 

Many of the themes discussed in this Section overlap directly with themes in the AGN (\ref{sec:agn}), Galaxy Clustering  (\ref{sec:static-cosmo}), and Primaeval Universe (\ref{sec:PU}) Sections. 

\subsection*{Science Drivers} \label{sec:galaxy-evolution-drivers}

The key areas in galaxy evolution that will benefit from the creation of joint data products are:

\begin{itemize}[noitemsep]
    \item Galaxy photometric redshifts
    \item Galaxy environment
    \item Robust estimation of galaxy physical properties e.g. stellar mass, dust extinction, AGN content
    \item Galaxy morphology and structure
    \item Joint analysis of quasars and their host galaxies
    \item Spatially resolved analysis of galaxy physical properties
    
\end{itemize}

\subsection*{Specific Derived Data Products} \label{sec:galaxy-evolution-products}

\rec{Matched depths across both surveys}{Investigate u-band and z-band observing strategy to ensure optical and infra-red depths across LSST and Euclid are well matched in terms of known galaxy SEDs}{The LSST u-band can help break many known degeneracies in galaxy photometric redshift and spectral energy distribution fitting. For example, passive and dusty star-forming galaxies are undetected or low signal-to-noise detections in the bluest bands while highly star-forming unobscured galaxies and AGN are very luminous at these wavelengths. Deep u-band data from LSST in complement with the Euclid near infra-red bands, will enable more robust estimation of galaxy properties. The z-band is important to identify the red sequence and estimate the richness of clusters at 1.0\,<\,$z$\,<\,1.5. It is therefore vital that the depths are well matched across the two surveys to ensure meaningful detections/upper limits on the fluxes of passive and dusty galaxy populations as well as galaxy clusters across a wide range of redshifts.}{GE}

\rec{De-blender}{Investigate an appropriate de-blending tool that will work across the full wavelength range and spatial resolution of the joint LSST and Euclid datasets}{A key challenge in the production of almost all the DDPs relevant for galaxy evolution will be de-blending the Euclid sources in the LSST images.}{GE}

\ddp{gal1}{Multi-band Catalogs}{Aperture and total fluxes across the LSST and Euclid bands for all objects detected in both surveys and derived via joint pixel-level analysis of both survey datasets. Essential for the production of high-quality photometric redshifts, object classifications, detection of galaxy clusters and characterisation of environment out to $z$\,$=$\,2 and robust characterisation of the physical properties of galaxies and AGN.}{Tiered approach through the lifetime of both surveys gradually increasing in area/depth/complexity.}{GE}

\ddp{gal2}{Postage Stamp Cutouts for Representative Samples}{Multi-wavelength images can aid secure identification of galaxy populations selected in colour space. Images can also be directly fed into machine learning and deep-learning algorithms for ``pixel'' photometric redshift estimates. There might be specific sub-populations of galaxies e.g. both star-forming and passive galaxies at high-redshifts as well as high-redshift AGN candidates where postage stamps would be beneficial to aid candidate identification. A subset of postage stamp images from both surveys for a representative sample of galaxies encompassing the full range in colour, spectral type and morphology would also be useful for training algorithms for galaxy classification.}{Tiered approach through the lifetime of both surveys.}{GE}

\ddp{gal3}{Galaxy Photometric Redshifts}{Photometric redshift estimates produced using a range of algorithms incorporating both empirical/training-set based methods and template-fitting and run on the joint multiwavelength catalogs (c.f. \ddpgeref{gal1}). Full probability distributions for the photometric redshift estimates might be required for some science use cases.}{After first data releases.}{GE}

\ddp{gal4}{Galaxy Physical Properties}{Physical properties of galaxies estimated using spectral energy distribution fitting to the multi-wavelength data (\ddpgeref{gal1}) to include (but not necessarily restricted to) stellar mass, star formation rate, dust extinction and AGN content.}{After first data releases.}{GE}

\ddp{gal5}{Galaxy Morphology \& Structure}{Bulge/disk decomposition, results of Sersic fits across the full multi-wavelength range and results of running machine learning classifiers jointly on the LSST\&Euclid images for a representative subset of galaxies (c.f. \ddpgeref{gal2}).}{After first data releases.}{GE}

\ddp{gal6}{Pixel-level decomposition of quasars and their host galaxies}{Pixel level decomposition of the images into a point source and extended component can be used to jointly measure quasar and host galaxy properties. Variability in the LSST bands can be further used to associate pixels to the quasar and separate these from the pixels associated with the host galaxy. Creation of this DDP will require access to image cutouts from both LSST and Euclid around known spectroscopically confirmed quasars (e.g. \ddpgeref{gal2}).}{After first Euclid wide data release.}{GE}



\subsection*{Algorithms} \label{sec:galaxy-evolution-algos}

Many algorithms are being developed across the science working groups and science collaborations of both LSST and Euclid both for pixel-level combination of the optical and near infra-red data and for the subsequent derivation of galaxy physical properties and environment. For example, Shirley et al. (in preparation)\footnote{ \url{https://github.com/lsst-uk/lsst-ir-fusion} }  are adapting the LSST software stack to jointly process optical images from LSST with near infrared images from the ESO VISTA Public Surveys. The joint pixel-level analysis has already been run on HSC and VISTA data over several hundred square degrees of sky to produce multi-wavelength catalogs. Within Euclid algorithms for combining ground-based data and Euclid data (currently applied to DES and KiDS) have been developed by, in particular, OU-EXT and OU-MER. Postage stamp cutouts can be directly obtained through the LSST Butler and similar functionality for Euclid can be provided. 

While these developments provide much of the necessary functionality, work is needed to optimise these tools for a combination of LSST and Euclid data, for instance ensuring correlated uncertainties resulting from any significant resampling of the pixel scales is robustly propagated to the photometric uncertainties. Larger simulations have already been constructed within Euclid to test the detection of galaxy clusters and large scale structures \citep{euclid/etal:2019}. 

Software to perform galaxy morphology (\ddpgeref{gal5} ) within Euclid is developed within OU-MER and through a dedicated working group. For \ddpgeref{gal6} similar studies have been conducted using HSC observations \citep{Li:21}. These DDPs require work in adaption and testing for  multi-survey/multi-facility datasets. 

Photometric redshift codes are reasonably mature and many comparisons of these codes have been made in the context of wide-field surveys (e.g. \citealt{Hildebrandt:10, Sanchez:14, graham/etal:2020}). We will rely on the photometric redshift working groups and coordination groups of both collaborations to make recommendations regarding the codes to be used. The main recommendation is that these are tested and run on the full multi-wavelength catalogs produced as a result of \ddpgeref{gal1}.



\subsection*{Technical Considerations} \label{sec:galaxy-evolution-resources}

Multi-core CPUs will be critical for efficiently processing the multi-wavelength images to produce joint photometric catalogs. GPUs may also be necessary for algorithms that take images as inputs. Storage requirements will scale with the size of the image cutouts and will be significant if the multi-dimensional probability distribution functions of galaxy photometric redshifts as well as all derived physical properties of galaxies need to be stored. Methods to reconstruct these PDFs a-posteriori are however being investigated \citep{Mucesh:21}.

\subsection*{Timescale}  \label{sec:galaxy-evolution-timescale}

We strongly advocate a tiered approach to producing the DDPs with the production of DDPs starting as soon as there is common area observed between the two surveys. The DDPs can build up in terms of area/depth (which in turn will translate into CPU time and storage requirements) and also complexity. For example, simple aperture based fluxes can be produced for the first iteration of multi-wavelength catalogs with more complex model-fitting techniques to produce joint photometry being investigated subsequently. The multi-wavelength catalogs will initially be used as inputs for galaxy photometric redshift and spectral energy distribution fitting with image inputs considered later. 

Image stamp inputs will be essential for the quasar host galaxy studies, which will begin after the first Euclid wide data release when coadded images from both surveys as well as light-curves from LSST should be in place. However, these images will only be needed for a small subset of objects. 
\subsection{Active Galactic Nuclei} \label{sec:agn}

\contributors{
\hyperref[author:fbauer]{Franz E. Bauer (WG)}, \hyperref[author:mbanerji]{Manda Banerji (WG)}, Viola Allevato, Sotiria Fotopoulou, Hermine Landt, Xin Liu, Maurizio Paolillo, Ilsang Yoon}

A by-product of gas accretion onto massive black holes (MBHs) is the highly efficient generation of electromagnetic radiation across all observable wavebands, which serve as powerful beacons of light seen across the entire observable universe. The various physical processes that produce this light, coupled with the compact size, additionally drive variability on all observable timescales. These two characteristics allow relatively rare, actively accreting MBHs (hereafter, active galactic nuclei, or AGN) to be distinguished from the far more numerous stars and galaxies in the optical and near-infrared bands. The planned surveys to be carried out by Rubin and Euclid will both push the discovery, characterisation, and utilisation of such accreting MBHs to unprecedented levels, impacting an exceptionally broad range of science areas. Each survey has its strengths and weaknesses, which can be enhanced and minimized respectively by employing both surveys in concert via the production of the specific DDPs, as outlined below. Many of the themes discussed in this Section impact and/or share synergies with those discussed elsewhere, such as in the Galaxies (\ref{sec:galaxy-evolution}), Galaxy Clustering (\ref{sec:static-cosmo})  and Primaeval Universe (\ref{sec:PU}) Sections.

\subsection*{Science Drivers} \label{sec:agn-drivers}

We provide here a non-exhaustive list of science topics that stand to benefit from joint data products to allow the identification of the largest and broadest possible population of AGN and their subsequent characterisation:

\begin{itemize}[noitemsep]
\item AGN structure,
\item AGN accretion and jet physics,
\item MBH formation and growth across cosmic time,
\item MBH feedback mechanisms and impact on host and larger scale environments,
\item MBH-host correlations and evolution across cosmic time,
\item Improved constraints on cosmological parameters and understanding of systematics (via strong-lensing, standardizeable candles, and clustering),
\item Large-scale structure across cosmic time,
\item Intergalactic matter across cosmic time,
\item Improved precision of the celestial reference frame.
\end{itemize}

AGN emit across the electromagnetic spectrum, but can be obscured to varying degrees by line-of-sight gas and dust. Obtaining broad optical\&near-infrared wavelength coverage by combining the LSST and Euclid datasets (in line with \fndccref{ddp}) will allow us to pick out AGN-related emission from underlying host galaxy light for a wider range of AGN, in particular both more obscured and fainter AGN.
Because AGN can contaminate and potentially negatively impact both stellar and galaxy samples, and vice versa, it is extremely important to identify as large a fraction of the AGN population as possible and quantify/separate their contributions as well as possible. For example, improved separation of AGN and hosts (see also \ddpgeref{gal6}) will produce cleaner samples with which to probe AGN physics, MBH-galaxy co-evolution, and for use as standardizable candles for high-redshift cosmological constraints.  Additionally, AGN variability can add significant dispersion to the colors and SEDs of galaxy nuclei, ultimately affecting the classification and physical parameter estimation of both AGN and their hosts; these can be mitigated by obtaining near-simultaneous observations across a large wavelength range. 

\rec{AGN-SimultaneousObs}{Maximize coordinated, near-simultaneous observations between LSST and Euclid.}{To best mitigate uncertainties related to AGN variability, we recommend that the Rubin and Euclid scheduling teams prioritise and optimise the level of near-simultaneous observations, both in the two southern Euclid deep fields (e.g. \recccref{widesurveys}) as well as across the full overlapping survey area to the extent feasible (e.g., implement rolling cadences for LSST which will improve overlap with Euclid's general observing plan). This stands to enhance significantly both AGN and galaxy science.}{AN}

Ultimately, the accurate assessment of AGN will have appreciable knock-on benefits for many of the science drivers in other Sections.

\subsection*{Specific Derived Data Products} \label{sec:agn-products}

To tackle the above, we must employ efficient and complete methods to pinpoint the locations of AGN on the sky, accurately deblend them from their host galaxies, and characterise their distances and intrinsic properties, ideally contemporaneously in time to avoid the effects of variability. In terms of specific DDPs that will aid and benefit from this process, we recommend the following.

The merging of independent LSST and Euclid list-driven forced photometry catalogs (\ddpccref{listdriven-catalogs}) will be useful initial products to enhance several important AGN science cases that benefit from total fluxes. These can be generated rapidly with relatively low computational overhead, although the mismatch in pixel scale, PSF and/or apertures will lead to potentially large systematic errors that must be accounted for by users.  

\ddp{agn-joint-phot-cat}{Multi-band Photometric catalogs}{Optical and near-infrared based PSF, aperture and total flux catalogs across the LSST and Euclid bands for all objects detected in either survey, based on co-aligned images with matched pixel scales. Minimum-level DDP, essential for initial classification of AGN candidates (\ddpanref{agn-joint-class-cat}), production of initial high-quality photometric redshifts (\ddpanref{agn-joint-photoz-cat}), and robust characterisation of the physical properties of AGN and host galaxies (\ddpanref{agn-joint-prop-cat}).}{Tiered approach over the lifetime of both surveys, synchronized with data releases.}{AN}


\ddp{agn-joint-forced-deblend-phot-cat}{Deblended Multi-band, Multi-epoch and Time-averaged Photometric Catalogs}{Forced decomposition of the images into variable point source and static extended components to measure multi-band AGN light curves and time-averaged AGN and host galaxy properties over full overlapping survey area. Creation of this DDP will require access to co-aligned, matched-pixel scale image cutouts from both LSST and Euclid. This is best enabled with a joint pixel-level processing of the images, with consistent pixel sizes, applying appropriate DCR corrections to the LSST images (\recccref{DCR-joint-model}). The higher intrinsic angular resolution of Euclid, combined with the broader 0.3--2\,$\mu$m images will greatly enhance host and nuclear photometry/SEDs and derived parameters (e.g. superior star/galaxy separation, independent host and nuclear/AGN photometric redshifts, stellar masses, dust extinction, star formation rates and history, and AGN properties). A better assessment of the host contribution to the nuclear emission will allow better assessment of nuclear variability fractions and behavior. Although the Euclid data sampling is sparse, multi-epoch nuclear photometry will be a useful resource to quantify near-infrared variability further out in accretion disk and obscured/reprocessed variability from the inner disk and broad line region. Minimum requirement would be for all AGN candidates (\ddpanref{agn-joint-class-cat}), while goal would be \textit{all} galaxies, to place limits on possible low-level AGN emission.}{Tiered and iterative approach over the lifetime of both surveys, synchronized with data releases.}{AN}

\ddp{agn-joint-class-cat}{AGN Candidate Catalog}{Higher level, iterative DDP to be used as input prior for production of deblended photometric catalogs and high-quality photometric redshift catalogs.
Morphological priors and/or features (e.g., shapes, sizes, residuals, cutouts) may be useful to enhance SED and variability based AGN classification efforts.}{Tiered and iterative approach throughout the lifetime of both surveys, synchronized with data releases.}{AN}


\ddp{agn-joint-photoz-cat}{AGN and Host Photometric Redshifts}{Photometric redshift estimates produced using a range of algorithms incorporating both empirical/training-set based methods and template-fitting, specifically tuned for AGN and host galaxies (\ddpanref{agn-joint-class-cat}). Based on multi-band photometry DDPs above (\ddpanref{agn-joint-phot-cat}, \ddpanref{agn-joint-forced-deblend-phot-cat}). Full probability distributions for the photometric redshift estimates will be essential for several science cases.}{Tiered and iterative approach throughout the lifetime of both surveys, synchronized with or immediately following data releases.}{AN}

\ddp{agn-joint-prop-cat}{AGN Physical Properties}{Physical properties of AGN can be estimated using variability in conjunction with spectral energy distribution fitting to the multi-band photometry and photometric redshift catalogs (\ddpanref{agn-joint-phot-cat}, \ddpanref{agn-joint-forced-deblend-phot-cat},
\ddpanref{agn-joint-class-cat}, \ddpanref{agn-joint-photoz-cat}), including (but not necessarily restricted to) nuclear dust extinction and multi-band luminosity, AGN/galaxy fraction and AGN fractional variablity, presence of relativistic beaming, and, if possible, black hole mass and accretion rate.}{Tiered approach throughout the lifetime of both surveys, synchronized with or immediately following data releases.}{AN}

\ddp{agnvar-joint-LCs}{Prompt Light Curves of Extreme AGN Variability Events}{Joint time-series photometric catalogs (including upper limits) for extreme AGN events such as TDEs, Changing State AGN, and Blazar flares, combining both Rubin (dense) and Euclid (sparse) data.}{From the start, updated nightly (provided an observation of the location was made) to enable and inform follow-up observations. Updates of the entire lightcurves should be made periodically (every month initially,  then every season) as the depth of the template (reference) images increase.}{AN}

For specific DDPs related to AGN host galaxy and environmental characterization, we avoid repetition and simply refer interested readers to the previous \secref{sec:galaxy-evolution}.

\subsection*{Algorithms} \label{sec:agn-algos}

Some multi-band deblended photometry tools and algorithms already exist, such as Tractor and TPHOT/TFIT \citep[e.g.,][]{Lang2016}, and have been employed on large scales (e.g., NOAO Legacy imaging, Euclid simulations).  Other multi-band deblenders like SCARLET \citep{2018A&C....24..129M} are still in development. They appear to work reasonably well, but still need substantial battle-testing amongst the community, particularly at the performance levels desired for joint LSST and Euclid analyses.

\rec{AGNdeblender}{Identify a robust multi-band de-blending tool that functions across the full wavelength range and spatial resolution of the joint LSST and Euclid datasets}{Separation of AGN and host galaxy light is a key challenge to produce reliable AGN-centric DDPs. Although products such as forced deblended nuclear photometry/limits seem straightforward, no past survey has produced one. Moreover, it will be critical to calculate proper/appropriate error estimates, understand limitations, etc.}{AN}

Some AGN classification algorithms based on SED and variability properties exist (e.g., prescriptions, ML-based), but these have primarily been used to select QSOs. The depth and precision of LSST and Euclid will push AGN studies to fainter levels, where host contamination will surely become a factor.

\rec{AGNclassifier}{Investigate AGN classification at fainter magnitudes, quantity selection effects \& completeness}{}{AN}

\subsection*{Technical Considerations} \label{sec:agn-resources}

In terms of human and computing resources, producing images on a common scale and generating a basic total photometry catalog seems feasible with only a modest scaling beyond what is already envisioned, but is the least interesting option. More sophisticated deblending will scale up computing costs substantially, but is already being planned and scoped out within both the Rubin-LSST and Euclid communities, separately, so the actual ``cost'' may only be a factor of a few higher.

\subsection*{Timescale}  \label{sec:agn-timescale}

The generation of DDPs will benefit from a tiered approach, starting from year 1 data products for both LSST and Euclid to test and hone techniques, as well as identify bottlenecks and problems. Clearly the calibration and understanding of both datasets and understanding will evolve greatly over time, and we expect that maximum benefits will only be realized near the end of both surveys. As one example, LSST DCR (\recccref{DCR-joint-model}) benefits from multiple observations over a wide range of airmass, which will need to be built up over many years. This needs to be balanced against the fact that intermediate products will be essential as input for ongoing follow-up efforts during the surveys.

\subsection{Cosmology from weak gravitational lensing, galaxy clustering and galaxy clusters} \label{sec:static-cosmo}

\contributors{\hyperref[author:cheymans]{Catherine Heymans (WG)}, \hyperref[author:kkuijken]{Konrad Kuijken (WG)}, 
\hyperref[author:jbosch]{James Bosch (WG)}, 
\hyperref[author:hhoekstra]{Henk Hoekstra (WG)}, \hyperref[author:flanusse]{Francois Lanusse (WG)}, \hyperref[author:pmelchior]{Peter Melchior (WG)}, \hyperref[author:mmoresco]{Michele Moresco (WG)}, \hyperref[author:spaltani]{St\'ephane Paltani (WG)}, \hyperref[author:mtroxel]{Michael Troxel (WG)}, Stefano Andreon, Adam Amara, Sandro Bardelli, Micol Bolzonella, Stefano Camera, Francisco Castander, Ranga Ram Chari, N. Elisa Chisari, Olga Cucciati, Melissa Graham, Daniel Gruen, Hendrik Hildebrandt, Olivier Ilbert, Benjamin Joachimi, R\'emy Joseph, C. Danielle Leonard, Anja von der Linden, Matteo Maturi, Lauro Moscardini, Emiliano Munari, Roser Pello, Mario Radovich, Barbara Sartoris, Tim Schrabback, Isaac Tutusaus, Jochen Weller, Angus Wright}

Cosmology is a primary science driver for both Euclid and Rubin, with both missions combining information from a set of static observations of the evolution of large scale structures, weak gravitational lensing, galaxy clustering and galaxy clusters.   These observables probe the impact that dark energy has on the growth of structure in the Universe, providing a distinct and complementary analysis to the expansion history transient cosmology observables discussed in Sections~\ref{sec:transients} and~\ref{sec:stronglensing}.  Forecasts combining Euclid or LSST with observations of the cosmic microwave background (CMB) from the Planck satellite \citep{planck/etal:2020} predict sub-percent level precision on joint measurements of the dark energy equation of state parameter, $w_0$, and a linear time evolution parameter, $w_a$ \citep{laureijs/etal:2011,lsst/etal:2009}.  As such Euclid and Rubin are expected to provide robust constraints to (in)validate the cosmological constant model where w$_0$\,$=$\,$-$1 and w$_a$\,$=$\,0.  The sum of the neutrino masses, $\sum $m$_\nu$, is also predicted to be constrained to better than 0.02\,eV enabling astrophysics to resolve the question of the neutrino hierarchy, one of the remaining undetermined features of the standard model of particle physics.  To achieve these forecasts, both Euclid and the LSST Dark Energy Science Consortium (DESC) have placed stringent requirements on the accuracy of their photometric redshift and lensing measurements. A joint analysis of Euclid with LSST is anticipated to significantly reduce the challenges of meeting these systematic error requirements, in addition to furthering and extending the cosmological capabilities of each survey alone \citep{jain/etal:2015, Rhodes_2017, schuhmann/etal:2019, capak/etal:2019, chary/etal:2020,graham/etal:2020}.

\subsection*{Science Drivers} \label{sec:static-cosmo-drivers}

The {\it weak gravitational lensing} of light by foreground structures introduces coherent distortions in the images of background galaxies.   This `cosmic shear' can be observed at different redshifts, directly probing the evolution of the projected total matter distribution \citep[see for example][]{heymans/etal:2013, asgari/etal:2021, amon/etal:2021, hikage/etal:2019}.   Lensing can also be observed around galaxies to measure the bias of the different galaxy types tracing the underlying density distribution \citep{hoekstra/etal:2002} and further our understanding of the galaxy-halo connection \citep{wechsler/tinker:2018}. Lensing observations rely on a combination of high fidelity galaxy shape measurements and multi-band photometry for photometric redshifts. There are many subtleties involved in these measurements, for example PSF effects, blending, detection biases and noise bias. The intrinsic alignment of galaxies and the uncertain impact of baryon feedback on the dark matter distribution provide additional astrophysical sources of systematic error in any lensing analysis \citep{mandelbaum:2018}.

Measurements of {\it galaxy clustering} probe the baryon acoustic oscillation peak, constraining the distance-redshift relation, in addition to redshift-space distortions, which constrain the growth of structure \citep{alam/etal:2021}.   Clustering observations typically rely on spectroscopic redshift surveys combined with a robust understanding of survey incompleteness to mitigate systematic errors \citep{ross/etal:2020}.  Photometric redshift galaxy clustering is less constraining than its spectroscopic counterpart, but nevertheless remains an important component of multi-probe large-scale structure analyses \citep{des/etal:2021} to self-calibrate measurement and model degeneracies. Cosmological constraints from galaxy clustering on small scales are limited by uncertainty in the non-linear behaviour of galaxy bias.

{\it Galaxy clusters} trace the highest density peaks in the Universe, with the abundance of clusters as a function of redshift providing tight constraints on the cosmic and structure formation history \citep{allen/etal:2011}.  Cluster identification relies on a combination of multi-band photometry, potentially spanning the full spectrum from the microwave (thermal Sunyaev-Zel'dovich effect), through the optical (red-sequence or profile matching), to the X-ray (hot cluster gas).  Cluster mass estimates are then calibrated using weak lensing measurements \citep{2019MNRAS.482.1352M}, as well as a combination of photometric and spectroscopic redshifts, to constrain the growth of cosmic structure and to determine the evolution of the cluster population \citep{bocquet/etal:2019}. To set robust cluster cosmological constraints, a full understanding of sample incompleteness and projection effects is required.   This is particularly relevant for optically selected cluster samples where accurate photometric redshifts are vital to optimise the analysis \citep{abbott/etal:2020}.

In the systematics-dominated Euclid-Rubin era, there is significant motivation to undertake simultaneous joint-probe analyses of lensing, clustering and clusters.   This approach has the ability to self-calibrate many sources of data-related and astrophysical systematics \citep{bernstein:2009, abbott/etal:2020,heymans/etal:2021}.   Rubin-Euclid Derived Data Products that directly benefit a single large-scale structure probe therefore indirectly benefit all probes and the overall cosmological constraints.   

There are many proposed DDPs to enhance cosmology constraints from Euclid and LSST, ranging from catalog-level products to full joint pixel analyses.  As the scientific gain of a joint Rubin-Euclid large-scale structure analysis is expected to be primarily driven through an overall enhancement in the control of systematic errors, it is challenging to quantify the expected improvement of the Rubin-Euclid constraining power on dark energy and other cosmology parameters accurately. This `inconvenient truth' leads to a clear initial recommendation based on the complex and realistic image simulation suites currently being developed independently in both consortia \citep{euclid/etal:2019,lsst/etal:2021} to validate and/or calibrate shear-photometric redshift selection and measurement biases \citep[see examples in][]{mandelbaum/etal:2018,kannawadi/etal:2019,maccrann/etal:2020,hoekstra:2021}. 

\rec{static-cosmo-sims}{Create a realistic joint Rubin-Euclid image simulation suite}{A detailed study of joint realistic image simulations would quantify the expected scientific gain from each DDP approach, and allow for the development of new DDP algorithms.  Without such a study, the joint survey selection bias function of any DDP scenario will remain unquantified, potentially leading to systematic errors in the final analysis that are outwith the required systematic tolerances for some science cases.  As such this simulation work is a necessary and vital prerequisite for static cosmology science using any of the proposed DDPs.  We recommend that a joint survey working group is rapidly formed to create this derived simulation product. This could build from existing NASA-funded efforts to produce joint Rubin--Roman image simulations \citep{eifler/etal:2021,troxel/etal:2021} and similar efforts by the Euclid Consortium (e.g. SIM-EXT for the simulation of external data).}{SC}
\subsection*{Specific Derived Data Products} \label{sec:static-cosmo-products}
Given the systematics-limited regime, there is much to be gained from a fully joint analysis of Rubin and Euclid pixels for lensing and photometric redshift measurements, optimally combining the full depth of Rubin ugrizy data with the spatial resolution of Euclid VIS and the Euclid YJH photometry. Anything less is considered by the community to be scientifically sub-optimal in terms of the cosmology we can learn. We therefore give a minimum `baseline' proposal that minimises data sharing, but advocate moving as soon as possible to a more ambitious and scientifically powerful full joint-pixel analysis. 

\ddp{static-cosmo-euclid-source}{Baseline `list-driven' Y1 ugrizy Rubin photometry for Euclid VIS sources}{Euclid requires Rubin multi-band optical photometry for photometric redshift estimates in regions of the Southern hemisphere. While most of the DES bands meet the requirements for Euclid DR1, Rubin data will be necessary for subsequent DR, and is essential for regions of the footprint not covered by DES. 
The minimum baseline DDP to facilitate this crucial measurement for many Euclid science cases is a `list-driven' forced Rubin-photometry catalog of VIS sources, at one year LSST depth.  For this DDP, PSF corrected multi-band photometry is extracted from the Rubin pixel data for a list of Euclid VIS sources, following the methodology of the Euclid OU-MER multi-band photometry pipeline that will also be used for multi-band space-ground photometry in the Northern Hemisphere. This will require bringing together the Rubin pixel data, the Euclid photometric aperture functions for the VIS sources, and the OU-MER photometry algorithm to deliver a multi-band DDP catalog. The Rubin pixel and Euclid model data that are used to derive these catalogs would not, however, be part of the released DDP.}{From the start, as this approach is the minimum necessary to facilitate Euclid Southern-sky lensing and cluster cosmology.}{SC}

\ddp{static-cosmo-rubin-source}{Baseline `list-driven' Euclid NISP YJH photometry for Rubin sources}{Supplementing Rubin multi-band optical photometry with Euclid near-infrared photometry is forecast to improve the photometric redshift estimates of $z_{\rm phot}>1$ galaxies, which carry a large share of the cosmic shear cosmology information.  \citet{graham/etal:2020} find a reduction in both the fraction of outliers and the scatter at the level of 10-20\%, with the inclusion of Euclid near-infrared photometry allowing Rubin to reach the DESC photometric redshift requirements beyond $z_{\rm phot}\,>\,$1.5 out to $z_{\rm phot}\sim\,$2.4.  To produce this DDP requires Euclid NISP pixel data to be used to extract calibrated PSF homogenised near-infrared photometry for a list of Rubin sources.  The improved redshift determination would then lead to enhanced cosmological constraints for Rubin, though it may be limited by the ability to accurately calibrate Rubin lensing measurements in this higher, more blended, redshift range. The setup for this DDP is thus similar to \ddpscref{static-cosmo-euclid-source} but with the roles of Rubin and Euclid reversed: it requires bringing together Euclid near-infrared pixels with a Rubin source list, Rubin photometric aperture functions and the Rubin multi-band photometry algorithms.   Note that the Rubin algorithms are not currently suitable for undersampled NISP data, so algorithm development would be required.   In addition Rubin's primary shape measurement method requires `metacalibrated photometry' which would place additional requirements on the NISP photometry extraction, if included \citep{zuntz/etal:2021}.}{From Rubin Y2 onwards, as the high photometric redshifts that benefit from NISP photometry will only reach DESC requirements for Rubin data beyond Y2 \citep{graham/etal:2020}.}{SC}

\ddp{static-cosmo-spectroscopy}{Rubin photometric redshift distributions calibrated with Euclid spectroscopy}{Euclid's NISP will produce a sparse catalog of galaxies with spectroscopic redshifts in the range 0.7\,$<z<\,$2.  These would complement spectroscopy from the Dark Energy Spectroscopic Instrument \citep{DESI/etal:2016}, and other spectroscopic surveys, that will be used to calibrate Rubin and Euclid photometric redshifts through a cross-correlation clustering analysis \citep[see for example][]{Newman:2008, Gatti/etal:2020,Hildebrandt/etal:2021}. This DDP would require the physical co-location of the Euclid spectroscopic redshift and Rubin photometric redshift catalogs, along with the relevant clustering algorithm to derive calibrated redshift distributions, $n(z)$, for the Rubin sources. The catalogs that are used to derive the calibrated $n(z)$ would not, however, be part of the DDP.}{From the start.}{SC} 

\ddp{static-cosmo-full-depth-rubin}{Deeper ugrizy Rubin photometry of Euclid VIS sources}{Weak lensing cosmology places tight restrictions on the accuracy of the measured redshift distribution of different tomographic source samples.  \citet{ilbert/etal:2021} demonstrate that these restrictions can be reached for Euclid by either utilising a representative spectroscopic calibration sample with $>\,$99.8\% purity, or by incorporating deep optical photometry.  Given the challenges of obtaining such a pure spectroscopic catalog, the combination of Euclid with full-depth Rubin photometry would provide an important validation and cross-check of the spectroscopic calibration within the Rubin footprint.   Deep Rubin photometry extends optical cluster detection out to $z=\,$2, which, when combined with
an expansion of the Euclid source sample to higher redshifts, facilitates high-redshift cluster science that has weaker requirements on the lensing measurement accuracy.  This DDP photometry could be list-driven forced photometry, as in \ddpscref{static-cosmo-euclid-source}, or joint-pixel photometry, as in \ddpscref{static-cosmo-ambitious-photom-wide}, dependent on algorithmic development timescale of the latter.}
{From Euclid $\sim$Y4 onwards, after which there will be a significant overlapping-area gain and accumulated depth in the Rubin bands.}{SC}

\ddp{static-cosmo-ambitious-photom-wide}{Joint-pixel Rubin ugrizy, Euclid VIS and Euclid NISP YJH photometry in the Wide Surveys}{Blending, where light from a galaxy contaminates the photometry of its neighbour, is expected to affect 6\% of Euclid source galaxies (J. Sanchez, private communication) and 63\% of Rubin source galaxies \citep{sanchez/etal:2021}.  This significant fraction of blends raises a challenge in determining accurate galaxy photometry.  Errors in the photometry leads to errors in photometric redshift estimates and consequently systematic bias in the final cosmological analysis.  Here Euclid and Rubin are highly complementary with Euclid identifying blends morphologically from VIS, and Rubin identifying blends with different colours.  \citet{joseph/etal:2021} determine the reconstruction accuracy of complex galaxy morphologies using a joint pixel analysis of idealised Euclid-Rubin image simulations.   Significant improvement is found, particularly for objects that are brighter in one survey than the other. Developments in machine learning pixel-level photometry provide a promising alternative route to mitigate blending errors \citep{boucard/etal:2020,arcelin/etal:2021}.  The expected improvement in the accuracy of the photometric redshifts, when using deep learning photometry, is at a similar level to that expected from the inclusion of the Euclid NISP photometry to Rubin, across all redshifts \citep{cabayol/etal:2021}. In this scenario a physical co-location of calibrated ugrizy, VIS and NISP YJH pixels would be required, in order to explore various alternative de-blending methodologies for photometry.   Algorithmic development will benefit from the \recscref{static-cosmo-sims} image simulations.}
{Dependent on algorithmic development.}{SC} 

\ddp{static-cosmo-ambitious-Shapewide}{Joint-pixel Rubin-Euclid galaxy shape analysis}{The majority of galaxies used for weak lensing analyses are small and at faint magnitudes.  As such Euclid and Rubin are highly complementary with the exquisite resolution of Euclid matched by the significant signal-to-noise imaging of Rubin.  Using simplified Euclid-Rubin simulations of isolated faint and small galaxies \citet{schuhmann/etal:2019} find a 20\% increase in galaxy shape measurement precision using a joint-pixel analysis.   A joint-pixel approach to galaxy shape measurement is also expected to benefit from improved deblending capabilities \citep{joseph/etal:2021}. A further benefit of this DDP is that it allows direct constraints to be put on the tidal, `intrinsic' alignment of galaxies that are physically close by comparing galaxy shapes at different isophotes \citep{leonard/etal:2018}, simultaneously taking advantage of the high-fidelity inner isophotes from Euclid VIS and the deep photometry from Rubin. In this scenario a physical co-location of calibrated shape-measurement-friendly Rubin bands in the gri and VIS pixels would be required.   We recommend that algorithmic development work is undertaken using the \recscref{static-cosmo-sims} image simulations.}
{Dependent on algorithmic development.}{SC}

\ddp{static-cosmo-ambitious-photom-deep}{Deep Survey joint-pixel photometry and shear analysis of Rubin ugrizy, Euclid VIS and YJH}{An alternative to mitigating deblending errors for cosmology on an individual source level, is to determine the bias introduced statistically, correcting for its presence at the population level \citep{melchior/etal:2021}.  This approach requires additional information at ultra faint magnitudes, to empirically calibrate the blend-bias to photometric redshifts and shear estimates at shallower Wide survey magnitudes \citep[see for example][]{buchs/etal:2019,hartley/etal:2020,Myles/etal:2021}. The Southern Euclid Deep Fields, covered by deep Rubin observations, are perfectly suited for such an analysis. To realize this DDP, a full exchange of calibrated pixel data (stacks and individual exposures) within a deep field would be required as well as a model PSF that is sufficient for the shape measurements for science cases based solely on data from this deep field. The EDF- South is a prime candidate field (\recsoref{surveys-3}).}{Dependent on algorithmic development.}{SC}

\subsection*{Algorithms} \label{sec:static-cosmo-algos}

Algorithms for forced photometry across multiple surveys exist enabling \ddpscref{static-cosmo-euclid-source} and \ddpscref{static-cosmo-rubin-source}.  Examples include, for example {\sc Tractor} \citep{Lang2016} and {\sc T-PHOT} \citep{TPhot}, which are sufficiently fast to be deployed at survey scale \citep{chary/etal:2020}.
For joint modeling and deblending, as recommended by \recscref{blending} and required by \ddpscref{static-cosmo-full-depth-rubin}, \ddpscref{static-cosmo-ambitious-photom-wide} and \ddpscref{static-cosmo-ambitious-Shapewide}, {\sc SCARLET} offers a non-parametric alternative \citep{2018A&C....24..129M}, which has recently been extended to pixel-level joint modeling and deblending and tested on simulations mimicking Euclid and Rubin images \citep{joseph/etal:2021}.
Despite being highly optimized on CPUs and run as current default deblender for Rubin, joint pixel level processing of Rubin and Euclid data may require the development of a GPU version of {\sc SCARLET}. Alternative routes using pixel-level machine learning are also explored. All of these approaches have no formal support for the undersampled PSFs of Euclid NISP. Dedicated developments (e.g. for internal oversampling) or systematics testing with \recscref{static-cosmo-sims} image simulations are necessary to assess the accuracy of the extracted near-infrared photometry.

Image simulations such as those mentioned in \recscref{static-cosmo-sims} already form part of the planning within the Euclid and Rubin projects.  Our recommendation is that the resources be used in a coordinated fashion to create joint simulations. Algorithms for photometric redshift calibration through cross-correlations are also core to the Euclid and Rubin plan enabling \ddpscref{static-cosmo-spectroscopy}.

Algorithms for combining calibrated well-sampled input images into a single `coadd' image that is a sufficient statistic for the static sky are well-developed mathematically \citep[e.g.][]{2017ApJ...836..188Z}, but have not yet been demonstrated to work for systematics-limited science.  Such algorithms are currently part of Rubin's internal data processing plans and development work, however, and if they succeed, even the most ambitious static-sky DDPs could be built from the six per-band Rubin coadds instead of the hundreds of single-epoch calibrated images, yielding a huge reduction in the storage and compute cost of producing the most ambitious DDPs.  Similar savings from coaddition on the Euclid side are much less likely to be possible; \citet{2011ApJ...741...46R} provides a method for coadding undersampled images that meet certain dithering criteria, but it is itself extremely expensive.  But coadding Euclid images is also much less important, given the degree to which Rubin data (without coaddition) dominates the joint processing compute resources.

\subsection*{Technical Considerations} \label{sec:static-cosmo-resources}

The development of software for the DDPs described have modest computing requirements, but should ideally operate on platforms where GPUs are accessible to facilitate neural network approaches. Initial development, performance tuning, and accuracy assessment will need access to image simulations (\recscref{static-cosmo-sims}). Further testing should be performed on actual survey images (\recccref{shared-data}) to evaluate the robustness in the presence of realistic image contaminants. This phase is most severely limited by the availability of experienced personnel.

For the creation of the DDPs themselves, in particular to more challenging ones listed above, substantial computing resources need to be utilized. \citet{chary/etal:2020} estimate the computation cost for joint-survey processing of Rubin, Euclid, and Roman images at about 30M CPU hours (for data processing alone; they estimate 300M when an extensive Monte Carlo simulation effort is included), but do not consider the possibility of using only coadded images from Rubin.  An estimate that takes into account the savings from coaddition and leaves out Roman data could easily be a factor of 10 smaller, but is hard to make rigorous estimates until algorithms are more mature.
In addition, the HPC center(s) that carry out the processing need to be able to either hold all the necessary data from both surveys (approximately 75 PB at the completion of both surveys if all Rubin calibrated images are used directly, but reduced to {\em ca.}~3 PB with only coadds from Rubin) or transfer the relevant portions over the network. Substantial expertise in computing center operations is required at this stage.

\subsection*{Timescale}  \label{sec:static-cosmo-timescale}

 See notes under the individual DDPs above. The most ambitious, joint pixel processing DDPs (\ddpscref{static-cosmo-ambitious-Shapewide} and \ddpscref{static-cosmo-ambitious-photom-deep}), will need to be rerun with each successive yearly DDP DR, incorporating the increasingly deeper overlapping Euclid with LSST data available at that time. 

\subsection{Strong Lensing} \label{sec:stronglensing}

\contributors{\hyperref[author:tcollett]{Tom Collett (WG)}, Timo Anguita, Simon Birrer, Fred\'eric Courbin, Tansu Daylan,  Jose Diego, Brenda Frye, Raphael Gavazzi, R\'emy Joseph, Phil Marshall, Ben Metcalf, Dominique Sluse, Graham Smith, Alessandro Sonnenfeld, Aprajita Verma, Giorgios Vernardos}

\subsection*{Science Drivers} \label{sec:stronglensing-drivers}
The impact of science derived from gravitational strong lensing is wide, including cosmology \citep{wong2020, birrer2020, collett2014, oguri2012}, testing of CDM predictions \citep{vegetti2012, wolfgang2021, daylan2018} and galaxy evolution phenomenology \citep{ritondale2019, shajib2021}. This science is currently hobbled by the shortage of suitable strong lenses to analyse. Only $\sim$\,1000 lenses are currently known and these come from heterogeneous surveys with a multitude of complicated selection functions. With the capacity to each discover over 100,000 lenses \citep{Collett2015}, Rubin and Euclid will drive studies of strong lenses into the statistical age. The unprecedented depth and resolution over large area are the key enablers of this staggering growth of the strong lens sample. 

Whilst LSST (depth, multi-band) and Euclid (resolution, near-infrared) have unique and complementary capabilities for strong lensing science, the union of LSST and Euclid is required to overcome the significant challenges that remain for discovery and exploitation, including:
\begin{itemize}
\item {\it How do we find 100,000 lenses in billion+ object surveys?}
Galaxy scale strong lens systems are typically a red elliptical galaxy lensing a blue background source into an Einstein ring of 0.5-1 arcsecond radius. Group and cluster scale lenses have Einstein radii up to tens of arcseconds. Because of the rarity of strong lenses on the sky, there are (relatively) a huge number of lens-like contaminants such as face-on spirals, mergers, and blends. These contaminants must be filtered out to assemble a pure sample of strong lenses. Even though machine learning methods have been developed that are able to recover lenses with  better than 99.9\% accuracy \citep{lanusse, jacobs2017, rojas2021, petrillo}, the rarity of lenses means that this still results in vastly more false positives than true lens systems.  Euclid faces a particular challenge: without multi-band optical imaging, Euclid cannot use color information in lens finding, resulting in lots more false positives. Further improvement in lens finding can be gained by simultaneously using all LSST and Euclid bands in the machine learning searches rather than running two separate search pipelines. Visual inspection is required as a final step to purify the lens sample and this requires the best possible high resolution, color images to enable efficient and high confidence human classification. Without this, the bulk of candidates will be ambiguous to the human eye \citep{jacobs2019, rojas2021, petrillo}. 
\item {\it How do we train a lens finder and how do we interpret the lenses that we find?}
Because lenses are incredibly rare, there are very few examples with which to train any machine learning based lens finder. This has typically been resolved by simulating observations of large numbers of strong lenses \citep{jacobs2019,rojas2021,petrillo}. To train a lens finder to discover strong lenses in both LSST and Euclid requires multi-band simulations of strong lenses. However, because lens finding is hard, the strong lens selection function is not always easy to write down. Understanding the selection function is critical since we are rarely scientifically interested in the population of lenses and lensed sources, rather the parent population of galaxies, groups and clusters that they are drawn from. For cosmological analyses, small biases in how we select lenses could ruin constraints if they are not correctly modelled \citep{2016MNRAS.462.3255C,birrer2020}.
Simulations of the population of strong lenses are also needed to calibrate the selection function. A joint set of simulations of the Universe's strong lens population should be developed for this purpose.
\item {\it How do we get redshifts for lens and source, without which we cannot do science?}
Once a sample of lens candidates has been assembled, science requires redshifts for both the foreground lens and the background source. Without redshifts there is no way to convert the observed angles into physically meaningful distances and masses. Spectroscopic redshifts are of course preferable, but cannot be obtained for all systems. Photometric redshifts for strong lensing are not like other photometric redshifts. The lens and the source are overlapping and not easily deblended. This means that high resolution and multiband data are both fundamental to doing science with strong lenses. Photometric redshifts for the lens are relatively easy: the lens typically dominates the light, and they are early type galaxies with magnitudes and colors that are well sampled in spectroscopic surveys. Photometric redshifts for the background source pose an entirely more challenging problem. Source colors are contaminated by light from the much brighter lens. Sources are typically also intrinsically faint, high surface brightness (typically star forming) sources that are beyond the reach of large scale spectrocopic surveys. As blue sources between redshifts 1 and 3, photometric redshifts can not be precise without accurate photometry and wide wavelength coverage. 
\item {\it How do we efficiently use finite follow up resources?}
For many science cases, additional data is required beyond Euclid and LSST. Spectroscopic redshifts, high resolution imaging, and faster cadenced imaging will all be in high demand for strong lens science in the LSST and Euclid era. Given that this finite followup resource must be spread over many thousands of lenses and lens candidates we must have a way to prioritise systems for followup. The first piece of followup gathered for almost all systems will be a spectroscopic redshift. To enable minimal failure rates requires the best possible photometric data, both to rank candidates by their likelihood of being a lens but also for having detectable emission or absorption features in the wavelength range of the available spectrograph. High resolution colour images and deblended lens and source photometric redshifts are critical for this. These can only be achieved through pixel level joint processing, or some measure of pan-sharpening.
\end{itemize}

\fnd{sl-1}{Combining Rubin and Euclid data is critical to strong lens discovery and exploitation}
{Whilst LSST (depth, optical multi-band) and Euclid (resolution, near-infrared multi-band) have unique powers for strong lens science, the union is required for discovery and scientific exploitation of strong lenses with Euclid \& LSST.}{SL}

\rec{sl-sims}{Conduct multi-band simulations of Euclid and LSST strong lenses}{Simulations of lenses as observed by both telescopes in their filter sets are necessary both to train lens finders, and to calibrate the lens discovery and followup selection functions. These simulations should be produced jointly.}{SL}

As the sample size grows rapidly in the 2020s, the strong lensing community will likely become followup and human-power limited. We have a rich history of collaborative effort (e.g. CLASS, SLACS, H0LIC0W, CLASH, HFF) and are keen to continue this into the LSST-Euclid era. There is already significant membership overlap between the Euclid strong lensing and LSST strong lensing groups. We propose to share ugrizy-VIS-YJH postage stamps of every strong lens candidate, including those already flagged in precursor surveys. This will enable us to do the maximum possible science, make the most efficient use of scientist time and to agilely respond to the changing scientific landscape of the next decade. Since, for strong lensing, both surveys are greatly enhanced by the addition of the other, full data sharing (including single visit data at low latency) for strong lensing candidates is also a politically fair and neutral outcome for both collaborations. We also stress that the rarity of strong lenses means that we are requesting to share data that represents a negligible fraction of the footprint of both Euclid and LSST, and thus sharing data to enable strong lensing science should have a negligible impact on colleagues with other science interests.

\subsubsection*{Importance of the DDPs for the strong lensing science} \label{sec:stronglensing-estimate}
Quantifying the level of improvement from combining LSST and Euclid is challenging since the strong lensing community has never faced challenges like Euclid and LSST. Further research may be able to ameliorate some of the weaknesses of the individual surveys, but there is no doubt that joint DDPs will significantly improve the scientific results of strong lensing from Euclid and LSST.

For lensing systems, source photometric redshifts will be almost impossible without a joint DDP. That means that only spectroscopic lenses will be usable for science and that spectroscopic followup resources will be stretched thinly by weaker candidates. If we assume that most of the lens systems are confirmed by OII emission detected in optical multi-object spectrographs like 4MOST and DESI, then good photometric redshifts can improve spectroscopic efficiency by a factor of 3\footnote{This assumes a photometric redshift uncertainty of 0.2, and is compared to not having source photometric redshift information.}.

A blind test has never been conducted to see how the fidelity of human visual classification would be improved by a multiband high resolution image compared to Euclid or LSST data alone. 
One analogue can be drawn from the HST followup of the CFHT lens sample of \citep{Gavazzi2012}: Almost all lenses that were classified as definitely lenses (rank 3) in CFHT were confirmed with HST, roughly a quarter of probable lenses (rank 2) were ruled out by HST with a similar fraction confirmed \footnote{Three quarters of the possible lenses (rank 1) were ruled out with one promoted to certain lens.}. 
Probable lenses initially outweighed the definite lenses by $\sim$10-to-1 so if the full CFHT sample had high resolution imaging the total `definite' lens sample should have been $\sim$4 times larger than the CFHT-only search. 
There are no analogues of searches in the opposite direction (adding lower resolution colors to single band HST data), but given the challenges of filtering out false positives without color information, Euclid is likely to also be swamped by low ranked candidates without the joint DDP.

Together the DDPs are critical for Euclid and LSST to deliver strong lensing science. It is quite likely that the DDPs will provide a sample of  confirmed, scientifically useful strong lenses that is an order of magnitude larger than if the two collaborations do not work together to produce and share these DDPs.

\subsection*{Specific Derived Data Products} \label{sec:strong-lensing-products}

\ddp{sl-1}{Pansharpened images of all strong lens candidates}{A tool to display high resolution color images of strong lens candidates (Pansharpening Euclid resolution with LSST colors) will enable visual purification of strong lens samples and for efficient followup prioritisation.}{From the start.}{SL}

\ddp{sl-2}{Deblended foreground lens and background source photometry for strong lens candidates}{A multiband deblender will enable photometric redshift estimates of both the foreground lenses and the background sources. This will allow all lenses to be included in science analyses, and for followup prioritisation.}{as soon as possible.}{SL}
\ddp{sl-3}{A joint color and morphology catalog for strong lens searches}{A combined multiband catalog of fluxes and basic morphological parameters will enable efficient preselection of objects within which to search for strong lenses.}{As soon as possible.}{SL}
\ddp{sl-4}{A strong lens probability for every massive galaxy, group and cluster}{A strong lens finder should be run on the combined Euclid + LSST dataset to find as many lenses as possible. This should be developed in a way that includes the infrastructure for future improved lens finders to repeat this search. }{Tiered approach: purify each surveys' highly ranked candidate sample as soon as there is a meaningfully large area of survey overlap ($>$\,1000\,deg$^2$, likely LSST Y1). Then the lens finder should be run on all LRGs and clusters, regardless of lens candidate probability assessed by the individual surveys (Euclid DR2/LSST Y3 or later).}{SL}
\ddp{sl-5}{ugrizy,VIS,YJH postage stamps of strong lens candidates}{The Euclid and Rubin strong lens communities wish to share multiband data of all high probability strong lenses selected at optical wavelengths and, ideally, all strong lens candidates selected at optical wavelengths, at the earliest possible date to enable the maximum possible science and the most efficient use of scientist time and followup resources. There is significant membership overlap and we believe that, for strong lensing, maximal data sharing is the optimal strategy to ensure the publication of timely, high impact science. Stamps must include uncertainties and point spread functions. The size of the postage stamps should be at least 15 arcseconds on a side for group and galaxy scale lenses and 3 arcminutes on a side for cluster scale candidates. Lenses that may contain high-redshift ($\gtrapprox$\,7) objects that are only detected in the near IR Euclid bands are excluded from this DDP.}
{As soon as possible for highly ranked candidates. Lower probability candidates could wait.}
{SL}

\subsection*{Algorithms} \label{sec:strong-lensing-algos}
Possible algorithms for all of theses strong lensing DDPs exist already, however there is substantial ongoing research into how to improve them for the Euclid and LSST era. 

Catalogs and postage stamps are routinely produced already, and thus no further development is required for their production from stacked data release data. However, a modest amount of development work may be required to serve postage stamps from single visits in low latency.
 Machine learning based lens finders are well developed, and applying simultaneously to Euclid and LSST is a modest change. Extending existing lens simulation for LSST or Euclid is trivial to also include the other survey.

The pansharpening and deblending DDPs require some further development, since only a small amount of work has been done by the strong lensing community in this area. Initial applications of pansharpening have been tried in \citep{faure2008} for strong lenses discovered in the COSMOS field by HST. Pansharpening is a common problem in Earth imaging and existing algorithms should be generalizable to Euclid and LSST, but further research is required to investigate the optimal approach and to see if there any foibles specific to this use case. Automated strong lens deblending is rapidly developing \citep{rojas2021, savary2021}, but development is needed to apply these approaches to simultaneous deblending of ground and space based imaging e.g. as in \citet{joseph/etal:2021}.

Deblending is an open problem in astrophysics. This problem is particularly challenging for strong lensing, where the lens and source are extremely close on the sky, and the source is distorted into complicated arc shapes. Automated lens modelling \citep{marshall2009,brault2015,nightingale2018} could provide a lensing specific solution to this problem. Alternatively a deblender might be successful exploiting the difference in color between or lens and source. 

Because these DDPs require processing of $\sim$\,5 million small postage stamps, there are no significant concerns with scalability. There is no need for access to datasets outside LSST and Euclid.

\subsection*{Technical Considerations} \label{sec:strong-lensing-resources}
These DDPs are not computationally challenging for modern compute clusters.

{\bf Storage and memory}:

We expect $\simeq\,$20 galaxy scale lenses per deg$^2$, $\simeq\,$5000 Luminous Red Galaxies (LRGs; i.e. candidate galaxy scale lenses) per deg$^2$, and $\simeq\,$5-10 candidate cluster lenses per deg$^2$. A 15$\times$15 arcsecond postage stamp centered on a LRG equates to $\simeq\,$0.2\,Mb from Euclid and $\simeq\,$0.3\,Mb from LSST stacked data, or $\simeq\,$30\,Mb for the full LSST time series. These data volumes per postage stamp are a factor of $\simeq\,$100 larger for the 3$\times$3 arcmin postage stamps required for clusters.

Static lens finding will be done on stacked data. Therefore to assign a lens probability to all LRGs and all clusters will require storage of order $\lesssim\,$0.6\,GB\,deg$^{-2}$.
For candidates with low lensing probability these data could be discarded if storage considerations are a concern.

Lensed transient finding will require time series data, with the highest impact discoveries expected to be sub-threshold for transient brokers. If storage capacity allows, we therefore request postage stamps based on Euclid stacks and LSST time series to be available on demand for all high probability galaxy scale lenses and all candidate cluster scale lenses. We estimate this equates to $\simeq\,$10\,GB\,deg$^{-2}$.

{\bf Compute time}:

For lens finding, less than a CPU second per LRG is likely to be required. Pansharpening and deblending will be much more intense, but is only required for the candidates assessed to have at least some chance of lensing (around 10 percent based on previous experience). 10 CPU minute per candidate is likely to be sufficient i.e. 100 CPU hours per square degree.

\subsection*{Timescale}  \label{sec:strong-lensing-timescale}
Timescales are primarily driven by 2 factors: 1) rapid discovery of lensed transients, 2) efficient spectroscopic followup of lens candidates with 4MOST and DESI.

Basic multiband photometry and morphology catalogs should be made available as soon as possible to increase the efficiency of lens searches by both collaborations (i.e. as soon as Euclid and LSST have each released overlapping sky patches to their respective communities (ideal timescale: LSST Y1).

Pansharpened images and deblended photometry for high probability lens candidates should be available as soon as possible to ensure that followup is prioritised efficiently (ideal timescale: as soon as both surveys have observed a candidate). For lower priority candidates this could be shared at a later stage.  There should be a further `on demand' pansharpened image service for potential lensed transients. Here, time is of the essence for triggering followup, but only a few hundred such candidates are expected per year. 

The joint lens finder should be run in two stages. Initially only to regrade candidates highly scored by one survey or the other (i.e. for sample purification), this should happen as soon as there is meaningfully large area of survey overlap ($>$\,1000\,deg$^2$, likely LSST Y1). At a later stage the lens finder should be run on all LRGs and clusters, regardless of lens candidate probability assessed by the individual surveys (ideal timescale: Euclid DR2/LSST Y3 or later).

We are only interested in sharing a tiny fraction of one percent of the total sky, which should have negligible impact on the interests of other science communities: the multiband cutout stamp server should be available as a DDP service. Whilst this exchange will enable maximum science with the combined strong lensing dataset, the timescale on which it happens will not impact followup of high ranked candidates, though it will impact discovery of lensed transients especially in the early years of the surveys. Initially the collaborations should share data on the highly ranked lens candidates (Ideal timescale is as soon as candidates are ranked highly), with sharing of all LRGs and clusters with lens probability meaningfully greater than 0 shared at a later date. 

The timescale is much more critical for lensed transients. As a minimum there should  be immediate postage stamp sharing for the handful of spectroscopically confirmed lensed transients each year, and preferably immediate sharing of postage stamps of all highly ranked candidate lensed transients so as to enable efficient selection for follow-up observations. With the order of 100 such transients expected a year this is a minuscule fraction of each survey.

Updates of these DDPs should be approximately annual, except for the lensed transients which should be updated as soon as there are new observations.
The proposed timelines should be flexible to accommodate the wishes of other science cases, whilst pragmatically recognising that strong lenses fill a negligible fraction of the Euclid/LSST sky. 
\subsection{Primaeval Universe} \label{sec:PU}

\contributors{\hyperref[author:afontana]{Adriano Fontana (WG)}, \hyperref[author:mbanerji]{Manda Banerji (WG)}, Rebecca Bowler, Marco Castellano, Jean--Gabriel Cuby, Daniel Mortlock, Sune Toft}

By the term "Primaeval Universe" we refer here to the earliest phase in the life of the Universe, when the first stars and galaxies formed in the earliest density fluctuation. This process is clearly continuous but its boundary is  marked by the last and more important phase transition occurred at large scale in the Universe: the reionization of the Inter Galactic Medium (IGM) by the UV radiation produced by early sources. 
This process progressively occurred in a redshift range that is  constrained by several observational between $z\simeq\,$9, when the Universe is still predominantly opaque to UV radiation (despite individual sources exist and ionize small region around them) and $z\simeq\,$6, where the IGM is eventually highly ionized. 

The evolution in time and space of this phase transition is driven by the evolution and nature of the sources that produced the UV ionizing photons. Studying the reionization process hence is a indirect albeit powerful tool to study the early evolution of galaxies and AGNs, even in an epoch where they are difficult to observe individually.  Determining the timeline and topology of the reionization process, and the physical processes involved, represent the latest frontier in observational cosmology, and one of the key goals of the next generation of extragalactic surveys. 

Current observational probes of the average neutral hydrogen fraction in the IGM point to a “late” reionization scenario where the IGM was half neutral at $z\simeq\,$7, and reionization completed at $z\simeq\,$5.5-6 \citep{DeBarros2017, Mason2019}. In this respect, a key finding has been the observation of a substantial decrease of the fraction of galaxies with a Ly$\alpha$ emission between $z\simeq\,$6 and $z\simeq\,$7, interpreted as the effect of scattering of the Ly$\alpha$ photons by the neutral medium \citep{Fontana2010,Stark2010,Pentericci2011,Pentericci2014,Pentericci2018,Ono2012,Schenker2012}. Furthermore, the analysis of Ly$\alpha$ visibility in independent lines-of-sight hints to a patchy reionization topology \citep{Treu2012, Pentericci2014}. While the global scenario is being settled, outstanding questions remain \citep{Choudhury2009}: 1) is the reionization process driven by bright or faint galaxies? 2) is it proceeding from high-density regions to lower density ones (“inside-out”)? 3) or viceversa (“outside-in”)? In order to answer these questions, we must collect large samples of galaxies at $z>\,$6 in order to study their large-scale statistical distribution and detailed physical properties. 

Large samples of high redshift galaxies are also needed to address the other main scientific question - understanding the mechanism that shaped galaxy formation and evolution in its earliest phases. The  observations that we have today suggest that galaxies in the first Gyr have  physical properties different from local ones - lower mass, lower metallicity but higher ionization and star-formation efficiency - but are too sparse to effectively constraint the theoretical models that describe their birth and evolution {\it ab initio}. The combined Euclid$+$Rubin surveys have the potential to  identify the few, extremely bright sources that we can study in great details with dedicated instrumentation (like {\it JWST}, {\it ALMA} \citep{Bowler2018} and in the long term ELT-class instrumentation), in order to shed light on their detailed physical and dynamical state.

Because of the IGM absorption, galaxies at $z>\,$6.5 are detected only in the bands Y and redward. The combination of sensitivity and image quality of Euclid (that will observe in the three bands Y, J and H down to magnitude $\simeq\,$24.5 in the Wide and magnitude $\simeq\,$26.5 in the Deep Fields), combined with  the relatively poor sensitivity in the y band of the Rubin telescope, conspire to make most of $z>\,$6.5  AGNs and galaxies detectable {\it only} by Euclid and not by Rubin--LSST. This has made the Euclid community much more engaged in the science cases relevant to the Primaeval Universe. 
However, Rubin--LSST data are precious to improve the reliability and efficiency of the selection procedure and significantly reduce the contamination from interlopers (brown dwarfs and early type galaxies at intermediate redshifts). 

In addition, the study of galaxies within the reionization epoch (i.e. at $z>$\,6.5) is obviously linked with the general study of galaxies at high redshift. The redshift range 3$\,<z<\,$6.5 will be widely accessible within the Rubin--LSST surveys, and largely benefit from the additional information that can be obtained from the near-infrared bands observed with Euclid. Although this science case has not received much attention during this joint discussion, we anticipate some of the relevant science cases and opportunities below. 

In the following we will discuss the observational strategy and the natural complement between the two data sets.

\subsection*{Science Drivers} \label{sec:pu-drivers}

In broad terms, the goal of the Euclid+Rubin data processing is the selection of a complete and reliable sample of high redshift galaxies and AGNs. This sample will complement the deeper surveys obtained with HST and JWST. These two instruments are optimal to detect the faintest, low luminosity objects, but lack the large area sensitivity to detect the intrinsically rarer brighter sources.

We envisage three different broad science goals:

\begin{itemize}
\item The combination of the wide and deep samples will be used to obtain a complete description of the galaxy $+$ AGN population, for instance compiling luminosity functions that extend from the faintest  low  mass sources to the brightest, massive systems hosting AGNs \citep{Barnett2021}.
\item{The brightest galaxies that will be detected by Euclid$+$Rubin are ideal targets to study their physical properties (e.g. metallicity, gas content, ionization state, internal velocity structure) in great detail with  dedicated follow--up with other facilities (like {\it JWST}, {\it ALMA} and in the long term ELT-class instrumentation) in order to test galaxy formation models}
\item{ The large scale distribution of the sources  can deliver important information on the mass of the underlying dark matter haloes and - in combination with the spectroscopic measurements of the Ly$\alpha$ emission - of the topology and history of reionization}.

\end{itemize}

\subsection*{Specific Derived Data Products} \label{sec:pu-products}

The selection of high redshift galaxies is obtained by a method that has been implemented and extensively adopted and tested in the last thirty years, the so-called Lyman Break technique.
It exploits the abrupt (redshifted) absorption due to the neutral hydrogen  either by the interstellar medium in the galaxy (at the wavelength corresponding to the redshifted Lyman-Limit) or in the Inter Galactic Medium located on the line of sight, that at $z>\,$5 produces the strongest signature at the redshifted Lyman-$\alpha$ wavelength. 

\begin{figure*}[t]
\centering\includegraphics[width=13cm]{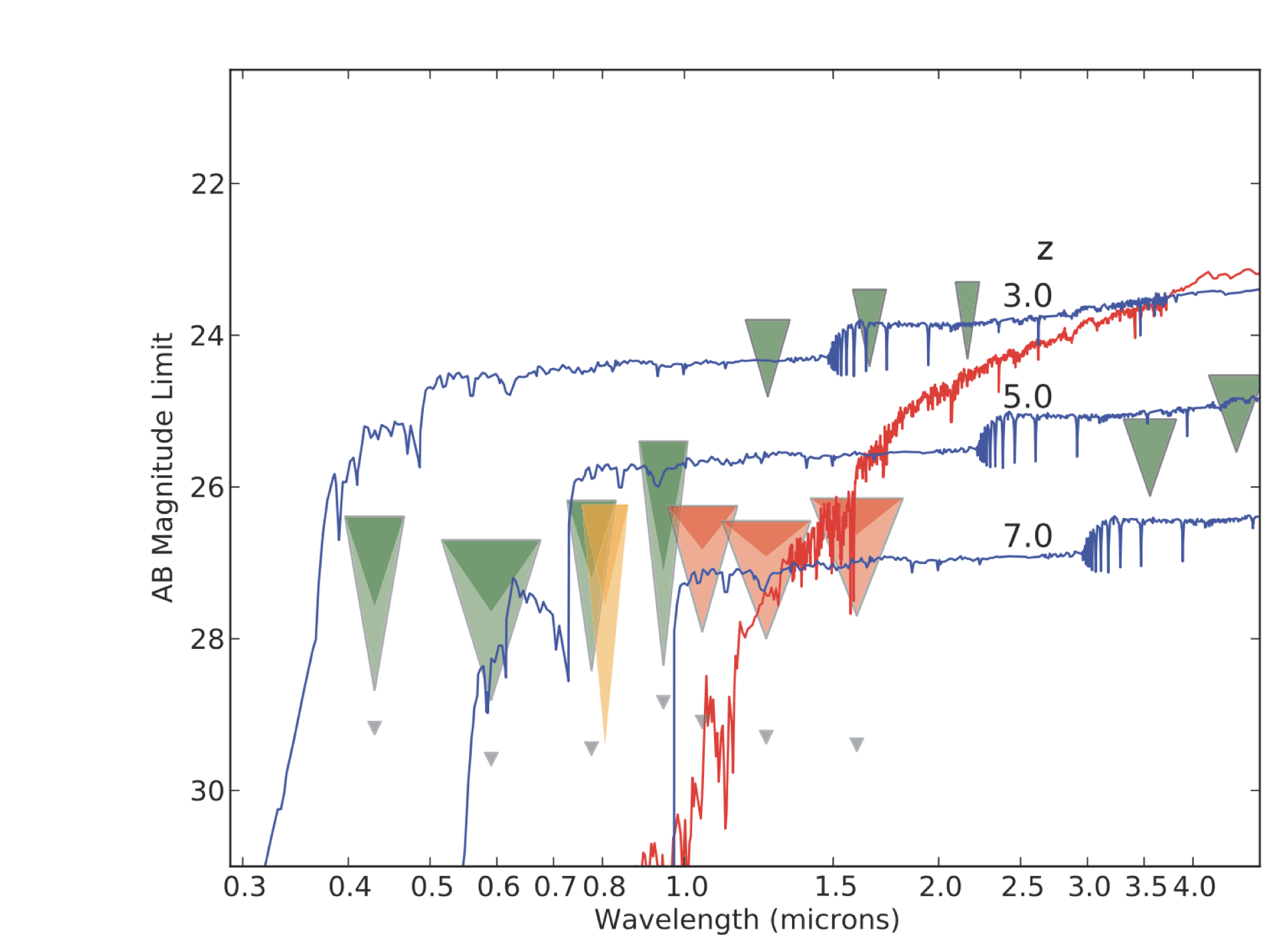} 
\caption{Spectral Energy Distribution of three template galaxies at $z$\,$=$\,3,5,7 (solid blue curve). 
Triangles show the typical  depth of HST deep surveys like CANDELS. 
\label{fig:PU_Figure1}}
\end{figure*}

\begin{figure*}[t]
\centering\includegraphics[width=13cm]{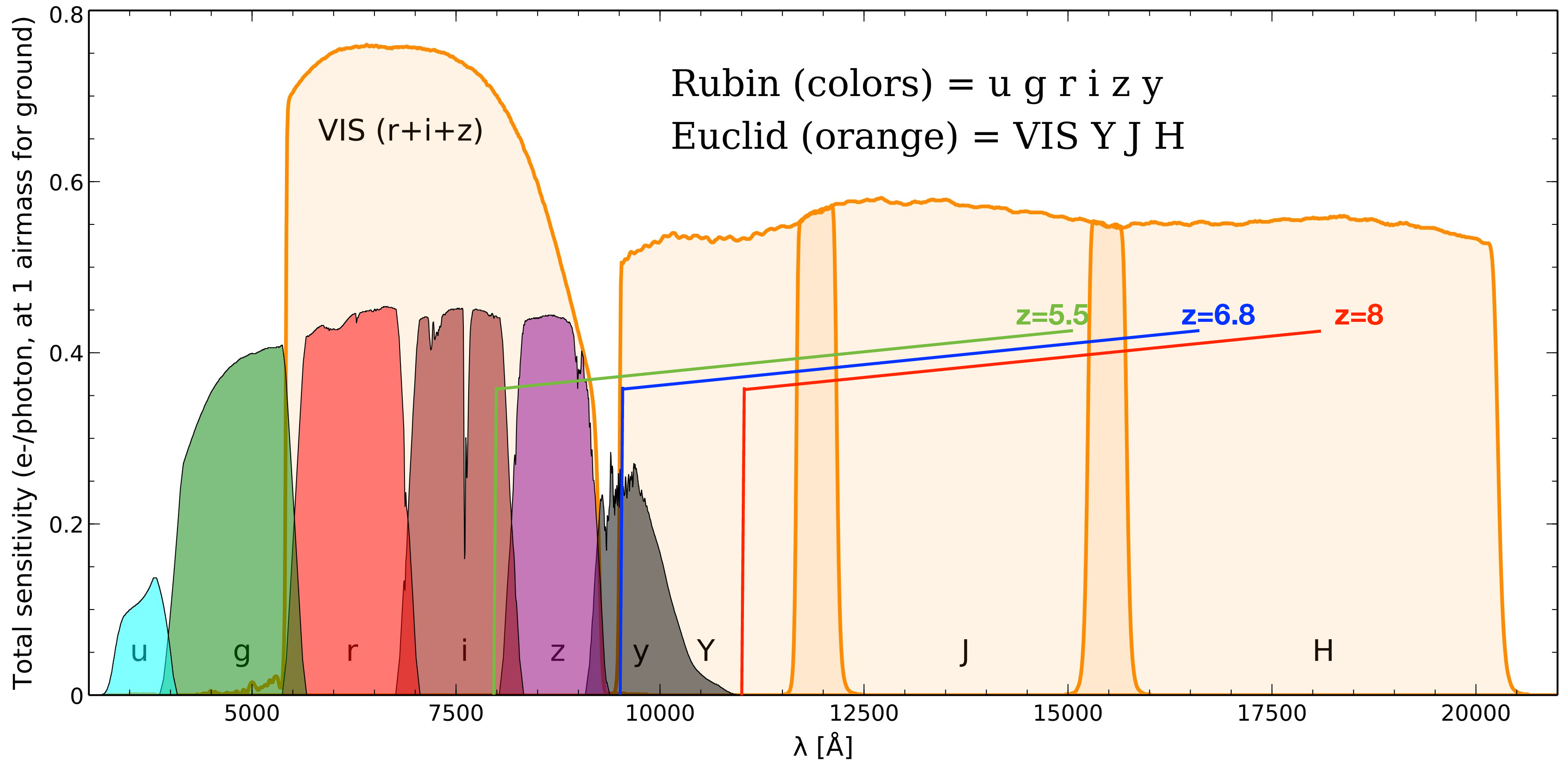} 
\caption{Schematic representation of the spectra of three galaxies at $z$\,$=$\,5.5, 6.8, 8 (the vertical lines show the position of the Ly$\alpha$ break, the nearly horizontal one the extreme UV spectrum) superimposed to the joint Rubin$+$Euclid filter set. \label{fig:PU_Figure2}}
\end{figure*}

The combined effect of redshift + HI absorption in galaxies at $z$\,$=$\,3-7 is shown in \figref{fig:PU_Figure1}, and how it translates into the LSST$+$Euclid filter system is shown in \figref{fig:PU_Figure2}. 
In particular, galaxies at $z>$\,6.8 can be detected only at $\lambda \geq$\,9100\,\text{\AA}, hence essentially only in Euclid YJH bands or in the (much less sensitive and less used) Rubin y band, and are identified by a large break in the VIS band and in the LSST ugriz images. 
As a result, candidate galaxies at high redshift are selected by the strong color break between the band sampling the UV flat spectrum long-ward of the Lyman-$\alpha$ and those sampling the absorbed part of the spectrum at shorter wavelengths.

 In principle, the selection can be done using only the Euclid system, but in practice the LSST images have the key role of enforcing the selection by confirming the non-detection in the shortest bands, thanks to their superior depth and number. Experience with the HST and ground--based surveys has demonstrated that adding further bands below the Lyman Limit significantly lowers the number of interlopers,  noise peaks and other artefacts making the selection more robust.

The DDP of interest for the detection and study of  objects in the reionization epoch ($z\geq\,$6.5) are therefore multi-wavelength photometric catalogs with these features:
\begin{itemize}
    \item Object Positions of objects detected only in the Euclid Y or J or H bands., resulting from detection performed in some combination of the Euclid YJH bands;
    \item Forced photometric measurements in the Euclid VIS and Rubin ugrizy images, at the Euclid YJH positions, correcting for different PSF in the various passbands, with a full suite of techniques (aperture, profile fitting, etc);
    \item A statistically robust estimate of the non-detections in the Euclid VIS and LSST ugriz images, obtained on individual bands as well as in sensible combination of different bands;
    \item Implementation of automated tools to reject image artifacts and other fake sources (trails, spikes etc) that could affect photometry and simulate high-$z$ sources in the resulting catalogs
\end{itemize}

Even if all these tools are implemented at the best state-of-the-art, a visual inspection and a dedicated processing  of the selected sources may be required. For this reason it is also necessary to have access to the image thumbnails of the selected sources in all bands, in order to remove with a final tailored analysis the remaining false detections.

\rec{evaluate}{Evaluate algorithms currently developed in the Euclid and LSST pipelines}{We recommend submitting the algorithms for photometric measurements currently developed within the Euclid and Rubin/LSST projects for evaluation by experts in the study of the high redshift Universe, to assess whether they are  compliant with our needs.}{PU}

\ddp{Optdetcat}{Euclid photometry for high-redshift galaxies detected in LSST}{Euclid YJH photometry for candidate $z\sim\,$4-6 galaxies detected in LSST. 
These galaxies will be drop-outs in the r and i bands. The Euclid near-infrared measurements will be invaluable for pinning down detailed physical properties of this high-redshift galaxy population via spectral energy distribution fitting. See also \ddpgeref{gal4}.}{From Year 1.}{PU} 

\ddp{IRdetcat}{Joint ugrizy,VIS Photometric Catalogs of Euclid high-${\bf z}$ candidates}{ugrizy,VIS forced photometry at the positions of objects only detected in the Euclid YJH bands (list driven only), after correcting for different PSF in the various passbands.}{From Year 1.}{PU}

\ddp{photostack}{Photometric Measurements on multi-band stacks of Euclid high-${\bf z}$ candidates}{For the same objects of \ddppuref{IRdetcat}, forced photometric measurements performed on images obtained by stacking all the ugriz images, to obtain a deeper confirmation of Lyman-break signature.}{From Year 1.}{PU}

\ddp{Stamps}{Stamps of high-${\bf z}$ candidates}
{To remove false detections and interlopers due to image artefacts, moving sources etc., LSST/ugrizy stamps will be provided for the objects of \ddppuref{IRdetcat} while Euclid/YJH stamps will be provided for the objects of \ddppuref{Optdetcat}. 
The size of the cutouts should be of the order of of 40x40 square arcseconds. This DDP prevails over \ddpslref{sl-5}, excluding strong lens cutouts in the presence of objects with high redshift (z$\gtrapprox$\,7) and over \ddpgeref{gal2}, excluding z-band cutouts according to \ddppuref{Optdetcat}. Each stamp should come with with proper local photometric calibration, background and PSF.}{From Year 1.}{PU}

\subsection*{Algorithms} \label{sec:pu-algos}
The algorithms needed to produce the DDPs are standard. They are based on those already developed and extensively tested in many extra-galactic surveys. They will be most likely derived by those adopted for the standard processing of both surveys. As such they do not need a large development and testing cycle.

\subsection*{Technical Considerations} \label{sec:pu-resources}
The processing described here does not require any additional or different resources to what is used for the standard processing of Euclid and LSST data. 

\subsection*{Timescale}  \label{sec:pu-timescale}
Timing is not crucial. These objects are not variable and can be observed at different epochs in the various bands. Variability in the same band can be useful to identify AGN sources, but this can be done only on Euclid Deep Fields data for objects at $z>\,$6.5.

\clearpage
\let\oldbibliography\thebibliography
\renewcommand{\thebibliography}[1]{
    \oldbibliography{#1} \setlength{\itemsep}{0pt}
    } 
{\small
\bibliography{report}
}
\appendix

\newpage
\section{Report authorship}

\subsection*{DDP Working Group members and guest specialists from the community$^*$}
\vspace{0.5cm}


\ddpauthor{0000-0003-0800-8755}{Leanne~P.~Guy}{lguy}
\ddpaffiliation{LSST Project Office, 950 N. Cherry Ave., Tucson, AZ 85719, USA}

\ddpauthor{0000-0002-3263-8645}{Jean--Charles~Cuillandre}{jcuillandre}
\ddpaffiliation{AIM, CEA, CNRS, Universit\'e Paris-Saclay, Universit\'e de Paris, F-91191 Gif-sur-Yvette, France}

\ddpauthor{0000-0002-6578-5078}{Etienne Bachelet}{ebachelet}
\ddpaffiliation{Las Cumbres Observatory, 6740 Cortona Drive, Suite 102, Goleta, CA 93117, USA}

\ddpauthor{0000-0002-0639-5141}{Manda Banerji}{mbanerji}
\ddpaffiliation{University of Southampton, SO17 1BJ, Southampton, UK}

\ddpauthor{0000-0002-8686-8737}{Franz E. Bauer}{fbauer}
\ddpaffiliation{Instituto de Astrof{\'{\i}}sica and Centro de Astroingenier{\'{\i}}a, Facultad de F{\'{i}}sica, Pontificia Univ. Cat{\'{o}}lica de Chile, C. 306, Santiago 22, Chile}
\ddpaffiliation{Millennium Institute of Astrophysics, Nuncio Monse{\~{n}}or S{\'{o}}tero Sanz 100, Of 104, Providencia, Santiago, Chile}
\ddpaffiliation{Space Science Institute, 4750 Walnut Street, Suite 205, Boulder, Colorado 80301, USA}

\ddpauthor{0000-0001-5564-3140}{Thomas Collett}{tcollett}
\ddpaffiliation{Institute of Cosmology \& Gravitation, University of Portsmouth, Portsmouth, po1 3fx, UK}

\ddpauthor{0000-0003-1949-7638}{Christopher J. Conselice}{cconselice}
\ddpaffiliation{Jodrell Bank Centre for Astrophysics, University of Manchester, Oxford Road, Manchester UK}

\ddpauthor{0000-0002-1398-6302}{Siegfried Eggl}{seggl}
\ddpaffiliation{Rubin Observatory / Department of Astronomy, University of Washington, WA, USA}
\ddpaffiliation{Department of Aerospace Engineering, University of Illinois at Urbana-Champaign, IL, USA}

\ddpauthor{0000-0001-7934-1278}{Annette Ferguson$^*$}{aferguson}
\ddpaffiliation{Institute for Astronomy, University of Edinburgh, Royal Observatory, Blackford Hill, Edinburgh EH9 3HJ, UK}

\ddpauthor{0000-0003-3820-2823}{Adriano Fontana}{afontana}
\ddpaffiliation{INAF, Osservatorio Astronomico di Roma, Via Frascati 33, I-00078 Monte Porzio Catone (RM), Italy}

\ddpauthor{0000-0001-7023-3940}{Catherine Heymans}{cheymans}
\ddpaffiliation{Institute for Astronomy, University of Edinburgh, Royal Observatory, Blackford Hill, Edinburgh, EH9 3HJ, UK.}
\ddpaffiliation{Ruhr University Bochum, Faculty of Physics and Astronomy, Astronomical Institute (AIRUB), German Centre for Cosmological Lensing, 44780 Bochum, Germany}

\ddpauthor{0000-0002-2960-978X}{Isobel M. Hook}{ihook}
\ddpaffiliation{Physics Department, Lancaster University, Lancaster, LA1 4YB, U.K.}


\ddpauthor{0000-0002-5592-023X}{\'{E}ric Aubourg}{eaubourg}
\ddpaffiliation{APC, Astroparticule et Cosmologie, Université Paris Diderot, CNRS/IN2P3, CEA/Irfu, Observatoire de Paris, Sorbonne Paris Cité, 10, rue Alice Domon \& Léonie Duquet, 75205 Paris Cedex 13, France}

\ddpauthor{0000-0002-3263-8645}{Herv\'e Aussel}{haussel}
\ddpaffiliation{AIM, CEA, CNRS, Universit\'e Paris-Saclay, Universit\'e de Paris, F-91191 Gif-sur-Yvette, France}

\ddpauthor{0000-0003-2759-5764}{James Bosch}{jbosch}
\ddpaffiliation{Department of Astrophysical Sciences, Princeton University, Princeton NJ 08544, USA}

\ddpauthor{0000-0001-5242-3089}{Benoit Carry$^*$}{bcarry}
\ddpaffiliation{Université Côte d'Azur, Observatoire de la Côte d'Azur, CNRS, Laboratoire Lagrange, France}

\ddpauthor{0000-0002-0641-3231}{Henk Hoekstra}{hhoekstra}
\ddpaffiliation{Leiden Observatory, Leiden University, PO Box 9513, NL-2300 RA Leiden, the Netherlands}

\ddpauthor{0000-0002-3827-0175}{Konrad Kuijken}{kkuijken}
\ddpaffiliation{Leiden Observatory, Leiden University, P.O. Box 9513, 2300RA Leiden, The Netherlands}

\ddpauthor{0000-0001-7956-0542}{Francois Lanusse}{flanusse}
\ddpaffiliation{AIM, CEA, CNRS, Universit\'e Paris-Saclay, Universit\'e de Paris, F-91191 Gif-sur-Yvette, France}

\ddpauthor{0000-0002-8873-5065}{Peter Melchior}{pmelchior}
\ddpaffiliation{Department of Astrophysical Sciences, Princeton University, Princeton NJ 08544, USA}

\ddpauthor{0000-0002-6875-2087}{Joseph Mohr}{jmohr}
\ddpaffiliation{Faculty of Physics, Ludwig-Maximilians-Universität, Scheinerstr. 1, D-81679 Munich, Germany}

\ddpauthor{0000-0002-7616-7136}{Michele Moresco}{mmoresco}
\ddpaffiliation{Dipartimento di Fisica e Astronomia ``Augusto Righi'', Alma Mater Studiorum Universit\`{a} di Bologna, via Piero Gobetti 93/2, I-40129 Bologna, Italy}
\ddpaffiliation{INAF - Osservatorio di Astrofisica e Scienza dello Spazio di Bologna, via Piero Gobetti 93/3, I-40129 Bologna, Italy}

\ddpauthor{0000-0001-8179-5965}{Reiko Nakajima}{rnakajima}
\ddpaffiliation{Argelander-Institut f\"ur Astronomie, Universit\"at Bonn, Auf dem H\"ugel 71, 53121 Bonn, Germany}

\ddpauthor{0000-0002-8108-9179}{St\'ephane Paltani}{spaltani}
\ddpaffiliation{University of Geneva, ch. d'\'Ecogia 16, 1290 Versoix, Switzerland}

\ddpauthor{0000-0002-5622-5212}{Michael Troxel}{mtroxel}
\ddpaffiliation{Department of Physics, Duke University Durham, NC 27708, USA}

\vspace{-0.5cm}
\subsection*{DDP Forum Community members}
\vspace{0.5cm}


\ddpauthor{0000-0001-7232-5152}{Viola Allevato}{allevato}\ddpaffiliation{INAF-OAS Bologna, via Piero Gobetti 93/3, I-40129 Bologna, Italy}
 
\ddpauthor{0000-0003-3481-3491}{Adam Amara}{amara}\ddpaffiliation{ Institute of Cosmology and Gravitation, University of Portsmouth,Portsmouth PO1 3FX, UK}
 
\ddpauthor{0000-0002-2041-8784}{Stefano Andreon}{andreon}\ddpaffiliation{INAF-Osservatorio Astronomico di Brera, via Brera, 28, 20121, Milano, Italy}
 
\ddpauthor{0000-0003-0930-5815}{Timo Anguita}{anguita}\ddpaffiliation{Departamento de Ciencias Fisicas, Universidad Andres Bello Fernandez Concha 700, Las Condes, Santiago, Chile}
 
\ddpauthor{0000-0002-8900-0298}{Sandro Bardelli}{bardelli}\ddpaffiliation{INAF-OAS Bologna, via Piero Gobetti 93/3, I-40129 Bologna, Italy}
 
\ddpauthor{0000-0001-8156-0429} {Keith Bechtol} {bechtol}\ddpaffiliation{Physics Department, University of Wisconsin–Madison, 1150 University Avenue, Madison, WI 53706-1390, USA}\
 
\ddpauthor{0000-0003-3195-5507}{Simon Birrer}{birrer}\ddpaffiliation{Kavli Institute for Particle Astrophysics and Cosmology and Department of Physics, Stanford University, Stanford, CA 94305, USA}
 
\ddpauthor{0000-0003-0492-4924}{Laura Bisigello}{bisigello}\ddpaffiliation{Dipartimento di Fisica e Astronomia, Università di Padova, Vicolo dell'Osservatorio, 3, I-35122, Padova, Italy}
 
\ddpauthor{0000-0003-3278-4607}{Micol Bolzonella}{bolzonella}\ddpaffiliation{INAF-OAS Bologna, via Piero Gobetti 93/3, I-40129 Bologna, Italy}
 
\ddpauthor{0000-0002-3938-692X}{Maria Teresa Botticella}{botticella}\ddpaffiliation{INAF, Osservatorio Astronomico di Capodimonte, salita Moiariello 16, 80131, Napoli Italy}
 
\ddpauthor{0000-0002-7084-487X}{Herv\'e Bouy}{bouy}\ddpaffiliation{Laboratoire d'Astrophysique de Bordeaux, Univ. Bordeaux, CNRS, B18N, all{\'e}e Geoffroy Saint-Hilaire, 33615 Pessac, France}
 
\ddpauthor{0000-0003-4359-8797}{Jarle Brinchmann}{brinchmann}\ddpaffiliation{Instituto de Astrofísica e Ciências do Espaço, Universidade do Porto, CAUP, Rua das Estrelas, PT4150-762 Porto, Portugal}
 
\ddpauthor{0000-0001-7316-4573}{Sarah Brough}{brough}\ddpaffiliation{School of Physics, University of New South Wales, NSW 2052, Australia}
 
\ddpauthor{0000-0003-3399-3574}{Stefano Camera}{camera}\ddpaffiliation{Dipartimento di Fisica, Univ. degli Studi di Torino, Via P.\ Giuria 1, 10125 Torino, Italy}\ddpaffiliation{INFN -- Istituto Nazionale di Fisica Nucleare, Sezione di Torino, Via P.\ Giuria 1, 10125 Torino, Italy}\ddpaffiliation{INAF -- Istituto Nazionale di Astrofisica, Osservatorio Astrofisico di Torino, Strada Osservatorio 20, 10025 Pino Torinese, Italy}
 
\ddpauthor{0000-0003-2072-384X}{Michele Cantiello}{cantiello} \ddpaffiliation{INAF -- Osservatorio Astronomico d'Abruzzo, via Maggini snc, 64100 - Teramo, Italy}
 
\ddpauthor{0000-0001-5008-8619}{Enrico Cappellaro}{cappellaro}\ddpaffiliation{INAF, Osservatorio Astronomico di Padova, vicolo dell'Osservatorio 5, I-35122, Padova Italy}
 
\ddpauthor{0000-0002-3936-9628} {Jeffrey L. Carlin} {carlin}\ddpaffiliation{NSF's NOIRLab/Rubin Observatory Project Office, 950 North Cherry Avenue, Tucson, AZ 85719, USA}\
 
\ddpauthor{0000-0001-7316-4573}{Francisco J Castander}{castander}\ddpaffiliation{Institute of Space Sciences (ICE, CSIC), Carrer de Can Magrans s/n, E-08193 Bellaterra (Barcelona), Spain} \ddpaffiliation{Institut d'Estudis Espacials de Catalunya (IEEC), E-08034 Barcelona, Spain}

\ddpauthor{0000-0001-9875-8263}{Marco Castellano}{mcastellano}
\ddpaffiliation{INAF, Osservatorio Astronomico di Roma, Via Frascati 33, I-00078 Monte Porzio Catone (RM), Italy}

\ddpauthor{0000-0001-7583-0621}{Ranga Ram Chari}{chari}\ddpaffiliation{Infrared Processing and Analysis Center, MS314-6, California Institute of Technology, Pasadena, CA, 91125, USA}
 
\ddpauthor{0000-0003-4221-6718}{Nora Elisa Chisari}{chisari}\ddpaffiliation{Institute for Theoretical Physics, Utrecht University, Princetonplein 5, 3584 CC, The Netherlands}
 
\ddpauthor{0000-0002-43372-38466}{Christopher Collins}{collins}\ddpaffiliation{Astrophysics Research Institute, Liverpool John Moores University, IC2, Liverpool Science Park, 146 Brownlow Hill, Liverpool L3 5RF, UK}
 
\ddpauthor{0000-0003-0758-6510}{Fr\'ed\'eric Courbin}{courbin}\ddpaffiliation{Institute of Physics, Laboratory of Astrophysics, Ecole Polytechnique F\'ed\'erale de Lausanne (EPFL), Observatoire de Sauverny, 1290 Versoix, Switzerland}
 
\ddpauthor{0000-0002-8767-1442}{Jean-Gabriel Cuby}{cuby}\ddpaffiliation{Aix-Marseille Univ, CNRS, CNES, LAM, Marseille, France}
 
\ddpauthor{0000-0002-9336-7551}{Olga Cucciati}{cucciati}\ddpaffiliation{INAF-OAS Bologna, via Piero Gobetti 93/3, I-40129 Bologna, Italy}
 
\ddpauthor{0000-0002-6939-9211}{Tansu Daylan}{daylan}\ddpaffiliation{Department of Physics and Kavli Institute for Astrophysics and Space Research, Massachusetts Institute of Technology, Cambridge, MA 02139, USA}\ddpaffiliation{Department of Astrophysical Sciences, Princeton University, 4 Ivy Lane, Princeton, NJ 08544}
 
\ddpauthor{0000-0001-9065-3926}{Jose M. Diego}{diego}\ddpaffiliation{Instituto de F\'isica de Cantabria (CSIC-UC), Santander, Spain}
 
\ddpauthor{0000-0003-3343-6284}{Pierre-Alain Duc}{duc}\ddpaffiliation{Universit\'e de Strasbourg, CNRS, Observatoire astronomique de Strasbourg, UMR 7550, F-67000 Strasbourg, France}
 
\ddpauthor{0000-0002-9686-254X}{Sotiria Fotopoulou}{fotopoulou}\ddpaffiliation{HH Wills Physics Laboratory, Tyndall Avenue, Bristol BS8 1TL, UK}
 
\ddpauthor{0000-0002-7496-3796}{Dominique Fouchez}{fouchez}\ddpaffiliation{Aix Marseille Univ, CNRS/IN2P3, CPPM, Marseille, France}
 
\ddpauthor{0000-0002-5540-6935}{Rapha\"el Gavazzi}{gavazzi}\ddpaffiliation{Laboratoire d'Astrophysique de Marseille, Aix-Marseille Université, CNRS, CNES, UMR7326, F-13388 Marseille, France}\ddpaffiliation{Institut d'Astrophysique de Paris, CNRS \& Sorbonne Université, UMR7095, F-75014 Paris, France}\ddpaffiliation{Institute of Astronomy, University of Cambridge, Madingley Road, Cambridge CB30HA, UK}
 
\ddpauthor{0000-0003-3270-7644}{Daniel Gruen}{gruen}\ddpaffiliation{Ludwig-Maximilians-Universität, Faculty of Physics, University Observatory, Scheinerstr. 1, 81679 Munich, Germany}\ddpaffiliation{Kavli Institute for Particle Astrophysics \& Cosmology, Stanford University, PO Box 2450, Stanford, CA 94305, USA}
 
\ddpauthor{0000-0002-3065-457X}{Peter Hatfield}{hatfield}\ddpaffiliation{Oxford Astrophysics, Denys Wilkinson Building, University of Oxford, Keble Rd, Oxford OX1 3RH, UK}
 
\ddpauthor{0000-0002-9814-3338}{Hendrik Hildebrandt}{hildebrandt}\ddpaffiliation{Ruhr University Bochum, Faculty of Physics and Astronomy, Astronomical Institute (AIRUB), German Centre for Cosmological Lensing, 44780 Bochum, Germany}
 
\ddpauthor{0000-0002-8150-9786}{Hermine Landt}{hlandt}\ddpaffiliation{Centre for Extragalactic Astronomy, Durham University, Durham DH1 3LE, UK}
 
\ddpauthor{0000-0001-9162-2371}{Leslie K. Hunt}{hunt}\ddpaffiliation{INAF - Osservatorio Astrofisico di Arcetri, Largo E. Fermi 5, 50125 Firenze, Italy}
 
\ddpauthor{0000-0002-3292-9709}{Rodrigo Ibata}{ibata}\ddpaffiliation{Universit\'e de Strasbourg, CNRS, Observatoire astronomique de Strasbourg, UMR 7550, F-67000 Strasbourg, France}
 
\ddpauthor{0000-0002-7303-4397}{Olivier Ilbert}{ilbert}\ddpaffiliation{Aix Marseille Université, CNRS, CNES, LAM, UMR 7326, F-13388 Marseille, France}
 
\ddpauthor{0000-0002-4677-5843}{Jens Jasche}{jasche}\ddpaffiliation{The Oskar Klein Centre, Department of Physics, Stockholm University, Albanova University Center, SE 106 91 Stockholm, Sweden}
 
\ddpauthor{0000-0001-7494-1303}{Benjamin Joachimi}{joachimi}\ddpaffiliation{Department of Physics and Astronomy, University College London, Gower Street, London WC1E 6BT, UK}
 
\ddpauthor{0000-0002-2704-5028}{Rémy Joseph}{joseph}\ddpaffiliation{Department of Astrophysical Sciences, Princeton University, 4 Ivy Lane, Princeton, NJ 08544, USA}\ddpaffiliation{The Oskar Klein Centre, Department of Physics, Stockholm University, AlbaNova, SE-10691 Stockholm, Sweden}
 
\ddpauthor{0000-0001-5455-3653}{Rubina Kotak}{kotak}\ddpaffiliation{Dept. of Physics \& Astronomy, University of Turku, Vesilinnantie 5, Turku, FI-20500, Finland}
 
\ddpauthor{}{Clotilde Laigle}{laigle}\ddpaffiliation{Sorbonne Universit\'{e}, CNRS, UMR 7095, Institut d'Astrophysique de Paris, 98 bis bd Arago, 75014 Paris, France}
 
\ddpauthor{0000-0002-7214-8296}{Ariane Lançon}{lancon}\ddpaffiliation{Universit\'e de Strasbourg, CNRS, Observatoire astronomique de Strasbourg, UMR 7550, F-67000 Strasbourg, France}
 
\ddpauthor{0000-0003-0069-1203}{S{\o}ren S. Larsen}{larsen}\ddpaffiliation{Department of Astrophysics/IMAPP, Radboud University, P.O. Box 9010, 6500 GL Nijmegen, The Netherlands}
 
\ddpauthor{0000-0003-0143-8891}{Guilhem Lavaux}{lavaux}\ddpaffiliation{CNRS \& Sorbonne Université, UMR7095, Institut d'Astrophysique de Paris, F-75014 Paris, France}
 
\ddpauthor{0000-0002-9339-1404}{Florent Leclercq}{leclercq}\ddpaffiliation{Imperial Centre for Inference and Cosmology (ICIC) \& Astrophysics group, Imperial College London, Blackett Laboratory, Prince Consort Road, London SW7 2AZ, UK}\ddpaffiliation{CNRS \& Sorbonne Université, UMR7095, Institut d'Astrophysique de Paris, F-75014 Paris, France}
 
\ddpauthor{0000-0002-7810-6134}{C. Danielle Leonard}{leonard}\ddpaffiliation{School of Mathematics, Statistics and Physics, Herschel Building, Newcastle University, NE1 7RU, Newcastle-upon-Tyne, UK}
 
\ddpauthor{0000-0002-3881-7724}{Anja von der Linden}{linden}\ddpaffiliation{Department of Physics and Astronomy, Stony Brook University, Stony Brook, NY 11794, USA}
 
\ddpauthor{0000-0003-0049-5210}{Xin Liu}{liu}\ddpaffiliation{Department of Astronomy, University of Illinois at Urbana-Champaign, Urbana, IL 61801, USA}\ddpaffiliation{National Center for Supercomputing Applications, University of Illinois at Urbana-Champaign, Urbana, IL 61801, USA}
 
\ddpauthor{0000-0002-9182-8414}{Giuseppe Longo}{longo}\ddpaffiliation{Dipartimento di Fisica “Ettore Pancini”, Università di Napoli Federico II,via Cintia, 80126, Italy}\ddpaffiliation{INFN - Unità di Napoli, via Cintia 9, 80126, Napoli, Italy}
 
\ddpauthor{0000-0001-9158-4838} {Manuela Magliocchetti} {magliocchetti}\ddpaffiliation{INAF-IAPS, Via Fosso del Cavaliere 100, 00133, Roma, Italy}\
 
\ddpauthor{0000-0001-7711-3677}{Claudia Maraston}{maraston}\ddpaffiliation{ Institute of Cosmology and Gravitation, University of Portsmouth,Portsmouth PO1 3FX, UK}
 
\ddpauthor{0000-0002-0113-5770}{Phil Marshall}{marshall}\ddpaffiliation{SLAC National Accelerator Laboratory, 2575 Sand Hill Road, MS29, Menlo Park, CA 94025, USA}

\ddpauthor{0000-0002-1208-4833}{Eduardo\ L.\ Mart\'in}{martin}\ddpaffiliation{Instituto de Astrof\'isica de Canarias (IAC), Calle V\'ia L\'actea s/n, E-38200 La Laguna, Tenerife, Spain} \ddpaffiliation{Departamento de Astrof\'isica, Universidad de La Laguna (ULL), E-38206 La Laguna, Tenerife, Spain}\ddpaffiliation{Consejo Superior de Investigaciones Cient\'ificas (CSIC), E-28006 Madrid, Spain}
 
\ddpauthor{0000-0001-7497-2994}{Seppo Mattila}{mattila}\ddpaffiliation{Dept. of Physics \& Astronomy, University of Turku, Vesilinnantie 5, Turku, FI-20500, Finland}
 
\ddpauthor{0000-0002-3517-2422}{Matteo Maturi}{maturi}\ddpaffiliation{Zentrum f\"ur Astronomie, Universitat\"at Heidelberg, Philosophenweg 12, D-69120 Heidelberg, Germany}\ddpaffiliation{Institute for Theoretical Physics, Philosophenweg 16, D-69120 Heidelberg, Germany}
 
\ddpauthor{0000-0002-9489-7765}{Henry Joy McCracken}{mccracken}\ddpaffiliation{Sorbonne Universit\'{e}, CNRS, UMR 7095, Institut d'Astrophysique de Paris, 98 bis bd Arago, 75014 Paris, France}
 
\ddpauthor{0000-0003-3167-2574}{R. Benton Metcalf}{metcalf}\ddpaffiliation{Dipartimento di Fisica e Astronomia "A. Righi", Alma Mater Studiorum Univ. di Bologna, via Piero Gobetti 93/2, I-40129 Bologna, Italy}\ddpaffiliation{INAF-OAS Bologna, via Piero Gobetti 93/3, I-40129 Bologna, Italy}
 
\ddpauthor{0000-0001-7847-0393}{Mireia Montes}{montes}\ddpaffiliation{Space Telescope Science Institute, 3700 San Martin Drive, Baltimore, MD 21218, USA}
 
\ddpauthor{0000-0002-0041-3783}{Daniel Mortlock}{mortlock}\ddpaffiliation{Astrophysics Group, Department of Physics, Imperial College London, London SW7 2AZ, UK}
 
\ddpauthor{0000-0002-3473-6716}{Lauro Moscardini}{moscardini}\ddpaffiliation{Dipartimento di Fisica e Astronomia "A. Righi", Alma Mater Studiorum Univ. di Bologna, via Piero Gobetti 93/2, I-40129 Bologna, Italy}\ddpaffiliation{INAF-OAS Bologna, via Piero Gobetti 93/3, I-40129 Bologna, Italy} \ddpaffiliation{INFN - Sezione di Bologna, viale Berti Pichat 6, I-40127 Bologna, Italy}
 
\ddpauthor{0000-0001-6022-0484}{Gautham Narayan}{narayan}\ddpaffiliation{Department of Astronomy, University of Illinois at Urbana-Champaign, Urbana, IL 61801, USA}\ddpaffiliation{Center for AstroPhysical Surveys, National Center for Supercomputing Applications, University of Illinois at Urbana-Champaign, Urbana, IL 61801, USA}
 
\ddpauthor{0000-0003-4210-7693}{Maurizio Paolillo }{paolillo}\ddpaffiliation{Dipartimento di Fisica “Ettore Pancini”, Università di Napoli Federico II,via Cintia, 80126, Italy}\ddpaffiliation{INAF, Osservatorio Astronomico di Capodimonte, Via Moiariello 16, 80131, Naples, Italy}\ddpaffiliation{INFN - Unità di Napoli, via Cintia 9, 80126, Napoli, Italy}
 
\ddpauthor{0000-0002-3733-8174}{Polychronis Papaderos}{papaderos}\ddpaffiliation{Instituto de Astrof\'{i}sica e Ci\^{e}ncias do Espaço - Centro de Astrof\'isica da Univ. do Porto, Rua das Estrelas, 4150-762 Porto, Portugal}
 
\ddpauthor{0000-0003-0858-6109}{Roser Pello}{pello}\ddpaffiliation{Aix Marseille Université, CNRS, CNES, LAM, UMR 7326, F-13388 Marseille, France}
 
\ddpauthor{0000-0001-7085-0412}{Lucia Pozzetti}{pozzetti}\ddpaffiliation{INAF-OAS Bologna, via Piero Gobetti 93/3, I-40129 Bologna, Italy}
 
\ddpauthor{0000-0002-3585-866X}{Mario Radovich}{radovich}\ddpaffiliation{INAF - Osservatorio Astronomico di Padova, via dell'Osservatorio 5, 35122, Padova, Italy}
 
\ddpauthor{0000-0002-6577-2787}{Marina Rejkuba}{rejkuba}\ddpaffiliation{European Southern Observatory, Karl-Schwarzschild Strasse 2, 85748 Garching, Germany}
 
\ddpauthor{0000-0002-3849-3467}{Javier Rom\'an}{roman}\ddpaffiliation{Instituto de Astrof\'{\i}sica de Canarias, c/ V\'{\i}a L\'actea s/n, E-38205, La Laguna, Tenerife, Spain}\ddpaffiliation{Departamento de Astrof\'{\i}sica, Universidad de La Laguna, E-38206, La Laguna, Tenerife, Spain}
 
\ddpauthor{0000-0003-4945-0056}{Rub\'en S\'anchez-Janssen}{sanchezjanssen}\ddpaffiliation{UK Astronomy Technology Centre, Royal Observatory, Blackford Hill, Edinburgh, EH9 3HJ, UK}
 
\ddpauthor{0000-0002-1256-655X}{Elena Sarpa}{sarpa}\ddpaffiliation{Dipartimento di Fisica e Astronomia ``Galileo Galilei'', Univ. degli studi di Padova, Via F. Marzolo, 8, I-35131 Padova, Italy }
 
\ddpauthor{0000-0003-1337-5269}{Barbara Sartoris}{sartoris}\ddpaffiliation{Universit{\"a}ts-Sternwarte M{\"u}nchen, Fakult{\"a}t f{\"u}r Physik, Ludwig--Maximilians Universit{\"a}t, Scheinerstrasse 1, 81679 M{\"u}nchen, Germany}
 
\ddpauthor{0000-0002-6987-7834}{Tim Schrabback}{schrabback}\ddpaffiliation{Argelander-Institut f\"ur Astronomie, Universit\"at Bonn, Auf dem H\"ugel 71, 53121 Bonn, Germany}

\ddpauthor{0000-0001-6116-2095}{Dominique Sluse}{sluse}\ddpaffiliation{STAR Institute, Quartier Agora - All\'ee du six Ao\^ut, 19c B-4000 Li\`ege, Belgium}
 
\ddpauthor{0000-0002-8229-1731}{Stephen J. Smartt}{smartt}\ddpaffiliation{Astrophysics Research Centre, School of Mathematics and Physics, Queen's University Belfast, BT7 1NN, UK}
 
\ddpauthor{0000-0003-4494-8277}{Graham P.\ Smith}{smith}\ddpaffiliation{School of Physics and Astronomy, University of Birmingham, Edgbaston, B15 2TT, UK}
 
\ddpauthor{0000-0001-9328-2905}{Colin Snodgrass}{snodgrass}\ddpaffiliation{Institute for Astronomy, University of Edinburgh, Royal Observatory, Edinburgh, UK}
 
\ddpauthor{0000-0003-4352-2063}{Margherita Talia}{talia}\ddpaffiliation{Dipartimento di Fisica e Astronomia "A. Righi", Alma Mater Studiorum Univ. di Bologna, via Piero Gobetti 93/2, I-40129 Bologna, Italy}\ddpaffiliation{INAF-OAS Bologna, via Piero Gobetti 93/3, I-40129 Bologna, Italy}
 
\ddpauthor{0000-0001-7961-8177}{Charling Tao}{tao}\ddpaffiliation{Centre de Physique des Particules de Marseille, Université Aix-Marseille, 163, avenue de Luminy, 13288, Marseille, cedex 09, France}
 
\ddpauthor{0000-0002-3263-8645}{Sune Toft}{toft}\ddpaffiliation{Cosmic Dawn Center (DAWN), Niels Bohr Institute, University of Copenhagen, Jagtvej 128, DK-2200, Copenhagen, Denmark }
 
\ddpauthor{0000-0001-7958-6531}{Crescenzo Tortora}{tortora}\ddpaffiliation{INAF -- Osservatorio Astronomico di Capodimonte, Salita Moiariello 16, 80131 - Napoli, Italy}
 
\ddpauthor{0000-0002-3199-0399}{Isaac Tutusaus}{tutusaus}\ddpaffiliation{Universit\'e de Gen\`eve, D\'epartement de Physique Th\'eorique and Centre for Astroparticle Physics, 24 quai Ernest-Ansermet, CH-1211 Gen\`eve 4, Switzerland}
 
\ddpauthor{0000-0002-7383-7106}{Christopher Usher}{usher}\ddpaffiliation{The Oskar Klein Centre, Department of Astronomy, Stockholm University, AlbaNova, SE-106 91 Stockholm, Sweden}
 
\ddpauthor{0000-0002-3859-8074}{Sjoert van Velzen}{velzen}\ddpaffiliation{Leiden Observatory, Leiden University, PO Box 9513, 2300 RA Leiden, The Netherlands}
 
\ddpauthor{0000-0002-0730-0781}{Aprajita Verma}{verma}\ddpaffiliation{Oxford Astrophysics, Denys Wilkinson Building, University of Oxford, Keble Rd, Oxford OX1 3RH, UK}
 
\ddpauthor{0000-0001-8554-7248}{Georgios Vernardos}{vernardos}\ddpaffiliation{Institute of Physics, Laboratory of Astrophysics, Ecole Polytechnique F\'ed\'erale de Lausanne (EPFL), Observatoire de Sauverny, 1290 Versoix, Switzerland}
 
\ddpauthor{0000-0001-6215-0950}{Karina Voggel}{voggel}\ddpaffiliation{Universit\'e de Strasbourg, CNRS, Observatoire astronomique de Strasbourg, UMR 7550, F-67000 Strasbourg, France}
 
\ddpauthor{0000-0002-5854-8269}{Benjamin Wandelt}{wandelt}\ddpaffiliation{Sorbonne Universit\'{e}, CNRS, UMR 7095, Institut d'Astrophysique de Paris, 98 bis bd Arago, 75014 Paris, France}\ddpaffiliation{Center for Computational Astrophysics, Flatiron Institute, 162 5th Avenue, New York, NY 10010, USA}
 
\ddpauthor{0000-0003-4859-3290}{Aaron E. Watkins}{watkins}\ddpaffiliation{Centre for Astrophysics Research, School of Physics, Astronomy and Mathematics, University of Hertfordshire, Hatfield AL10 9AB, UK}
 
\ddpauthor{0000-0002-8282-2010}{Jochen Weller}{weller}\ddpaffiliation{Universit\"ats-Sternwarte, Fakult\"at f\"ur Physik, Ludwig-Maximilians Universit\"at M\"unchen, Scheinerstr. 1, 81679 M\"unchen, Germany}\ddpaffiliation{Max Planck Institute for Extraterrestrial Physics, Giessenbachstrasse, 85748 Garching, Germany}
 
\ddpauthor{0000-0001-7363-7932}{Angus H Wright}{wright}\ddpaffiliation{Ruhr University Bochum, Faculty of Physics and Astronomy, Astronomical Institute (AIRUB), 44780 Bochum, Germany}
 
\ddpauthor{0000-0003-2874-6464}{Peter Yoachim}{yoachim}\ddpaffiliation{Department of Astronomy, University of Washington, Seattle, WA 98195, USA}\ddpaffiliation{DIRAC Institute, University of Washington, Seattle, WA 98195, USA }
 
\ddpauthor{0000-0001-9163-0064}{Ilsang Yoon}{yoon}\ddpaffiliation{National Radio Astronomy Observatory, 520 Edgemont Road, Charlottesville, VA 22033, USA}
 
\ddpauthor{0000-0002-5845-8132}{Elena Zucca}{zucca}\ddpaffiliation{INAF-OAS Bologna, via Piero Gobetti 93/3, I-40129 Bologna, Italy}

\newpage
\section{Acknowledgments}

This material is based on work supported in part by the
National Science Foundation through Cooperative Agreement
1258333 managed by the Association of Universities for
Research in Astronomy (AURA), and the Department of
Energy under Contract No. DE-AC02-76SF00515 with the
SLAC National Accelerator Laboratory. Additional LSST
funding comes from private donations, grants to universities,
and in-kind support from LSSTC Institutional Members. This
research has made use of NASA’s Astrophysics Data System
Bibliographic Services.

Acknowledgments from the DDP community:
\begin{itemize}

\item  Etienne Bachelet gratefully acknowledges support from NASA grant 80NSSC19K0291.

\item  Jarle Brinchmann acknowledges  support  by  Fundação  para  a  Ciência  e  a  Tecnologia  (FCT)  through  the  research  grantsUIDB / 04434 / 2020 and UIDP/04434/202  and through FCT project PTDC / FIS-AST / 4862 / 2020.

\item Thomas Collett acknowledges funding from a Royal Society University Research Fellowship.

\item Christopher Collins acknowledges support from the STFC grant ST/S006095/1.

\item Jose M. Diego acknowledges support from projects PGC2018-101814-B-100 and MDM-2017-0765.

\item Peter Hatfield acknowledges generous support from the Hintze Family Charitable Foundation through the Oxford Hintze Centre for Astrophysical Surveys.

\item Catherine Heymans acknowledges support from the European Research Council under grant number 647112, and support from the Max Planck Society and the Alexander von Humboldt Foundation in the framework of the Max Planck-Humboldt Research Award endowed by the Federal Ministry of Education and Research. 

\item Isobel Hook acknowledges support from STFC grants ST/V000713/1 and ST/R000514/1.

\item Ariane Lançon acknowledges support from project ANR-19-CE31-0022, which is funded by Agence Nationale de la Recherche, France.

\item Eduardo L. Martín acknowledges funding from the Spanish Ministry of Economy and Competitiveness (MINECO) and the Fondo Europeo de Desarrollo Regional (FEDER) under grants PID2019-109522GB-C53 and PID2019-107061GB-C66\@. 

\item Michele Moresco acknowledges support from MIUR, PRIN 2017 (grant 20179ZF5KS)and grants ASI n.I/023/12/0 and ASI n.2018-23-HH.0.

\item Polychronis Papaderos acknowledges support from FCT grants UID/FIS/04434/2019, UIDB/04434/2020, UIDP/04434/2020 and the project "Identifying the Earliest Supermassive Black Holes with ALMA (IdEaS with ALMA)" (PTDC / FIS-AST / 29245 / 2017).

\item Javier Rom\'an acknowledges support from the State Research Agency (AEI-MCINN) of the Spanish Ministry of Science and Innovation under the grant "The structure and evolution of galaxies and their central regions" with reference PID2019-105602GB-I00/10.13039/501100011033.

\item Stephen J. Smartt acknowledges funding from STFC grant ST/S006109/1.

\item Aprajita Verma acknowledges funding from INAF-PRIN 1.05.01.85.08.

\item Aaron Watkins acknowledges support from the STFC grant ST/S00615X/1.

\end{itemize}


\newpage
\section{DDP Forum Community}

350 scientists registered on the \href{https://community.rubin-euclid-ddp.org/}{Rubin-Euclid DDP forum} throughout 2021:

\begin{tabular}{ 
p{\dimexpr0.25\textwidth-2\tabcolsep-\arrayrulewidth\relax} p{\dimexpr0.25\textwidth-2\tabcolsep-\arrayrulewidth\relax} p{\dimexpr0.25\textwidth-2\tabcolsep-\arrayrulewidth\relax} p{\dimexpr0.25\textwidth-2\tabcolsep-\arrayrulewidth\relax} }
Natasha Abrams & Christina Adair & Edward Ajhar & Viola Allevato \\
Yusra AlSayyad & Bruno Altieri & Adam Amara & Irham Taufik Andika \\
Stefano Andreon & Timo Anguita & James Annis & Philip Appleton \\
Bob Armstrong & Eric Aubourg & Herv\'e Aussel & Carlo Baccigalupi \\
Etienne Bachelet & Ivan Baldry & Michael Balogh & Eduardo Banados \\
Manda Banerji & Fernando Atrio Barandela & David Barrado & James Bartlett \\
Oliver James Bartlett & Franz Bauer & Keith Bechtol & Matthew R Becker \\
Victor J. S. Bejar & Charles Bell & Karim Benabed & Federica Bianco \\
Maciej Bilicki & Simon Birrer & Laura Bisigello & Alain Blanchard \\
Jonathan Blazek & Robert Blum & Hans Boehringrer & Micol Bolzonella \\
Angela Bongiorno & Jim Bosch & Maria Teresa Botticella & Alexandre Boucaud \\
Quentin Le Boulc'h & Dominique Boutigny & Herv\'e Bouy & Rebecca Bowler \\
Malcolm Bremer & Max Brescia & Hubert Bretonni\`ere & Jarle Brinchmann \\
Sarah Brough & Amandine Le Brun & James Buchanan & Fernando Buitrago \\
Patricia Burchat & Colin Burke & Remi Cabanac & Stefano Camera \\
Enrico Cappellaro & Karina Caputi & Carmelita Carbone & Jeff Carlin \\
Jon Carrick & Benoit Carry & Santiago Casas & Francisco Castander \\
Gianluca Castignani & C\'ecile Cavet & Stefano Cavuoti & Ranga Ram Chary \\
Nora Elisa Chisari & Aleksandra Ciprijanovic & Will Clarkson & Benjamin Cl\'ement \\
Johann Cohen-Tanugi & Thomas Collett & Chris Collins & Christopher Conselice \\
Asantha Cooray & Matteo Costanzi & Pau Tallada Cresp\'\i & Jean-Gabriel Cuby \\
Jean-Charles Cuillandre & Hubert Degaudenzi & Ian Dell Antonio & Anastasio D\'\i az-S\'anchez \\
Hugh Dickinson & Jose Diego & Joao Dinis & Sluse Dominique \\
Darko Donevski & Simon Peter Driver & Pierre-Alain Duc & Siegfried Eggl \\
Jose A. Escartin & St\'ephanie Escoffier & Maximilian Fabricius & R\'emi Fahed \\
Xiaohui Fan & Ginevra Favole & Anna Feltre & Annette Ferguson \\
Henry Ferguson & Angelo Ferrari & Pedro Ferreira & Ryan Foley \\
Adriano Fontana & Pablo Fosalba & Sotiria Fotopoulou & Dominique Fouchez \\
Chris Frohmaier & Hisanori Furusawa & Louis Gabarra & Ken Ganga \\
Raphael Gavazzi & Eric Gawiser & Bryan Gillis & Carlo Giocoli \\
Leo Girardi & Pedro Gomez-Alvarez & Ariel Goobar & Melissa Lynn Graham \\
Alister Graham & Sebastian Grandis & Ben Granett & Mikael Granvik \\
Philippe Gris & Daniel Gruen & Julia Gschwend & Axel Guinot \\
Leanne Guy & Luigi Guzzo & Nico Hamaus & Nina Hatch \\
Peter Hatfield & Stein V H Haugan & Katrin Heitmann & Sergio Miranda La Hera \\
Catherine Heymans & Hendrik Hildebrandt & Henk Hoekstra & Isobel Hook \\
Mike Hudson & Nuria Huelamo & Marc Huertas-Company & Markus Hundertmark \\
Leslie Hunt & Rodrigo Ibata & Olivier Ilbert & Stephane Ilic \\
Angela Iovino & \v{Z}eljko Ivezi\'{c} & Jens Jasche & Lynne Jones \\
Olivia Jones & Roelof de Jong & R\'emy Joseph & Arun Kannawadi \\
Vanshika Kansal & JJ Kavelaars & Heather Kelly & Lee Kelvin \\
Somayeh Khakpash & Martin Kilbinger & Tom Kitching & Gijs Verdoes Kleijn \\
Leon Koopmans & Angelica Kovacevic & Martin Kuemmel & Konrad Kuijken \\

\end{tabular}

\begin{tabular}{
p{\dimexpr0.25\textwidth-2\tabcolsep-\arrayrulewidth\relax} p{\dimexpr0.25\textwidth-2\tabcolsep-\arrayrulewidth\relax} p{\dimexpr0.25\textwidth-2\tabcolsep-\arrayrulewidth\relax} p{\dimexpr0.25\textwidth-2\tabcolsep-\arrayrulewidth\relax} }
Clotilde Laigle & Ariane Lan\c{c}on & Hermine Landt & Francois Lanusse \\
S{\o}ren Larsen & Massimiliano Lattanzi & Ren\'e Laureijs & Guilhem Lavaux \\
Florent Leclercq & Bomee Lee & Louis Legrand & Danielle Leonard \\
Giorgio Lesci & Shun-Sheng Li & Shuang Liang & Kian-Tat Lim \\
Yen-Ting Lin & Anja von der Linden & Xin Liu & Nicolas Lodieu \\
Cristina Martinez Lombilla & Chris Lovell & Gabriella De Lucia & Georgios Magdis \\
Manuela Magliocchetti & Guillaume Mahler & Constance Mahony & Elisabetta Maiorano \\
Alex Malz & Rachel Mandelbaum & Bob Mann & Luis Manuel \\
Claudia Maraston & Lucia Marchetti & Ole Marggraf & Phil Marshall \\
Eduardo Martt\'\i n & Nicolas Martinet & Richard Massey & Daniel Masters \\
Matteo Maturi & Ben Maughan & Alan McConnachie & Henry Joy McCracken \\
Julie McEnery & Sean McGee & Simona Mei & Peter Melchior \\
Jean-Baptiste Melin & Yannick Mellier & Emiliano Merlin & Ben Metcalf \\
Hironao Miyatake & Joseph Mohr & Michele Moresco & Alberto Moretti \\
Daniel Mortlock & David Mota & Suvodip Mukherjee & Reiko Nakajima \\
Gautham Narayan & Christian Neissner & Jeff Newman & Luciano Nicastro \\
Ignacio Sevilla Noarbe & Mario Nonino & Dara Norman & Pascal Oesch \\
Florian Pacaud & Cristobal Padilla & Mat Page & Jorge Carretero Palacios \\
Eliana Palazzi & Stephane Paltani & Maurizio Paolillo & Francisco Paz-Chinch\`on \\
Reynier Peletier & Roser Pello & Antonio Perez & Vincenzo Petrecca \\
Valeria Pettorino & Francesco Piacentini & Sandrine Pires & Alice Pisani \\
Jennifer Pollack & Mikko P\"ontinen & Lucia Pozzetti & Andy Ptak \\
Markus Rabus & Alvise Raccanelli & Mario Radovich & Troy Joseph Raen \\
Maria Angela Raj & Thomas Reiprich & Marina Rejkuba & Jason Rhodes \\
Marina Ricci & Hans-Walter Rix & Brant Robertson & Santi Roca-F\`abrega \\
Benjamin Rose & Cyrille Rosset & Martin Sahl\'en & Ziad Sakr \\
Eusebio Sanchez & Alex Saro & Barbara Sartoris & Marc Sauvage \\
Roberto Scaramella & Claudia Scarlata & Mischa Schirmer & Sam Schmidt \\
Morgan A. Schmitz & Michael Schneider & Tim Schrabback & Meg Schwamb \\
Diana Scognamiglio & Aidan Sedgewick & Mauro Sereno & Stephen Serjeant \\
Francesco Shankar & Yue Shen & Raphael Shirley & Marko Shuntov \\
Stephen Smartt & Graham Smith & Colin Snodgrass & Enrique Solano \\
Alessandro Sonnenfeld & Jenny Sorce & Spencer Stanford & Daniel Stern \\
Veronica Strazzullo & Rachel Street & Robert Szabo & Margherita Talia \\
Charling Tao & Dan Taranu & Andy Taylor & WeiLeong Tee \\
Matthew Temple & Malte Tewes & Sune Toft & Francesc Torradeflot \\
Crescenzo Tortora & Michael Troxel & Ignacio Trujillo & Eleni Tsaprazi \\
Isaac Tutusaus & Chris Usher & Simona Vegetti & Hector Manuel Velazquez \\
Aprajita Verma & Willem-Jan Vriend & Nicholas Walton & Benjamin Wandelt \\
Feige Wang & Aaron Watkins & Arjen van der Wel & Niraj Welikala \\
Martin White & Imogen Whittam & Klaas Wiersema & Vivienne Wild \\
Roy Williams & Gerard Williger & Angus H Wright & Stijn Wuyts \\
Jinyi Yang & Guang Yang & Ilsang Yoon & Mijin Yoon \\
Weixiang Yu & Andrea Zacchei & Gianni Zamorani & Yuanyuan Zhang \\
Elena Zucca & & &

\end{tabular}

\newpage
\section{Charter for the Euclid/Rubin Derived Data Products Working Group} \label{sec:ddpcharter}

\textbf{Date}: Version May 29, 2020 

\textbf{Approved by}: The Euclid Consortium Board and The Vera C. Rubin Observatory Director

\textbf{Key Concepts}:
\vspace{-0.4cm}
\begin{itemize}[noitemsep]
\item Both the Rubin and the Euclid science communities would benefit from the Rubin and Euclid datasets being jointly processed to produce shared “derived data products” (DDPs).
\item A DDP-WG should recommend an initial set of DDPs, which would be shared promptly and simultaneously with both the Euclid Consortium and all Vera Rubin Observatories data rights holders (the “LSST Science Community”) for scientific use in a way that protects the unique science of each collaboration.
\item The initial set of DDPs, if approved, should form the basis of a Letter of Intent signed by both Rubin and Euclid leadership to create the DDPs.
\item The DDP-WG should be a standing committee that can recommend revisions to DDPs or further DDPs as both the Euclid and Rubin survey progress.
\item The DDP-WG is not the group that will decide who makes the DDPs, where they are made, how they are made, or what funding mechanism shall pay for that effort. This group should not focus on issues of data rights or potential scientific collaborations between Rubin/Euclid.
\end{itemize}

\textbf{Composition:} The DDP-WG will consist of an equal number of representatives from the Vera Rubin Observatory data rights holders (hereafter Rubin) and the Euclid Consortium (hereafter Euclid). There will be about 10 people from each set of data rights holders on the DDP-WG plus several ex officio members from project leadership on each side. DDP-WG members could be, but need not be, data rights holders for both Euclid and Rubin; DDP-WG members should represent the interests of Euclid or Rubin, depending on who nominated them to the DDP-WG; the members of the DDP-WG should broadly represent their respective projects in terms of science and data processing expertise. The DDP-WG should select two co-chairs, one representing Euclid and one representing Rubin. The Eucli members of the DDP-WG will be selected by the Euclid Consortium Board (ECB) based on nominations
from the Euclid Consortium Lead. The Rubin members of the DDP-WG will be selected by the Rubin Observatory Director based on consultations with the Rubin Science Advisory Committee. Each consortium can individually decide how to fill vacancies or whether to rotate their members of the DDP-WG.

\textbf{Charge:} The DDP-WG should help plan a virtual or in-person workshop that is open to interested Rubin and Euclid data rights holders. This workshop should be focused on gathering community input into the desired initial DDPs. Based on the input from that meeting, the DDP-WG will:
\vspace{-0.4cm}
\begin{itemize}[noitemsep]
\item Design an initial set of DDPs that could be shared promptly and simultaneously with both the Euclid Consortium and the LSST Science Community for scientific use, in a way that protects the unique science of each collaboration and is consistent with both communities’ data policies.
\item Outline the scientific justification and quantify, approximately, its impact for each proposed DDP.
\item Issue an initial set of recommendations within 9 months of the creation of the DDP-WG; these recommendations would be made to the ECB and Rubin Observatory Director.
\item Set a cadence for virtual and in person meetings and workshops that they feel is consistent with developing recommendations for revised or new DDPs and then make those recommendations to the ECB and Rubin Observatory Director.
\item Gather input from their respective communities about desired DDP. Focus only on designing DDPs, and not on issues of DDP creation or forming inter-project science collaborations.
\end{itemize}

\textbf{DDP creation:} The DDP-WG reports to and recommends DDPs to the ECB and Vera Rubin Observatory Director for approval. If approved, the respective consortia will then have to come to an eventual agreement about where, by whom, on what time scale, how, and with what funding the DDPs will be created.

\end{document}